\documentclass[12pt,a4paper]{article}
\usepackage{graphicx}
\usepackage{amsmath,amsfonts,amssymb}
\def\RE{\ensuremath{\mathop\mathrm{Re}\nolimits}}
\def\RT{\ensuremath{\RE_{\rm t}}}
\def\RG{\ensuremath{\RE_{\rm g}}}
\def\RC{\ensuremath{\RE_{\rm c}}}
\def\BF{\begin{figure}}
\def\EF{\end{figure}}
\def\BC{\begin{center}}
\def\EC{\end{center}}
\let\TW=\textwidth
\title{On the decay of turbulence\\
in plane Couette flow}
\author{Paul Manneville\\
\normalsize Hydrodynamics Laboratory,\\
\normalsize \'Ecole Polytechnique,
Palaiseau 91128, France}

\date{\small Accepted for publication in Fluid Dynamics Research}

\begin{document}

\maketitle
\sloppy

\begin{abstract}
Upon decreasing the Reynolds number, plane Couette flow first forms alternately turbulent and laminar oblique bands out of featureless turbulence below some upper threshold \RT.
These bands exist down to a global stability threshold \RG\ below which laminar flow ultimately prevails.
We study the fragmentation and decay of these bands in systems that are extended enough for several bands to exist.
We use direct numerical simulations appropriately tailored to deal with such large systems during long enough durations.
We point out a two-stage process involving the rupture of a band and next its slow
shrinking.
Previous interpretations of turbulence decay
in wall-bounded flows within the {\it chaotic transient\/} or {\it spatiotemporal intermittency\/} paradigms are discussed.      
\end{abstract}

\vspace{2pc}
\noindent{\it Keywords}: wall-bounded flow, turbulence decay,
laminar-turbulent coexistence

\section{Introduction\label{sec1}}

The transition to turbulence in wall-bounded flow is a long-standing problem of great theoretical and practical importance. Specific difficulties arise from its hysteretic character linked to the fact the nontrivial turbulent regime stands at a distance from the laminar regime and may coexist with it.
On general grounds, laminar flow prevails as the unique permanent regime, i.e. in the long-time limit and whatever the initial state, below some value  \RG\ of the Reynolds number \RE\ called the {\it global stability\/} threshold.
On the other hand, a {\it linear instability\/} threshold \RC\  can usually be obtained from standard stability analysis, beyond which the base profile is unconditionally unstable against infinitesimal perturbations.
Both the laminar (trivial) base flow and some nontrivial turbulent regime can coexist spatially in the interval $[\RG,\RC]$.
The flow through a straight pipe of circular section and the plane shear flow between two parallel plates moving in opposite directions (plane Couette flow, PCF for short)  represent extreme situations since the corresponding laminar base flow profiles are linearly stable for all \RE, i.e. $\RC=\infty$. A interesting recent review of current issues can be found in (Mullin \& Kerswell, eds., 2005).
Here we shall study PCF considered as a paradigmatic case of wall-bounded flow with potential relevance to less academic systems such as channel or boundary-layer flows.

Parameters characterising the experiment are the distance $2h$  between the plates driving the flow (wall-normal direction $y$) and the in-plane dimensions $L_x$ and $L_z$ in the streamwise and spanwise directions, respectively.
The physical properties of the fluid under shear are all contained in its kinematic viscosity $\nu$ and the Reynolds number is defined as $\RE=Uh/\nu$, where $U$ is the speed of the plates.
Laboratory experiments were initially focussed on the determination of \RG\ (Bottin, 1998; Bottin \& Chat\'e, 1998; Bottin {\it et al.}, 1998) in systems of aspect ratio $L_{x,z}/2h$ moderate to large.
\BF
\BC
\includegraphics[width=0.55\TW,clip]{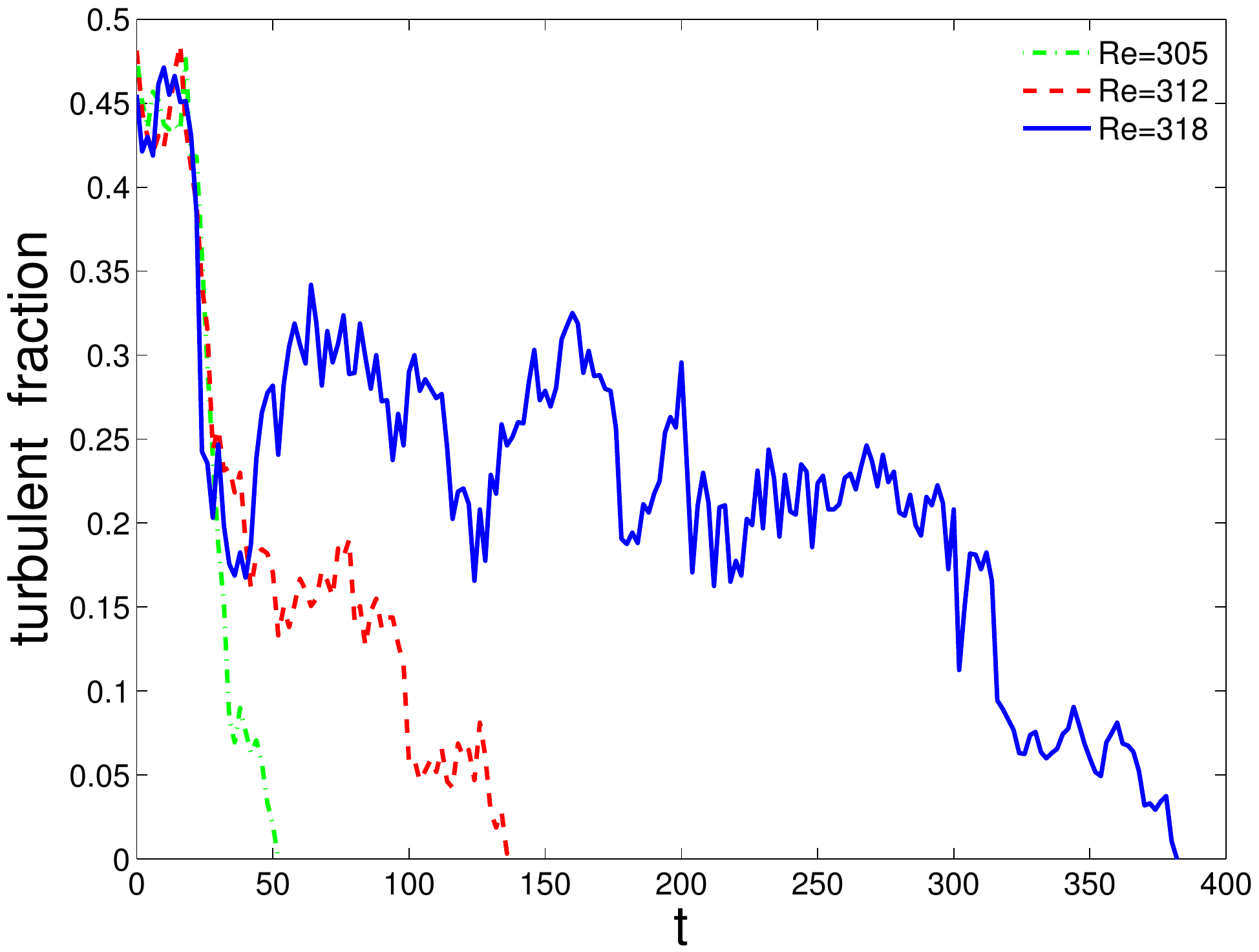}
\hskip0.5em
\includegraphics[angle=90,width=0.25\TW,clip]{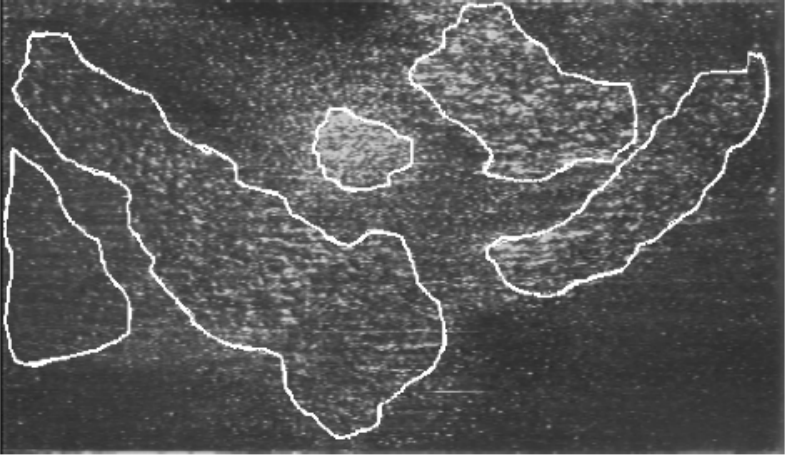}\\[3ex]
\includegraphics[width=0.70\TW,clip]{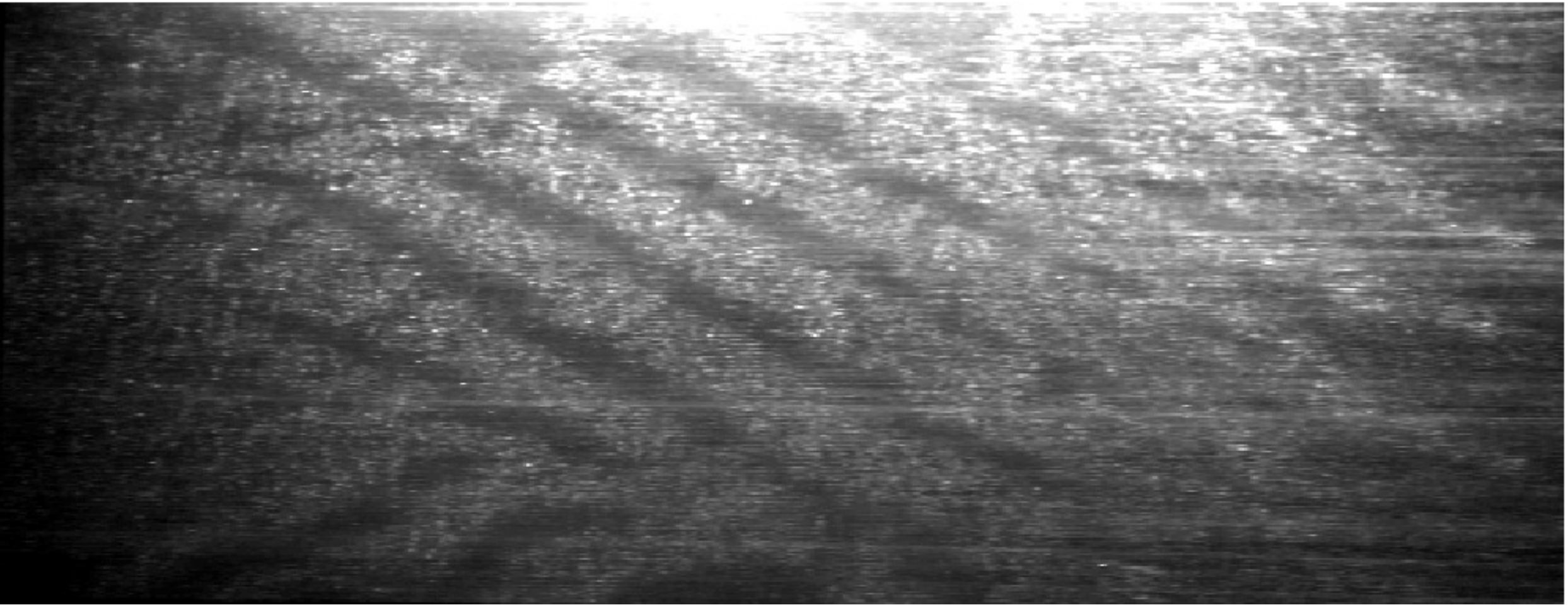}\\[1ex]
\includegraphics[width=0.70\TW,clip]{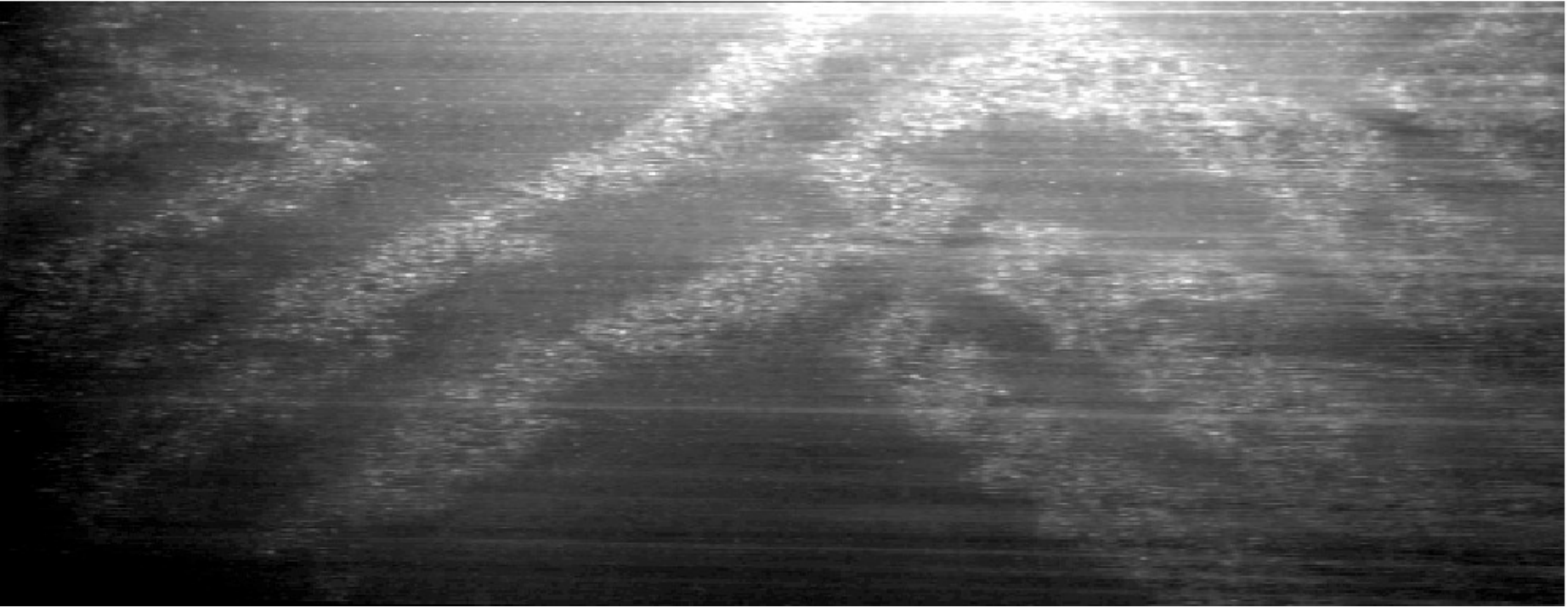}
\EC
\caption{Top: Experimental results by Bottin (1998).
Left: turbulent fraction as a function of time in experiments where turbulent flow is quenched down to the value indicated in the legend in a system of size $L_x=286h$, $L_z=72h$, $2h=7\>$mm; time $t$ is given in seconds.
Right: snapshot of a transient solution at $\RE=320$ in a wider system with $L_x=572h$, $L_z=145h$, $2h=3.5\>$mm, indicating the region detected as turbulent. The streamwise direction is vertical.
Original data and image courtesy of S.~Bottin.
Bottom: Prigent's very large aspect ratio experiment  ($L_x=770 h$, $L_z=340 h$, gap $2h=1.5\>$mm). In the middle of the transitional range ($\RE=376$, top) a near-ideal pattern of oblique bands is obtained but closer to \RG\, ($\RE=331$, bottom),  bands break down. The streamwise direction is horizontal. Images courtesy of A.~Prigent.\label{fig15}}
\EF
Figure~\ref{fig15}  (top, left) illustrates turbulence decay observed in experiments where a flow prepared at very high \RE\ is quenched down to some final $\RE\lesssim\RG$, using time series of the {\it turbulent fraction}, the relative surface of the region occupied by turbulence. Extensive statistics were gathered on the transients' lifetimes which were found to be roughly exponentially distributed at given \RE, with mean lifetimes increasing rapidly as \RE\ approached $\RG\simeq325$ from below.
Irregular patches with locally oblique shape were observed, see Fig.~\ref{fig15} (top, right).
Later experiments  (Prigent, 2001; Prigent {\it et al.}, 2003; for a review see Prigent \& Dauchot, 2005) showed that, at larger aspect ratio, an upper threshold $\RT\simeq410$  could be defined, above which turbulence was essentially featureless and below which it  was present in the form of alternately laminar and turbulent oblique bands.
Upon decreasing \RE, the upper part of transitional range $[\RG,\RT]$ was marked by a continuous increase of the oblique modulation of the turbulence intensity  (Fig.~\ref{fig15}, centre).
In the lower part, the bands were fragmented and less regularly organised (Fig.~\ref{fig15}, bottom).
PCF is the zero-curvature limit of counter-rotating  cylindrical Couette flow where the oblique bands take the form of spirals (Coles \& Van Atta, 1967) and have essentially the same characteristics with differences attributed to the fact that streamwise boundary conditions are periodic in the azimuthal direction.
Most of the quantitative results were obtained in the cylindrical case, while band formation was studied in PCF upon decreasing \RE\ by steps, taking snapshots of the pattern, and just measuring wavelengths (Prigent, 2001, p.~75--76).
In particular the decay of bands around \RG, which was not studied in detail, will be our main concern here.

Besides experiments of the type sketched above, many theoretical and numerical studies have been devoted to the decay of turbulence in wall-bounded flows, often using domains of the size of  the  {\it minimal flow unit\/} (MFU), the smallest domain in which turbulence can be sustained at low to moderate Reynolds number (Jim\'enez \& Moin, 1991).
MFU-sized domains have periodic boundary conditions in the direction(s) perpendicular to the wall at distances of the order of the relevant wall-normal length scale (tube's diameter, distance between plates, boundary layer thickness).
This approach has shed light on the general mechanisms sustaining the turbulent regime, with streamwise vortices generating streaks by lift-up, and the instability of streaks feeding back the vortices, see Hamilton, Kim \& Waleffe (1995) or Waleffe (1997).
In the transitional range,  the structure of the nontrivial state remains highly coherent at small scale owing to the moderate value of the Reynolds number.
The flow regime that develops is then not developed turbulence but a milder chaotic evolution which is in principle describable in terms of {\it low dimensional dynamical systems\/}, allowing the use of concepts and tools of {\it deterministic chaos\/} theory.
In this context the relevant objects are {\it unstable periodic orbits\/} that are special exact solutions of the Navier--Stokes equations in the corresponding MFU (Nagata, 1990; Waleffe, 2003; Toh \& Itano, 2003; Cvitanovi\'c \& Gibson, 2010;\dots), and the {\it homoclinic tangle\/} formed from the invariant manifolds attached to them.
Within this framework, the decay of turbulence is described as a {\it chaotic transient\/} for which a lifetime can be defined and the probability distribution of lifetimes is found to be exponentially decreasing (Eckhardt \& Faisst, 2005; Eckhardt {\it et al.}, 2008; Schneider {\it et al.}, 2010;\dots).

Up to now, it has however turned difficult to establish a clear correspondence between laboratory experiments and numerical simulations analysed in the dynamical systems setting.
This is because the spatial coherence introduced by periodic boundary conditions at small distances used in simulations supports the reduction to {\it low-dimensional\/} dynamical systems and the {\it temporal\/} chaos interpretation, but stays at odds with the physical situation that corresponds to flow domains {\it extended\/} in one dimension (pipe or duct flow) or two dimensions (PCF, channel and boundary layer flow), situating the transitional problem in a fully {\it spatiotemporal\/} perspective.
Puff intermittency in pipe flow (Moxey \& Barkley, 2010; Barkley, 2011)  and oblique turbulent patterning in PCF are manifestations of the spatiotemporal processes at work.
Like in the theory of critical phenomena in thermodynamic phase transitions (e.g. Stanley, 1978), the effective space dimension is expected to play an important role.
For example, weakly turbulent, streamwise localised, equilibrium structures with definite global structure called {\it puffs\/} are known to control turbulence decay in the pipe-flow case.
They do not seem to have any counterpart  in PCF where turbulent spots are always found to experience wild streamwise and spanwise shape fluctuations without apparent overall structure except for a tendency to obliqueness.
Also, no equivalent of the puff-slug transition in pipe flow  (Duguet {\it et al.}, 2010b; Barkley, 2011) has yet been identified in PCF and other 2D cases.

The oblique band regime in PCF is less well documented than the puff regime in pipe flow. 
It was numerically studied by Barkley \& Tuckerman (2005a,b, 2007) who recovered it by performing simulations in periodic domains elongated in the direction normal to the bands but narrow (a few MFU) in the complementary in-plane direction.
Though this approach gives a reliable account of streamwise correlations essential to the band formation (Philip \& Manneville, 2011), it has yet not given much information about the decay mechanisms. 
Others, in particular Duguet {\it et al.} (2010a), considered periodic extended domains of sizes of the order of Prigent's experimental set-up and performed fully resolved numerical simulations in quantitative agreement with the experiments. The corresponding computational load was however an obstacle to the consideration of the many configurations relevant to decay.
Our recent work (Manneville \& Rolland, 2010) showed that band formation is a highly robust phenomenon that withstands drastic resolution decrease, reliably reproducing the wavelength variations and qualitatively preserving most of the features of the flow, apart from a general but tolerable downward shift of the transitional range.
Here, we take advantage of this result to study domains sufficiently wide to fit several wavelengths of the oblique band pattern in long-lasting simulations but at moderate computational cost, expecting a realistic rendering of the breakdown of bands not targeted before.
Section~\ref{s2} below is devoted to the numerical conditions of the experiments, in particular the preparation of initial conditions, and the post-treatment.
Section~\ref{s3} describes our main result that band collapse involves two basic processes, first the formation of a  laminar gap breaking a band (\S\ref{s3.2}) and next a slow retreat of the band fragments left by the breaking (\S\ref{s3.3}).
The discussion in \S\ref{s4} begins with a review of the theoretical concepts introduced to account for the decay of turbulence and next situates our findings within this framework, showing that they give a reasonable understanding of experimental observations.

\section{Preliminaries\label{s2}}

\subsection{Numerical considerations}

We perform direct numerical simulations of the Navier--Stokes equations using Gibson's public domain code {\sc ChannelFlow} (Gibson, 2010).
It implements a pseudo-spectral scheme using Chebyshev polynomials in the wall-normal direction $y$ where no-slip boundary conditions are applied at $y=\pm1$ in units of  $h$, the half-gap between the plates driving the flow.
Fourier modes  are used to deal with the space dependance in the streamwise ($x$) and spanwise ($z$) directions with periodic boundary conditions at distances $L_x$ and $L_z$, again expressed in units of $h$.
The speed $U$ of the counter-translating plates serves to define the turn-over time $h/U$ as the time unit.

Preliminary work (Manneville and Rolland, 2010) has shown that the main features of transitional PCF are appropriately reproduced in under-resolved simulations, with apparently just a downward shift of the Reynolds number range where the bands are observed.
As can be anticipated for wall-bounded flows at moderate Reynolds numbers, the most significant effect of resolution lowering lies in the account of the wall-normal dependence, hence the number of Chebyshev polynomials used.
We showed that this number can be drastically reduced from 27--33, which is usually considered as sufficient in the transitional range $\RE\sim 300$--400 (Duguet {\it et al.} 2010a), down to 11 without losing the transitional range in extended geometry.
The price to be paid was a lowering of the interval $[\RG,\RT]$ from $[325,410]$ in experiments (Prigent {\it el al.} 2003) or well-resolved simulations (Duguet {\it et al.} 2010a) down to $\sim[210,270]$ for $N_y=11$. Increasing $N_y$ was shown to improve the situation readily and the value $N_y=15$ was pointed at as a good choice.
On the other hand, in the low-\RE\ range of interested here, the flow is highly coherent and rather smooth at the scale of the MFU, with size $\ell_x=12.8\,h$, $\ell_z=4.2\,h$ according to Waleffe (2003).
Taking effective space steps $\delta x'=1.5$ and $\delta z'=0.5$ makes about 8.5 points in each direction to describe fluctuations at the MFU scale. After aliasing removal according to the 2/3 rule, this gives $\delta x=1$ and $\delta z=1/3$ for the evaluation of the nonlinear terms in the pseudo-spectral scheme.
A good compromise between acceptable rendering of the  transitional range  and moderate computational load at large aspect ratio was thus suggested to be $N_y=15$, $N_x=L_x$, and $N_z=3L_z$, shifting $[\RG,\RT]$ down to $[\approx275,\approx350]$, below experimental observations by 15\% only, while preserving most features of the banded state qualitatively, in particular the wavelength selection properties.

The fact that a lowered resolution renders the transition reasonably well seems due to the fact that the self-sustaining process (Waleffe, 1997) involves a small number of wall-normal modes. An appropriate account of streamwise correlations in the featureless turbulent regime beyond \RT\  is however needed in view of a realistic restitution of the bands (Philip \& Manneville, 2011).
This is testified by the failure of the too drastic wall-normal modelling described in (Lagha \& Manneville, 2007a) which happens to be corrected upon slight improvement as shown in (Manneville \& Rolland, 2010).  

Numerical experiments discussed here have been performed with $L_x=432$ and $L_z=256$, which is the largest size that could easily be managed with our desk-top computer. Simulation during one time unit ($h/U$) takes about 1~min.
Knowing from experiments that the pattern's wavelengths are $\lambda_x\simeq110\,h$ and $\lambda_z\approx85\,h$ around \RG\  decreasing to about $40\,h$ around \RT, the values taken for $L_{x,z}$ lead us to expect a pattern with about three wavelengths close to \RG.
It can be argued that, if curvature effects could be neglected, using periodic boundary conditions would make the exactly counter-rotating CCF even closer to our computational implementation than actual PCF.
This would not change our expectation much since in the azimuthal ($\theta$, streamwise) and axial ($z$, spanwise) directions, the wavelengths vary by steps from $\lambda_\theta=100\,h$ and $\lambda_z=40\,h$ close to \RT\ to $155\,h$  and $70\,h$ around \RG, respectively (Prigent {\it et al.}, 2003). 

\subsection{Preparation of the initial state}

The main purpose of the paper is to study how turbulent bands decay in a large-aspect-ratio system.
The system is thus prepared in the featureless regime at $\RE=450$.
The Reynolds number is then decreased by steps down to the range of interest around $\RG\approx275$.
At each step, simulations are performed during limited time lapses.
For $\RE\ge\RT\approx350$, the flow is still featureless.
Figure~\ref{fig1} (left) displays the state obtained close to \RT\  at $\RE=340$ in the intermittent
\BF
\BC
\includegraphics[angle=90,width=0.18\TW,clip]{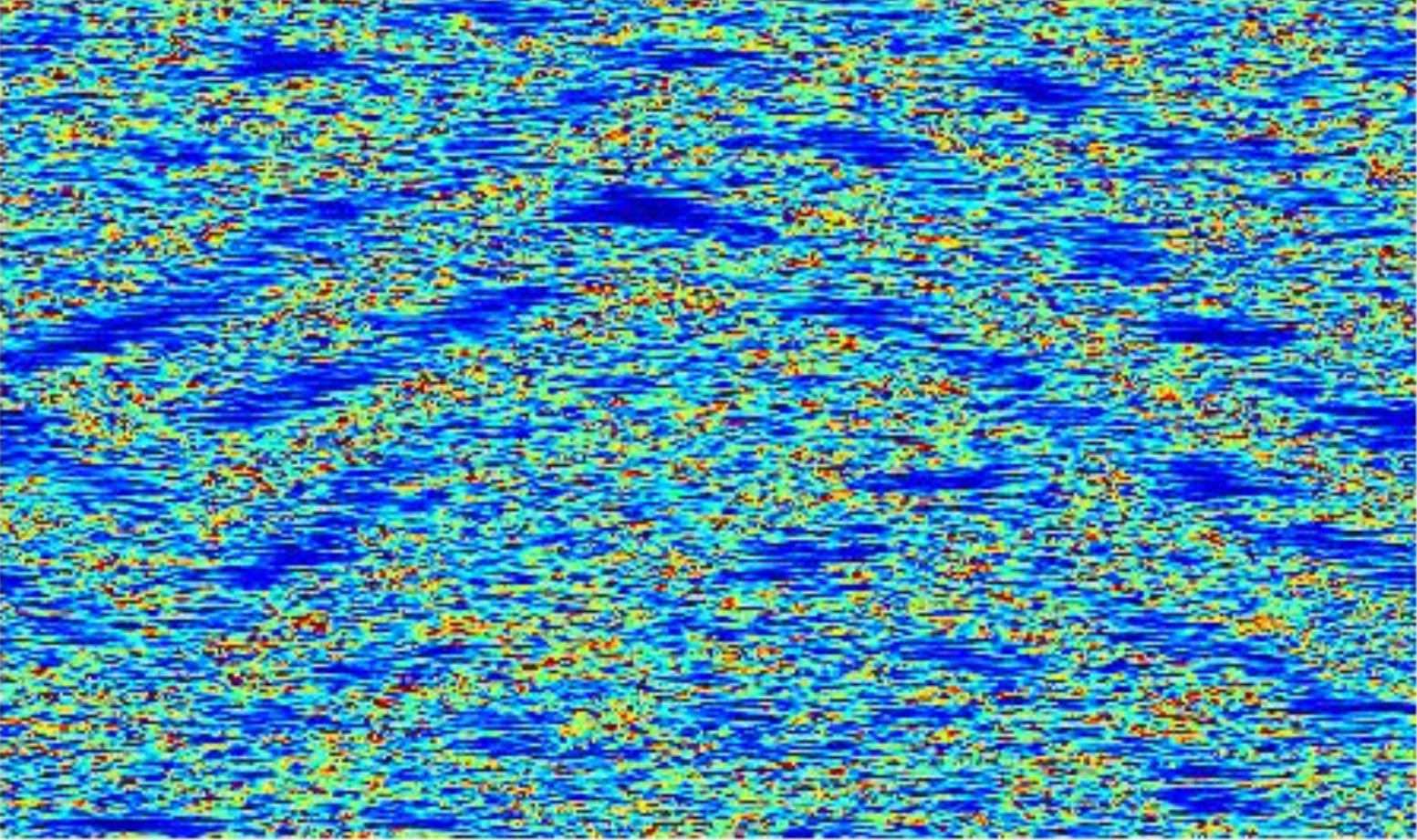}
\hskip0.5em
\includegraphics[angle=90,width=0.18\TW,clip]{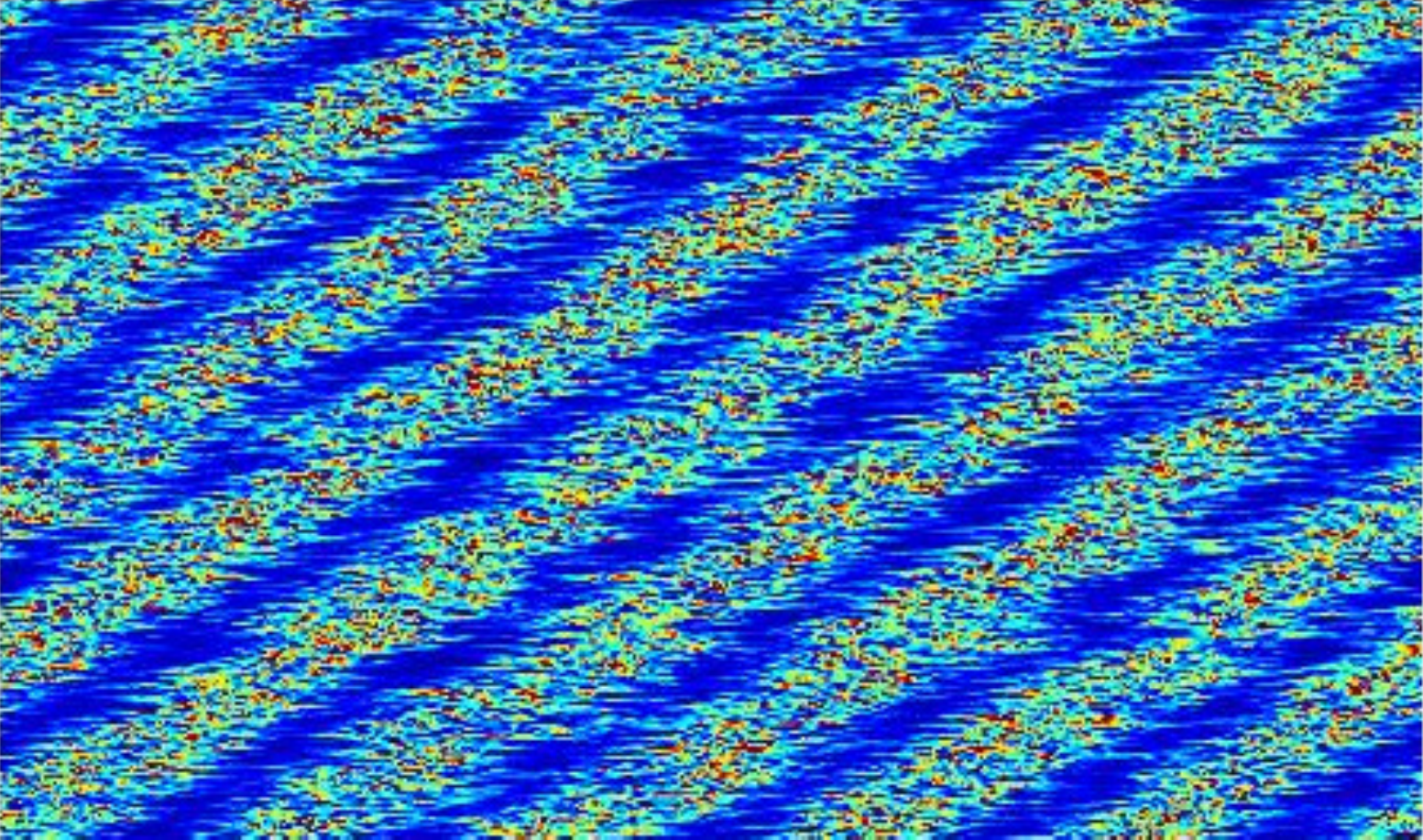}
\hskip0.5em
\includegraphics[angle=90,width=0.18\TW,clip]{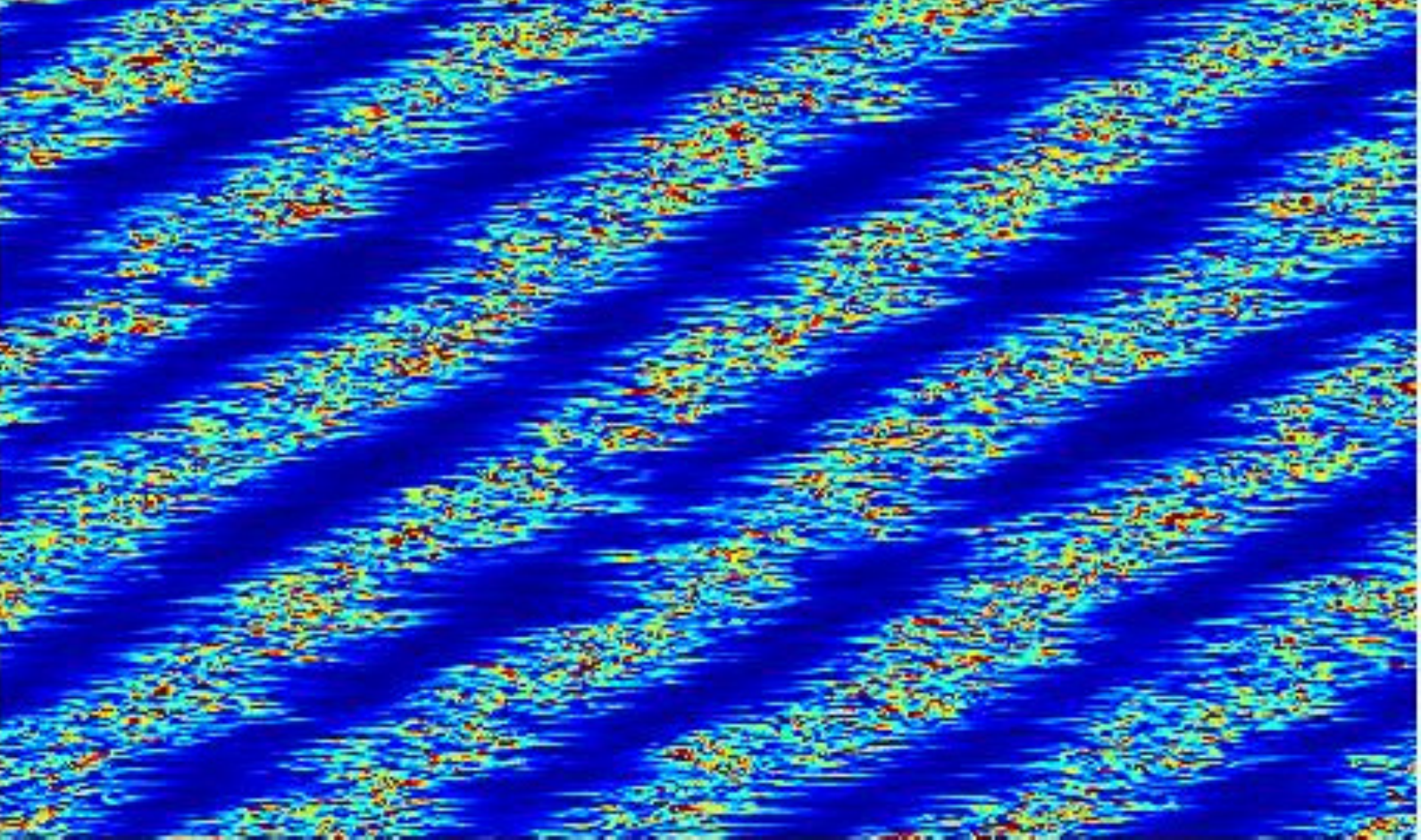}
\hskip0.5em
\includegraphics[angle=90,width=0.18\TW,clip]{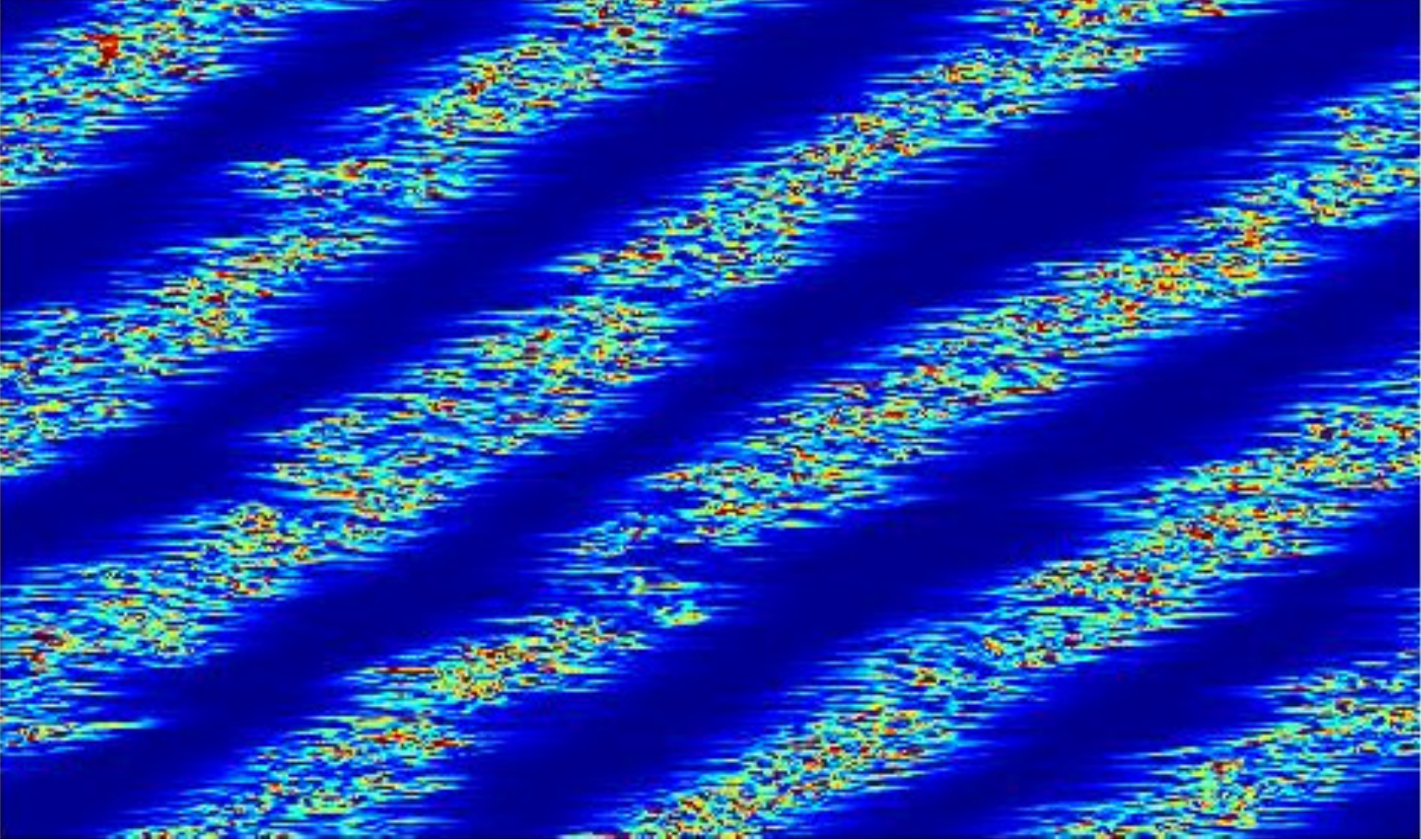}
\EC
\caption{Flow patterns  (colour online) obtained in the simulation at $\RE=340$, $t=20,\!000$ (left), $\RE=320$, $t=12,\!000$ (centre-left), $\RE=290$, $t=18,\!000$ (centre-right), and $\RE=280$, $t=12,\!000$ (right) represented using $e(x,z,t)$.
The streamwise direction is along the vertical axis, here and henceforth.
$L_x=432$, $L_z=256$, $N_x=432$, $N_y=15$, $N_z=768$.
Colour scale from 0 (blue) to 0.1 (red); in each figure $e(x,z,t)$ peaks to above 0.2 whereas,  from left to right, the means vary from 0.0363 to 0.0175. \label{fig1}}
\EF
regime (Prigent {\it et al.}, 2003; Barkley \& Tuckerman, 2005a,b).
Here, and in most of the figures showing patterns, we display the local perturbation energy field averaged over the thickness
$$
e(x,z,t)=\frac12\int_{-1}^{+1}
\mathrm d y\,\left[\mbox{$\frac12$}\mathbf{\tilde v}^2\right],
$$
where $\mathbf{\tilde v}=\mathbf v - y\,\mathbf{\hat x}$, $\mathbf v$ being the total velocity field and $y\,\mathbf{\hat x}$ the laminar flow. 
Averaging over the thickness neglects subtleties linked to laminar flow overhanging turbulent flow or the reverse at the laminar-turbulent interface (Coles \& van Atta, 1967; Barkley \& Tuckerman, 2007) but this does not change the qualitative conclusions.
At $\RE=340$, the two orientations allowed by symmetries are represented and the wavelengths corresponding to peaks in the power spectrum of $e(x,z,t)$ are $\lambda_x=86.4$ and $108$ and $\lambda_z=42.7$ and $51.2$, in good agreement with values expected close to \RT.
Farther away from \RT\ at $\RE=320$, the pattern is now well organised (Fig.~\ref{fig1}, centre-left) and locked on $\lambda_x=L_x/5=86.4$, and $\lambda_z=L_z/5=51.2$.
When \RE\ is further decreased, after a stage of profound reorganisation following Eckhaus-like instabilities and the subsequent elimination of dislocations, the system gets locked upon patterns with fewer wavelengths, first four bands at $\RE=290$ (Fig.~\ref{fig1}, centre-right), and next three bands at $\RE=280$.
Whereas in the latter case the value $\lambda_z=85.5$ agrees with experimental findings for PCF, $\lambda_x=144$ is somewhat larger than what might be expected but of the same order of magnitude as what is found in CCF.
Such slight discrepancies are not surprising in view of the commensurability conditions  imposed by the periodic boundary conditions fixing the dimensions of the system and the associated resonant locking of the wavelengths achieved in the pattern.
The Reynolds number is then decreased to $\RE=275$ for which the three-band pattern  obtained seems sustained, as far as one can tell from a simulation of finite duration (here $t_{\rm max}=25,\!000\,h/U$).

\subsection{Post-treatment of turbulent-laminar coexistence via filtering and thresholding\label{sft}}
As can be seen in Figure~1, the transitional regime is characterised by the coexistence of regions that are alternately turbulent and laminar. The identification of these two possible local states is performed by thresholding after filtering. Here the discrimination uses $e(x,z,t)$.
We take a Gaussian filter function in the form $G_\Delta(r)=(6/\pi\Delta^2)^{1/2} \exp(-6r^2/\Delta^2)$ yielding the transfer function $\hat G_\Delta(k)=\exp(-k^2\Delta^2/24)$ which has the same second moment as the box filter of width $\Delta$ (Pope, 2000, p.~561ff.).
To take into account the difference in coherence along the streamwise and spanwise directions, we choose an effective filtering box which is rectangular and scales as the size of the domain for which an exact nontrivial solution was found by Waleffe (2003), since such coherent states are expected to be the smallest structures relevant in the low-\RE\ spatiotemporal context we are dealing with. We thus take $\Delta_x=12.8\,\kappa$ and $\Delta_z=4.2\,\kappa$, where $\kappa$ is a factor of order one to be specified (usually larger than 0.5 and smaller than 4).
Several averaged quantities can then be computed, conditioned by the fact that the region is considered either {\it turbulent\/} or {\it laminar}, according to whether the local filtered perturbation energy is {\it larger\/} or {\it smaller\/} than some cut-off value $e^{\rm c}$. The discussion of the filtering and thresholding problems is straightforward.
\BF
\BC
\includegraphics[angle=90,width=0.18\TW,clip]{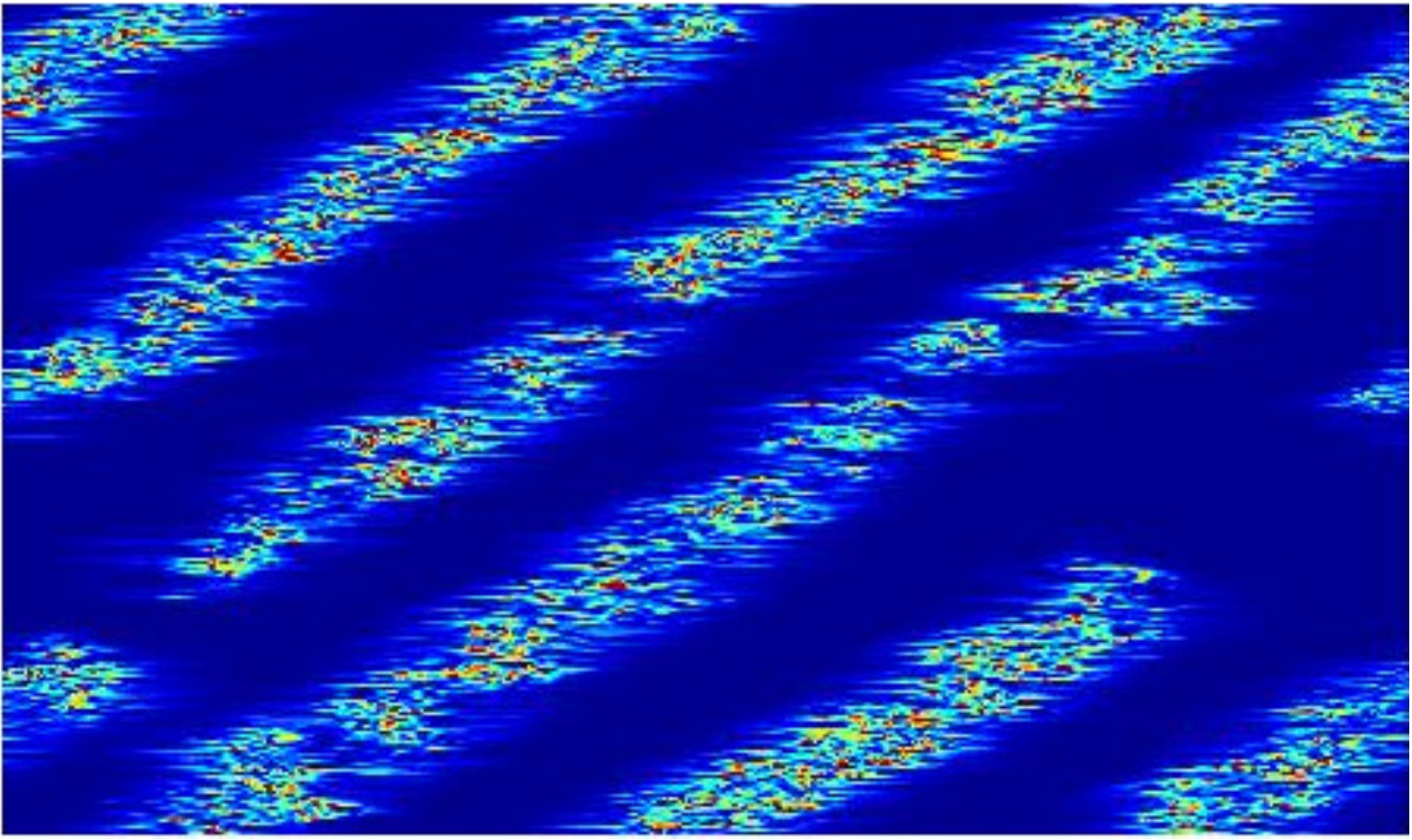}\\[2ex]
\includegraphics[angle=90,width=0.18\TW,clip]{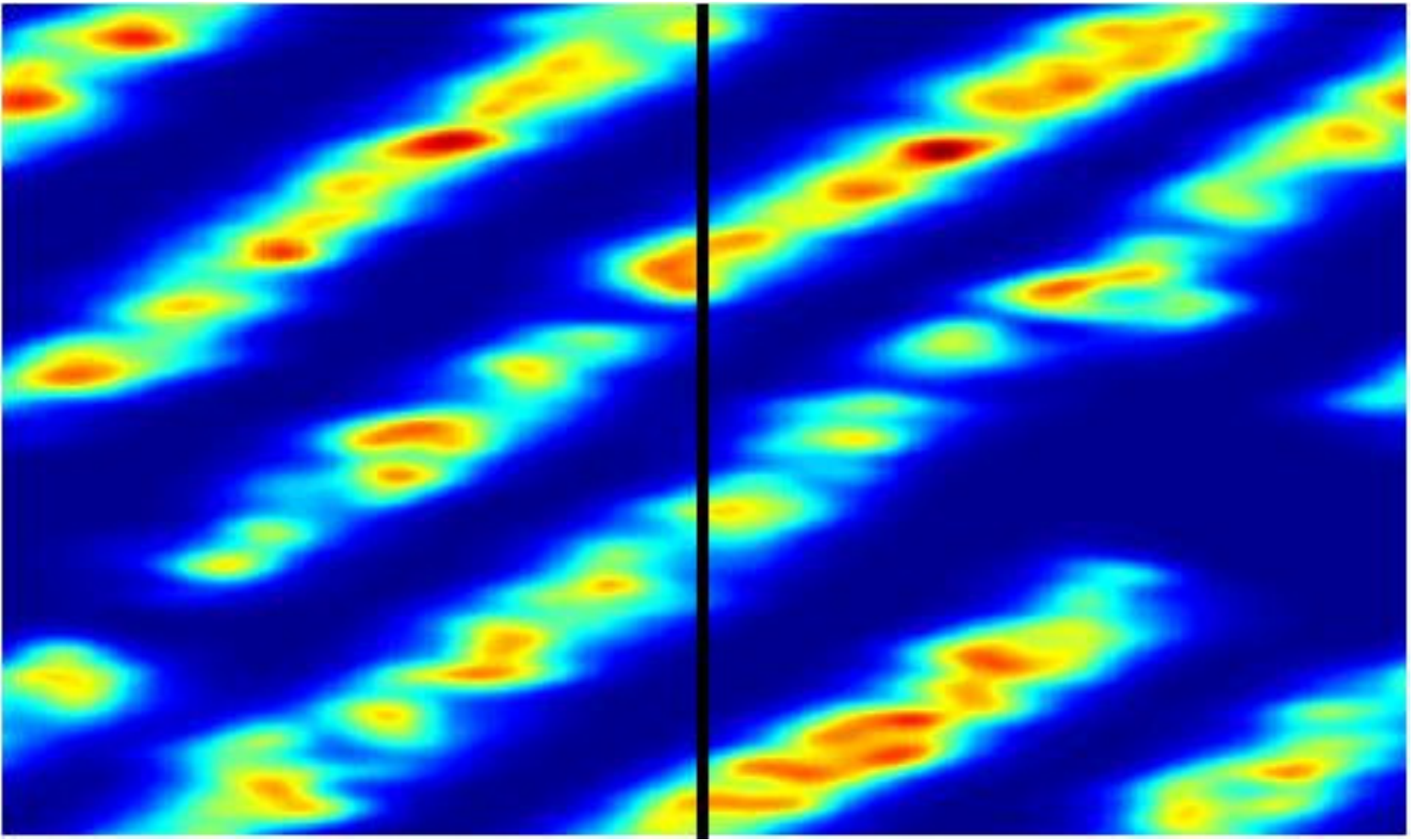}
\hskip3em
\includegraphics[angle=90,width=0.18\TW,clip]{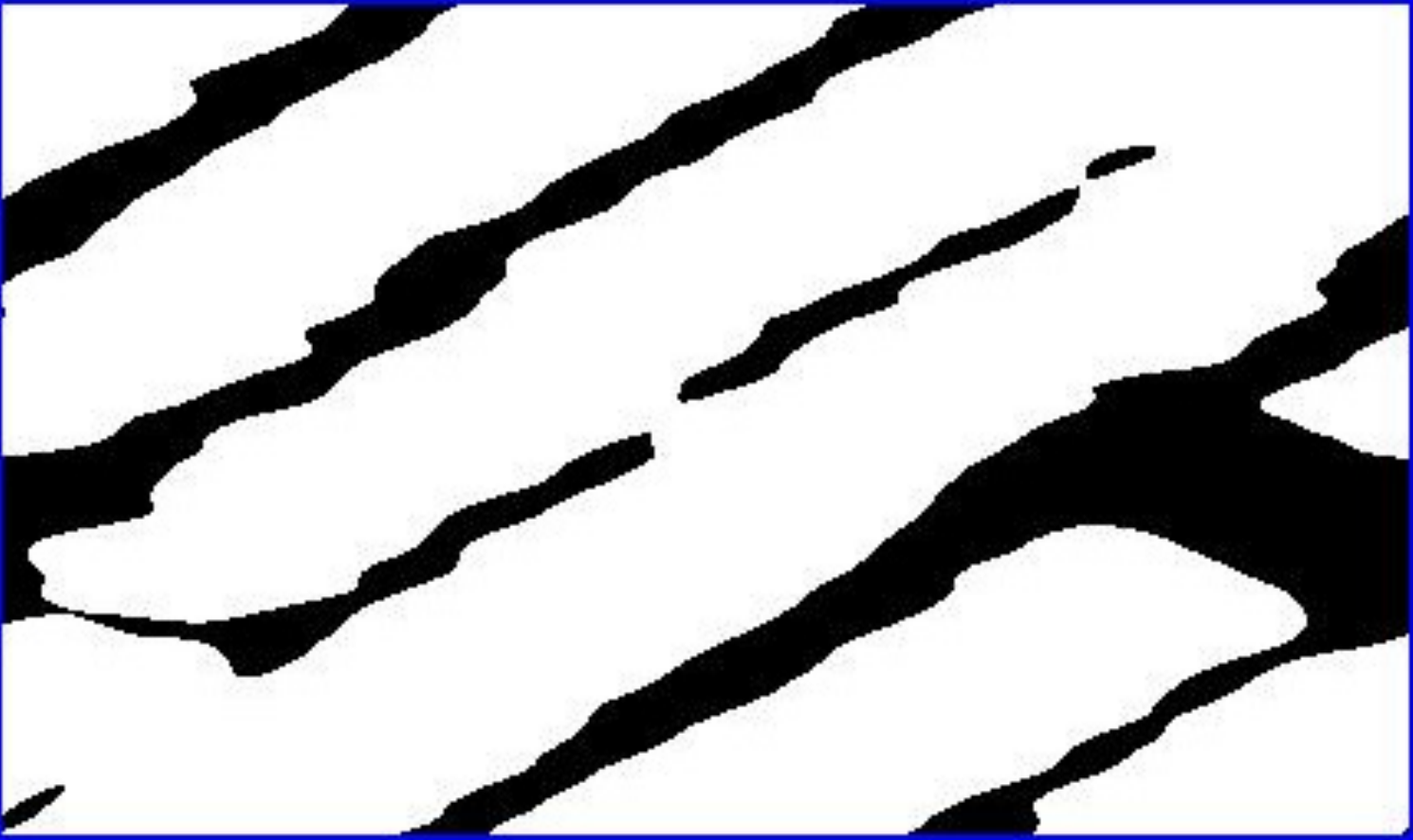}
\hskip0.5em
\includegraphics[angle=90,width=0.18\TW,clip]{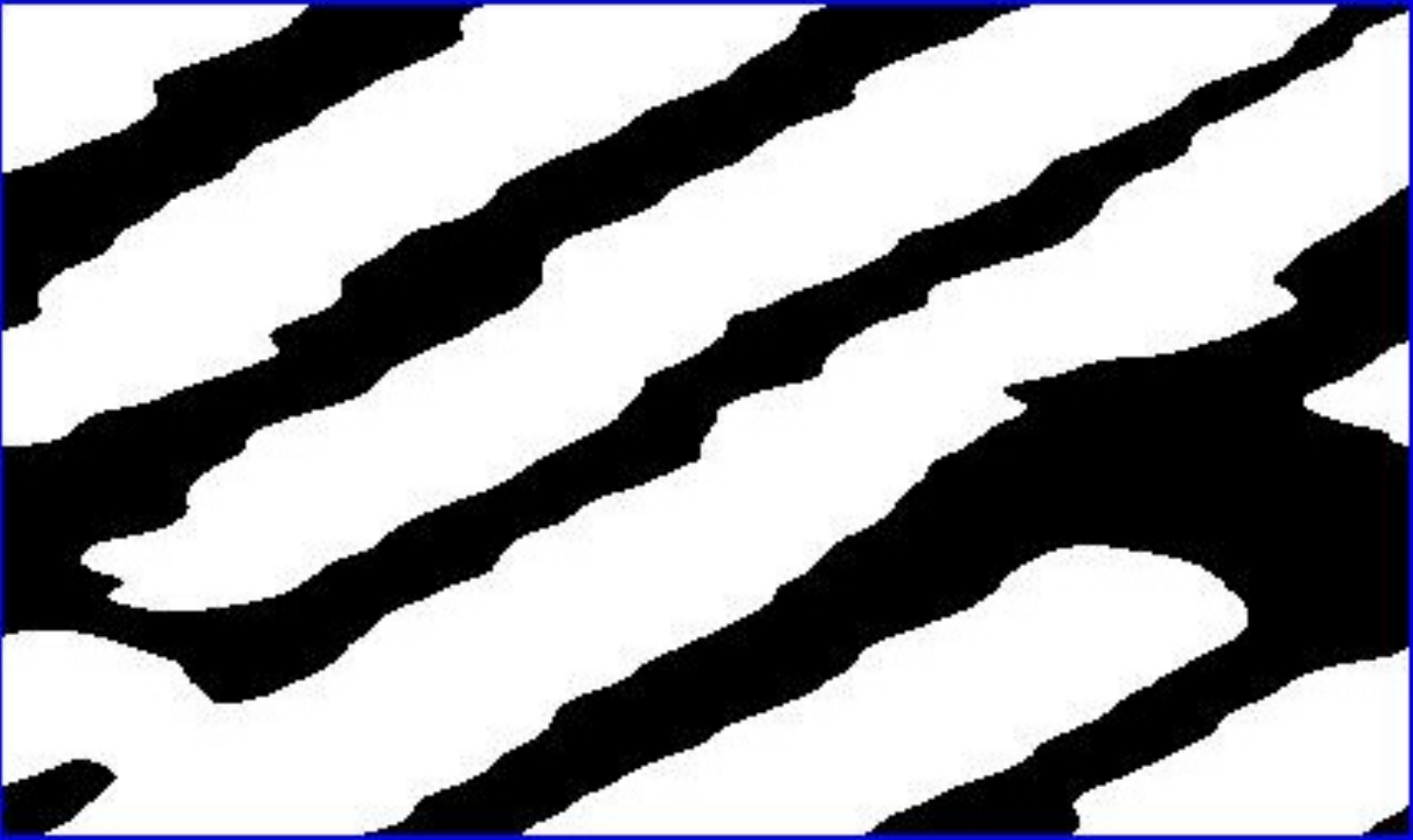}
\hskip0.5em
\includegraphics[angle=90,width=0.18\TW,clip]{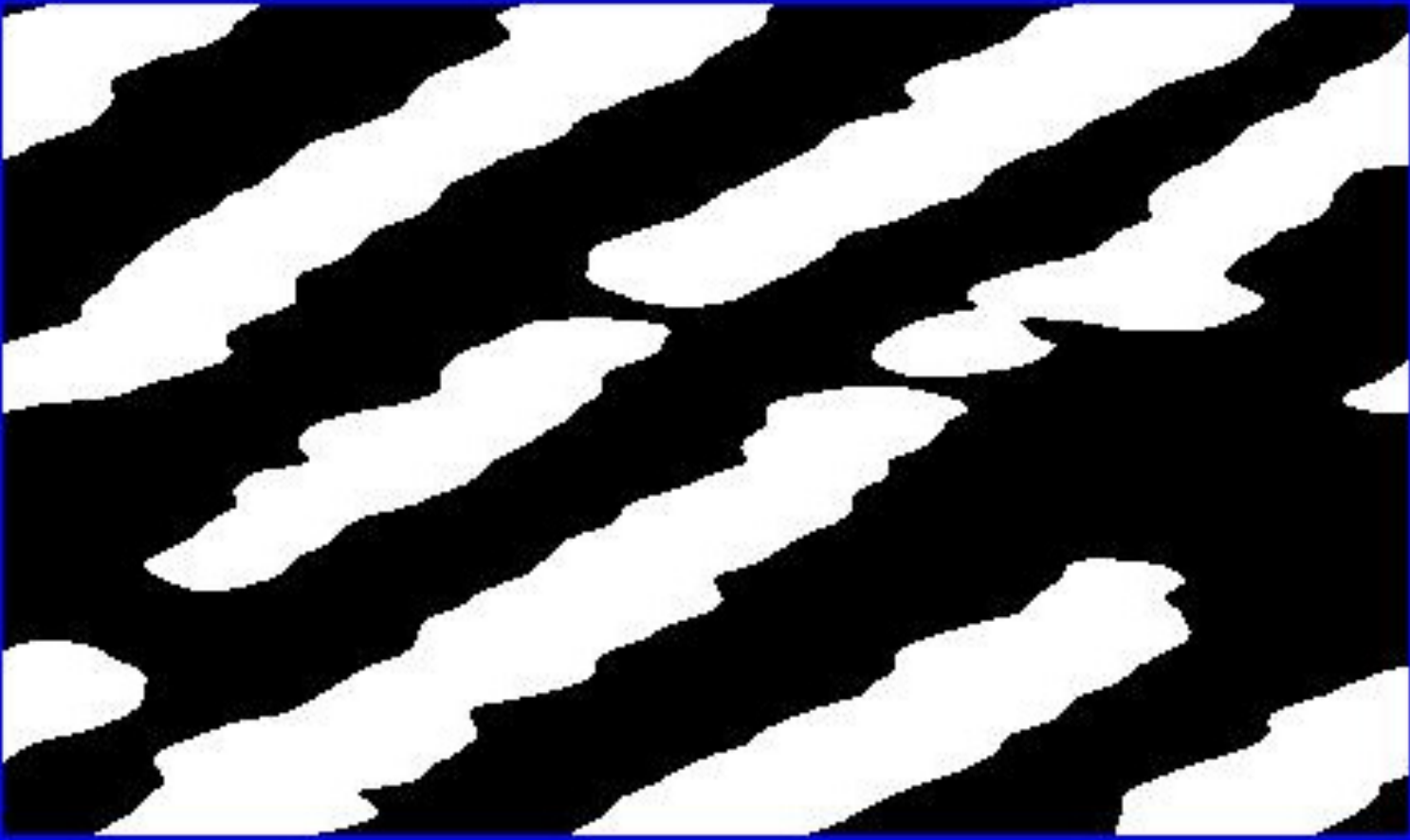}\\[2ex]
\includegraphics[angle=90,width=0.18\TW,clip]{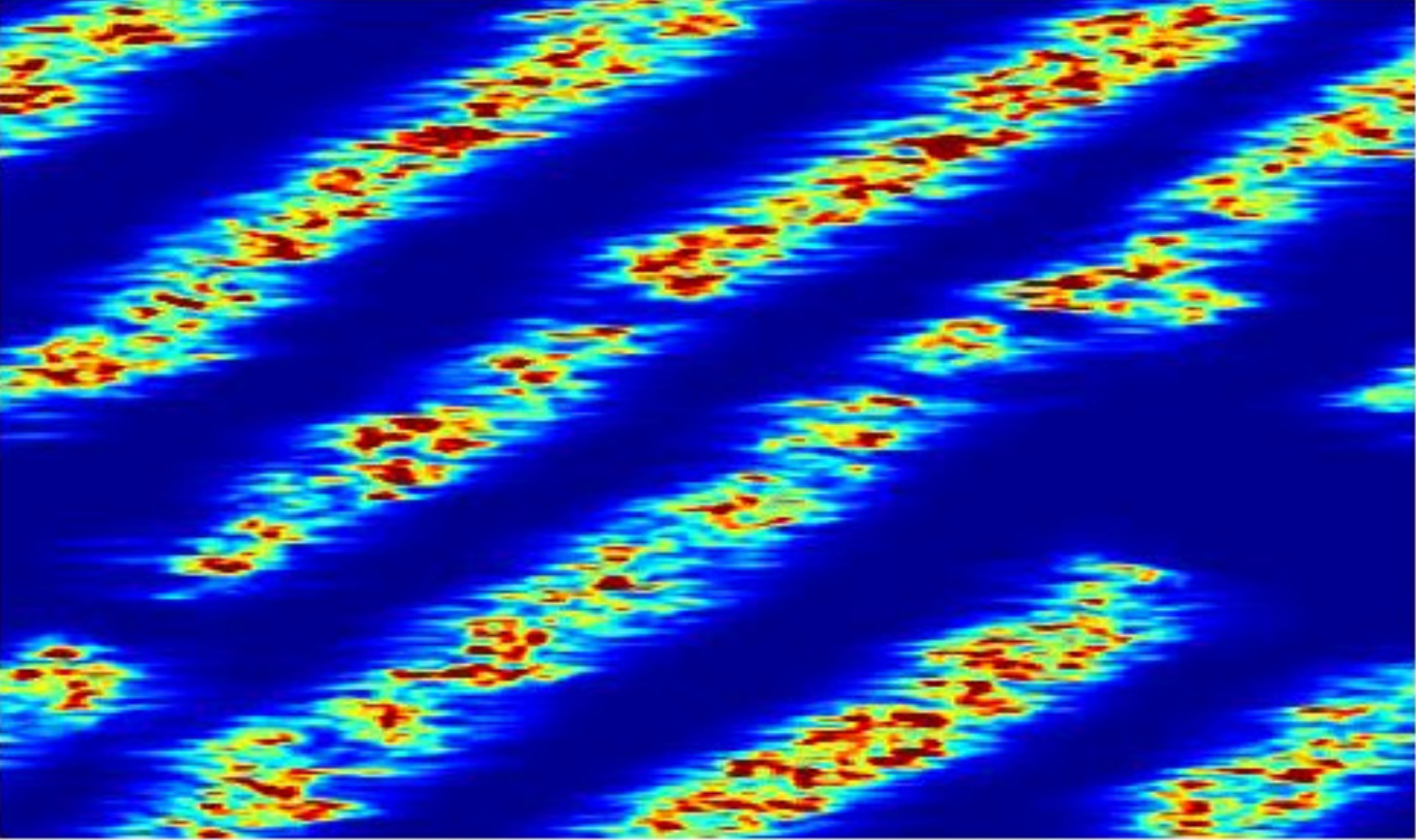}
\hskip0.5em
\includegraphics[angle=90,width=0.18\TW,clip]{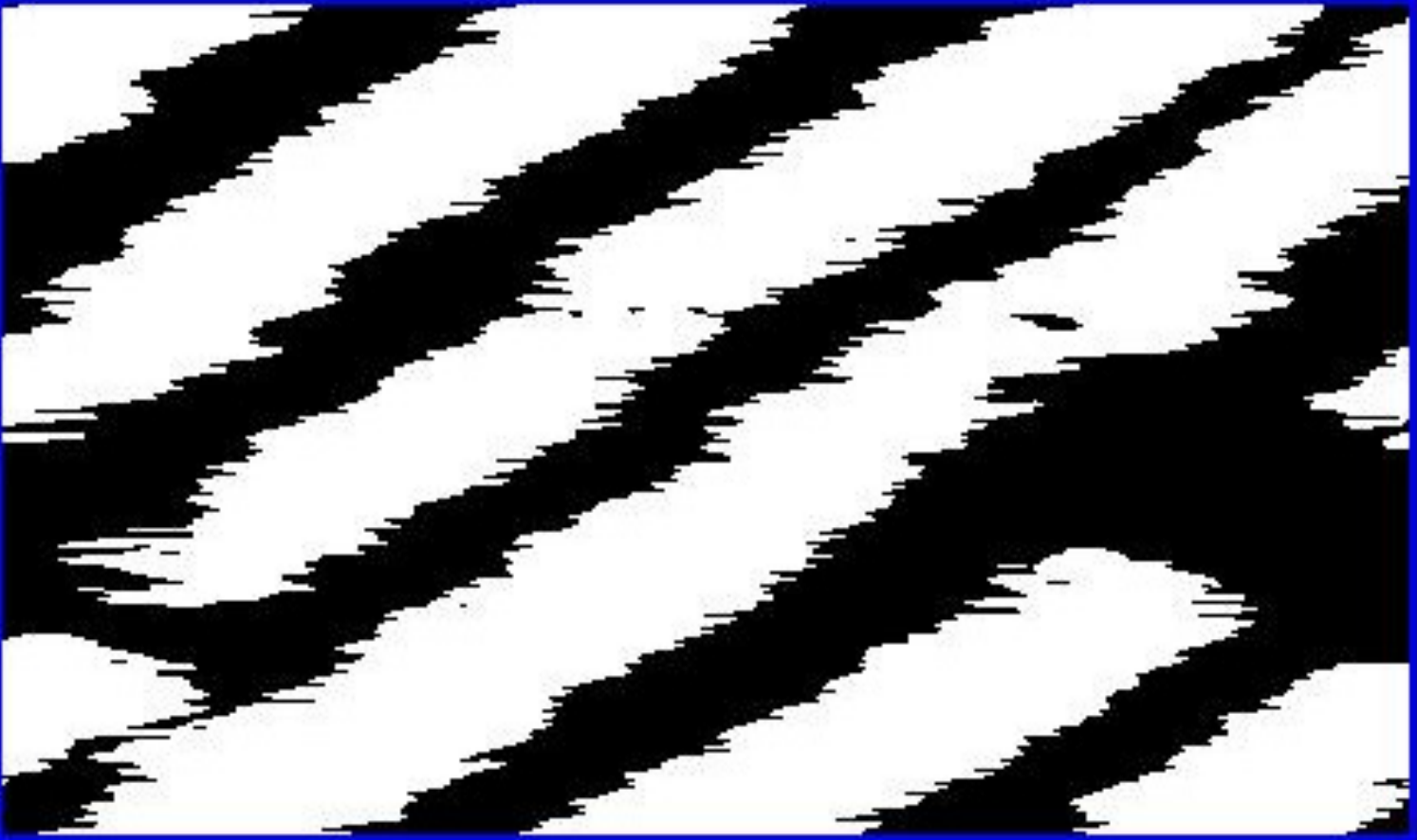}
\hskip3em
\includegraphics[angle=90,width=0.18\TW,clip]{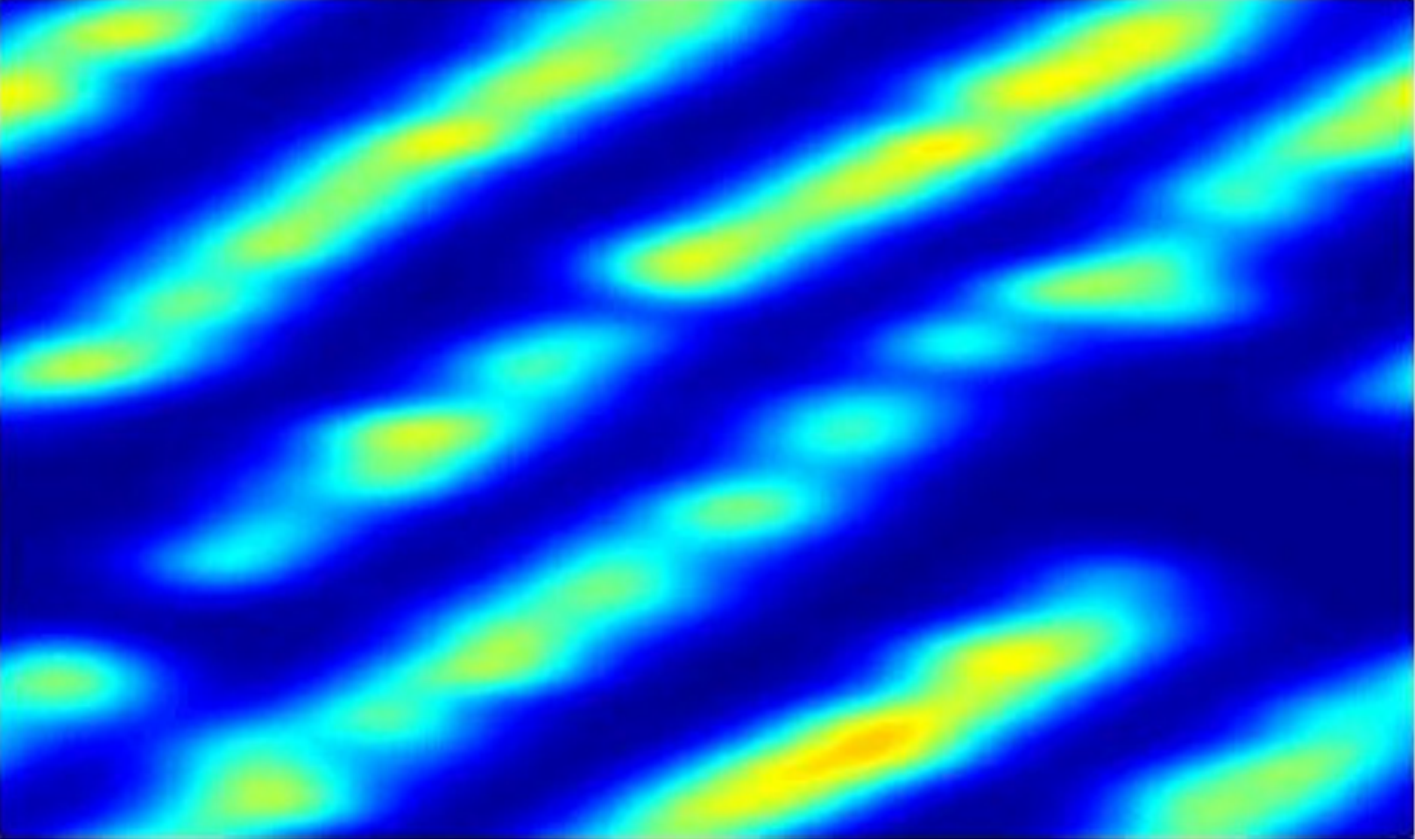}
\hskip0.5em
\includegraphics[angle=90,width=0.18\TW,clip]{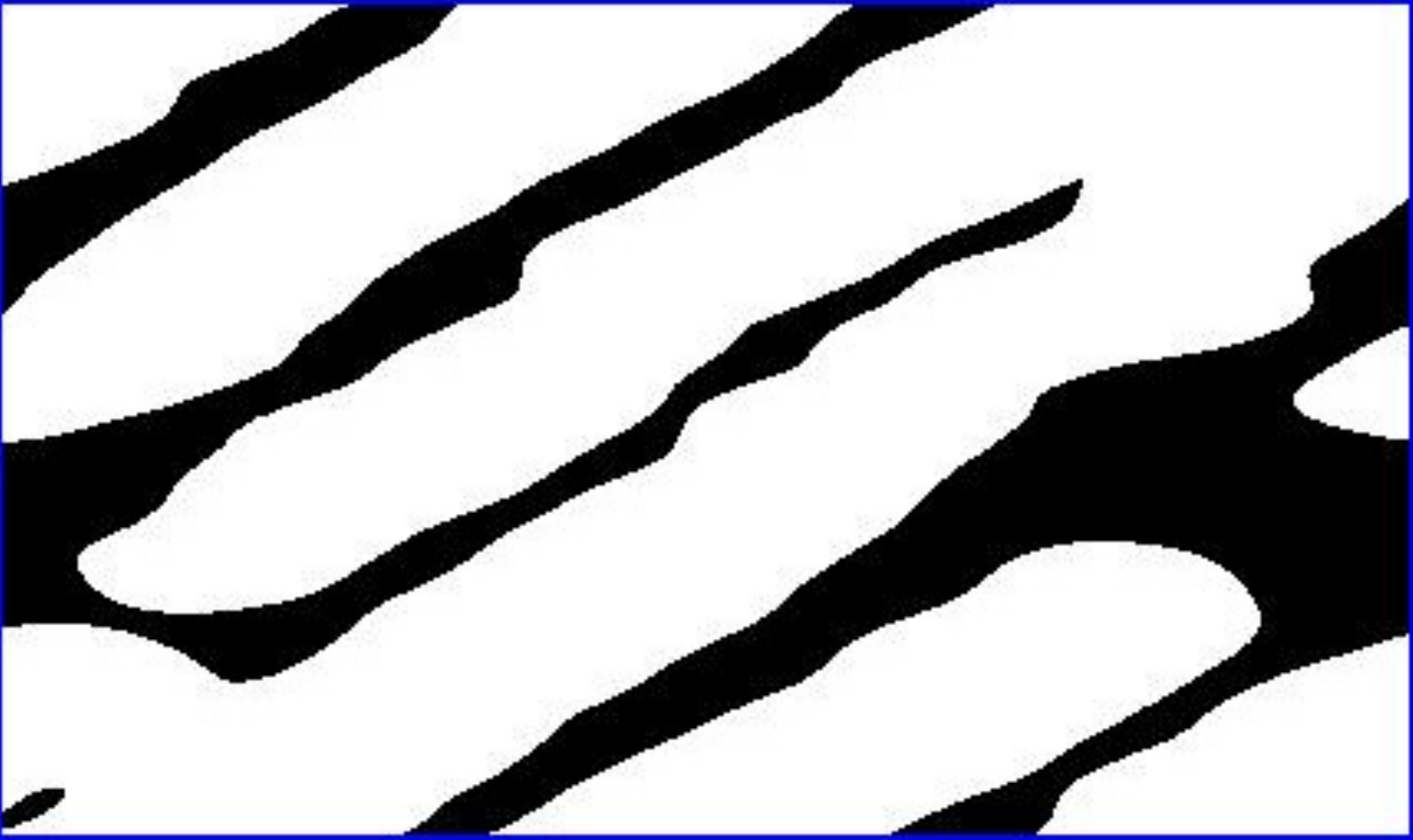}
\EC
\caption{Illustration of filtering and thresholding. Top: Flow pattern obtained in the simulation at $\RE=271.25$, $t=1250$.
Bottom: filtered state and  indicator functions of the turbulent region (white) for several values of $\kappa$ and energy cutoff  $e^{\rm c}$.
First line: $\kappa=2$ and $e^{\rm c}=0.001$, $0.003$, $0.010$;
in the leftmost panel the horizontal line indicates the location of the spanwise profiles in Figure~\ref{fig3}.
Second line:  $\kappa=0.5$ (left) and   $\kappa=4$ (right), in both cases $e^{\rm c}=0.003$. Except in the top image where the same colour scale is used as in Fig.~\ref{fig1}, i.e. from 0 to 0.1 with an energy peaking at 0.2450, here the colour scale ranges from~0 to~0.0555 as fixed by the image with $\kappa=2$ on \texttt{¥}he left of the first line. The filtered energy peaks at 0.1046 for $\kappa=0.5$ and at 0.0371 for $\kappa=4$.\label{fig2}}
\EF
\BF
\includegraphics[width=0.48\TW,height=0.36\TW,clip]{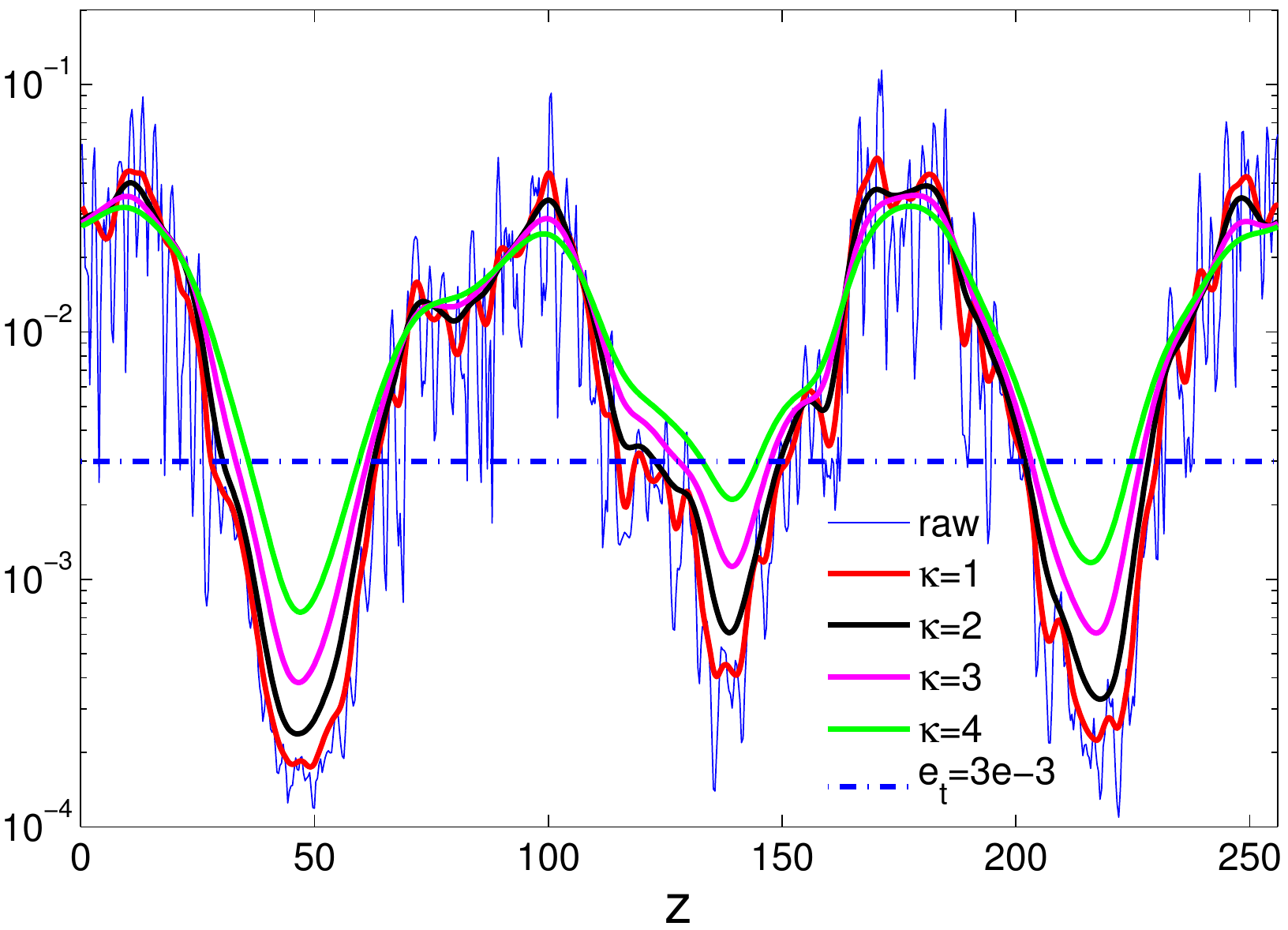}
\hfill
\includegraphics[width=0.48\TW,height=0.36\TW,clip]{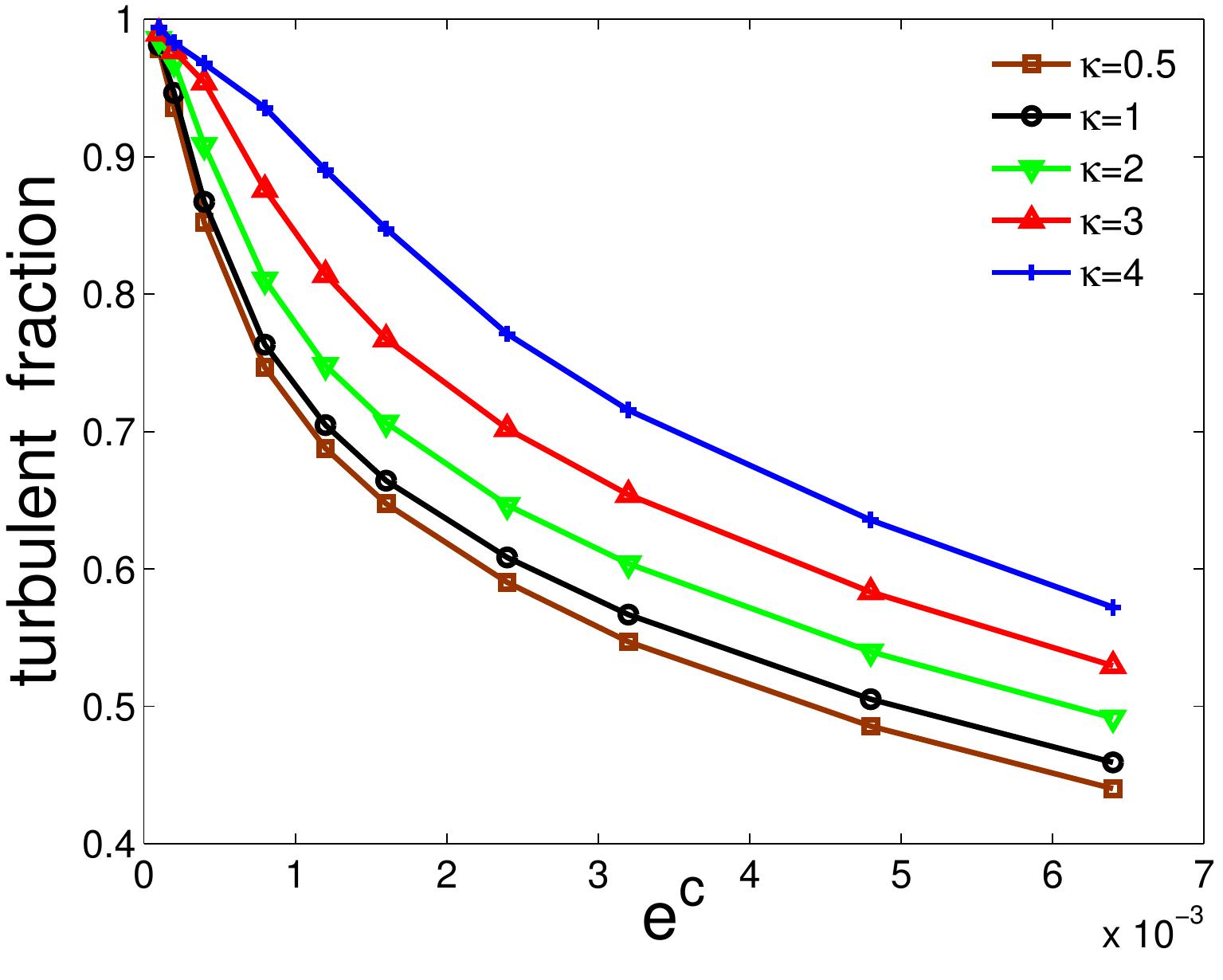}\\[2ex]
\includegraphics[width=0.48\TW,height=0.36\TW,clip]{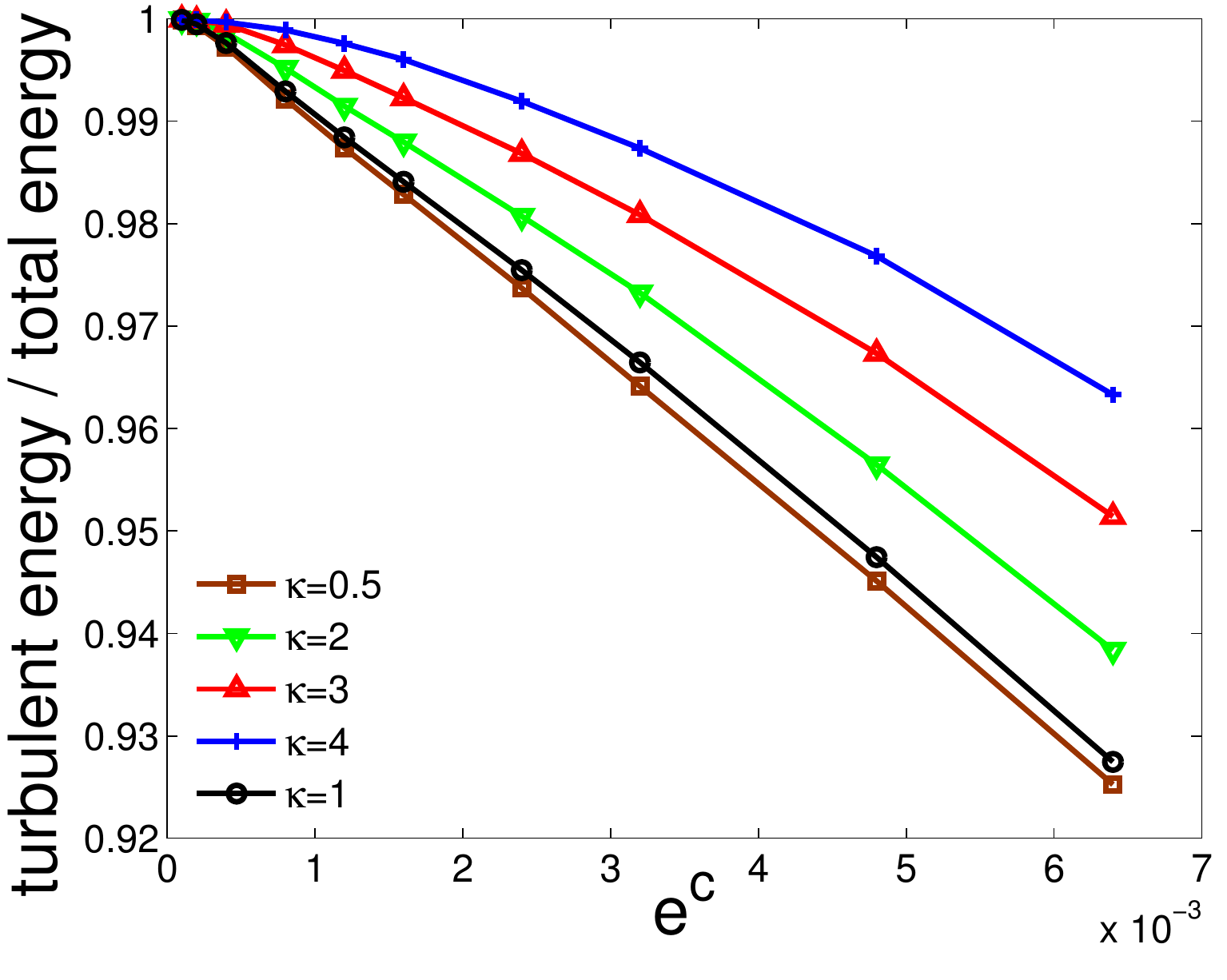}
\hfill
\includegraphics[width=0.48\TW,height=0.36\TW,clip]{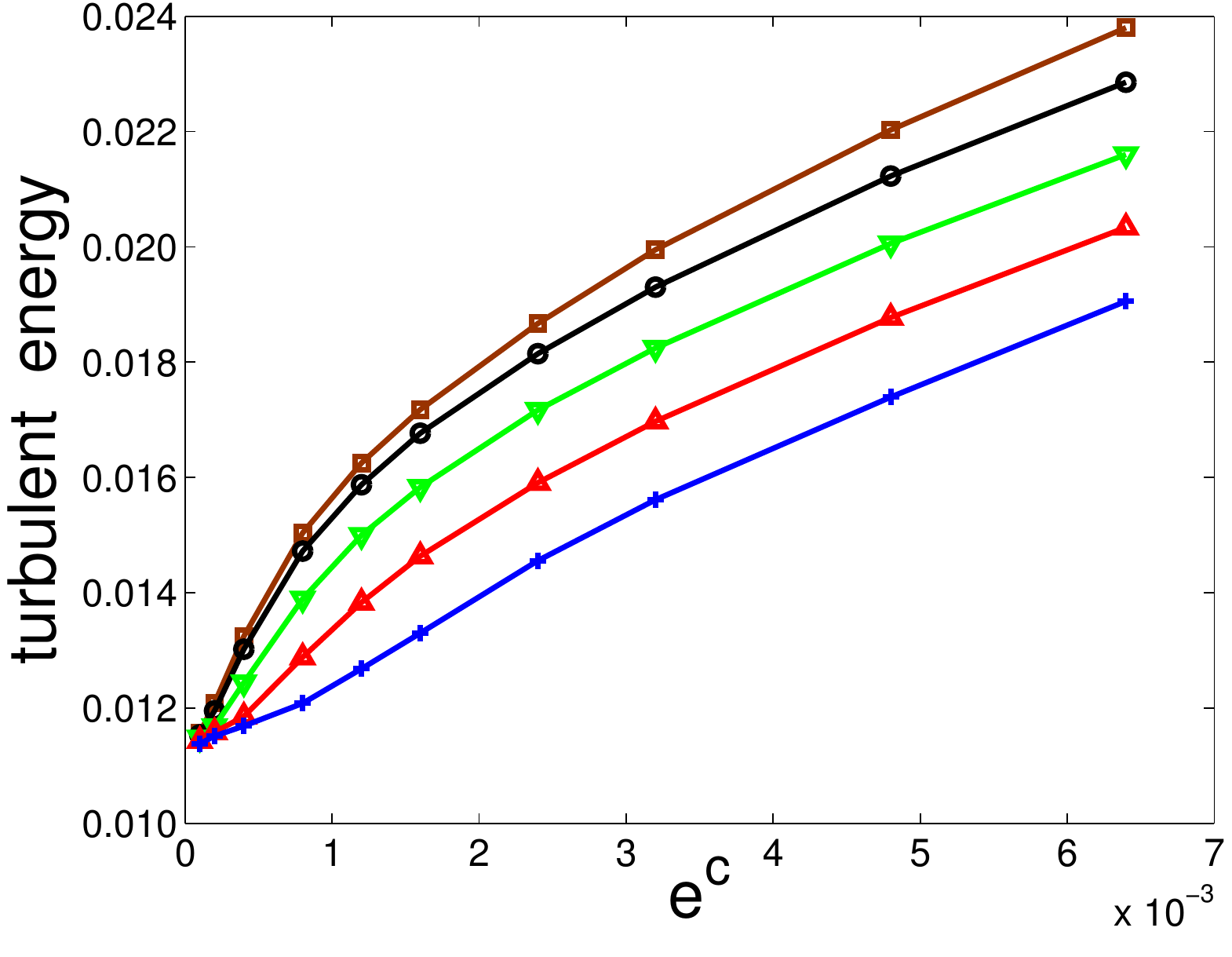}
\caption{Top-left: Energy profiles along $z=L_z/2$, raw and filtered with different values of $\kappa$ as indicated, same state as in Fig.~\ref{fig2}. Filtering is clearly insufficient for $\kappa=1$ and too strong for $\kappa=4$; $\kappa=2$ or 3 seem more satisfactory; in the following we keep $\kappa =2$.
Top-right: Variation of the turbulent fraction  as a function of  $e^{\rm c}$ for several values of  $\kappa$; as can easily be understood from the images in Fig.~\ref{fig2} and the profiles in the graphs on the left, the turbulent fraction decreases as $e^{\rm c}$ increases at given $\kappa$ and increases as $\kappa$ increases at given $e^{\rm c}$, as consequences of the nesting of domains with different filter widths and thresholds.
Bottom: Relative energy content of the turbulent fraction (left) and turbulent energy (right) as functions of  $e^{\rm c}$ for several values of  $\kappa$; observed trends derive from the previous observation in a straightforward way.
No significant break appear in these curves that would suggest an optimum value to  $e^{\rm c}$.
Wanting that most of the energy be concentrated in the domain labelled `turbulent', we choose  $e^{\rm c}=0.003$ (dash-dotted line in the top-left panel) so that about 97.5\% of the energy is allocated to the turbulent region for $\kappa=2$ in the case considered.\label{fig3}}
\EF

 The  {\it mean perturbation energy\/} is defined as
 $$
 E(t)=\frac{1}{L_xL_z}\int\!\!\!\int\mathrm d x\,\mathrm d z\, e(x,z,t).
 $$
 Let $\mathcal{I}_{\rm t}(x,z,t)$ be the indicator function of the turbulent region after thresholding.
The instantaneous {\it turbulent fraction\/} is then
$$
F_{\rm t}(t)=\frac{1}{L_xL_z}\int\!\!\!\int\mathrm d x\,\mathrm d z\,\mathcal{I}_{\rm t}(x,z,t),
$$
and one can further define a {\it mean turbulent energy\/}, i.e. the mean energy per unit surface in the region identified as turbulent:
$$
E_{\rm t} (t)= \frac{\int\!\!\!\int\mathrm d x\,\mathrm d z\, \mathcal{I}_{\rm t}(x,z,t)\,e(x,z,t)}{\int\!\!\!\int\mathrm d x\,\mathrm d z\,\mathcal{I}_{\rm t}(x,z,t)}.
$$
The time averages of these quantities $\langle F\rangle$, $\langle E\rangle$, and $\langle E_{\rm t}\rangle$ can then be computed when the system has reached its stationary regime.
The effects of changes in $\kappa$ and   $e^{\rm c}$ on the laminar-turbulent detection are illustrated and discussed in Figures~\ref{fig2} and~\ref{fig3}.
There is no ideal relative filter size $\kappa$  and energy cutoff $e^{\rm c}$ but once they are fixed, qualitative comparisons between different states are fully informative and quantitative variations of the observables are of significance.
In the following we take $e^{\rm c}=3\times10^{-3}$ and $\kappa=2$ most of the time.

\section{Turbulence decay of PCF in extended geometry\label{s3}}
\subsection{Phenomenology of band decay\label{s3.1}}
The flow field obtained at $\RE=275$ and $t=7,\!000$ is used as an initial condition to study the lower transitional range around \RG.
Always starting from this state, several simulations are performed by decreasing \RE\ down to a slightly smaller value of \RE\ and waiting until the laminar regime is reached. 
Figure~\ref{fig4} displays the variation with time of the distance $D$ to the laminar flow defined as $D=\sqrt{2E}$ during each experiment.
\BF
\BC
\includegraphics[width=0.70\TW]{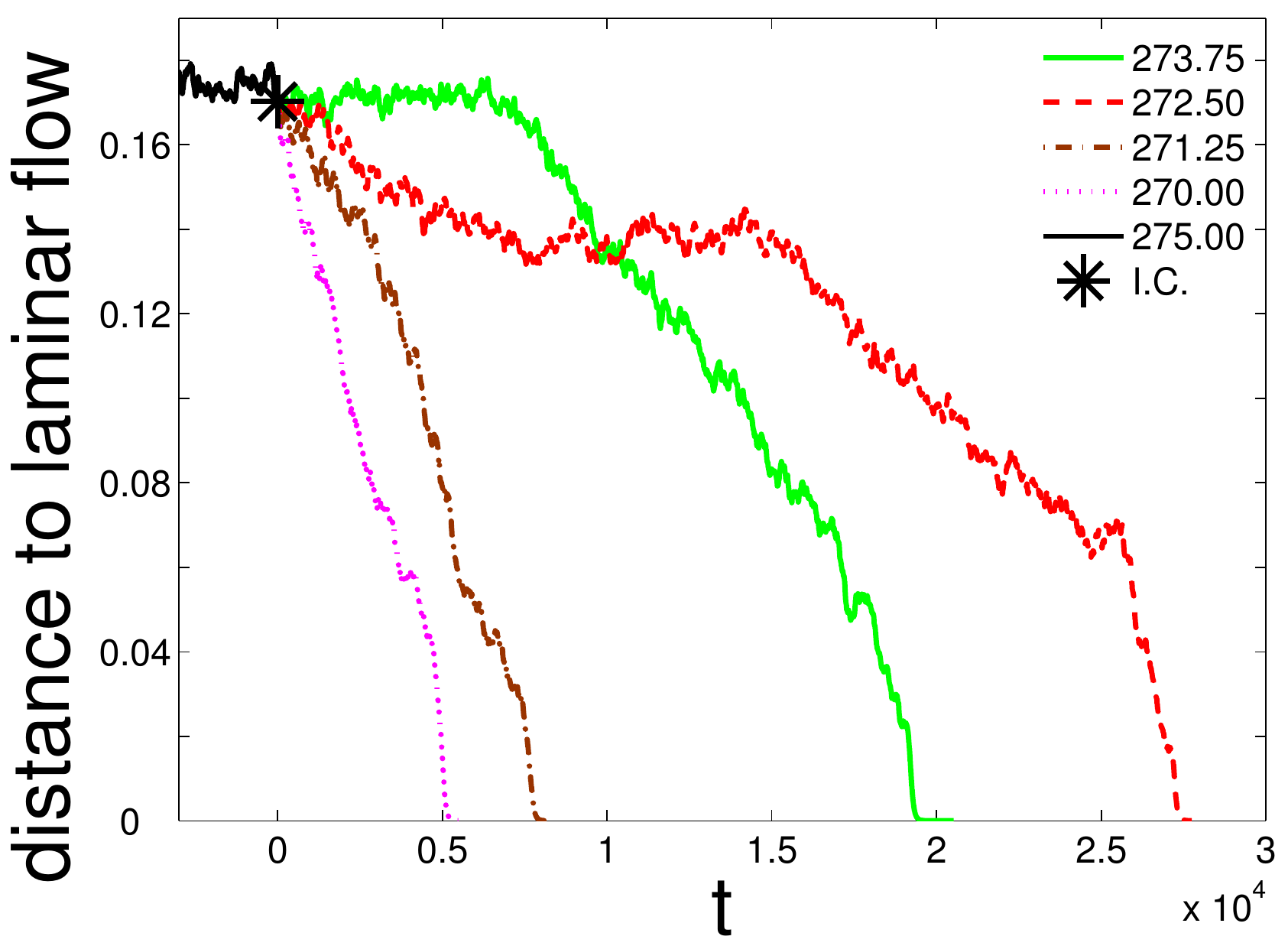}
\EC
\caption{Variation of distance $D$ to the laminar flow as a
function of time during decay starting from the three-band state at $\RE=275$ at $t=7,\!000$ (asterisk at the end of the continuous black line) for $ \RE=273.75$ (green, full line), $272.50$ (red, dashed), 271.25 (brown, dot-dashed), and $270.00$ (magenta, dotted).\label{fig4}}
\EF

The first experiment in this series is performed at $\RE=273.75$.
Each of the three bands forming the initial state break in turn but recover (an example of broken band is given for $t=3,\!000$ in Fig.~\ref{fig5}, left) until two bands break nearly simultaneously around $t=7,\!000$.
Each broken band recedes, yielding a more complex pattern with several band segments separated by a single continuous band.
The situation is illustrated at $t=10,\!250$ in the centre panel in Fig.~\ref{fig5}. Next the band portions recede and disappear while the continuous band breaks, retracts along its length (snapshot at $t=15,\!250$) and breaks (snapshot at $t=17,\!000$). Fragments left next rapidly decay and at $t\approx19,\!250$ the flow is entirely laminar.
\BF
\BC
\includegraphics[angle=90,width=0.18\TW,clip]{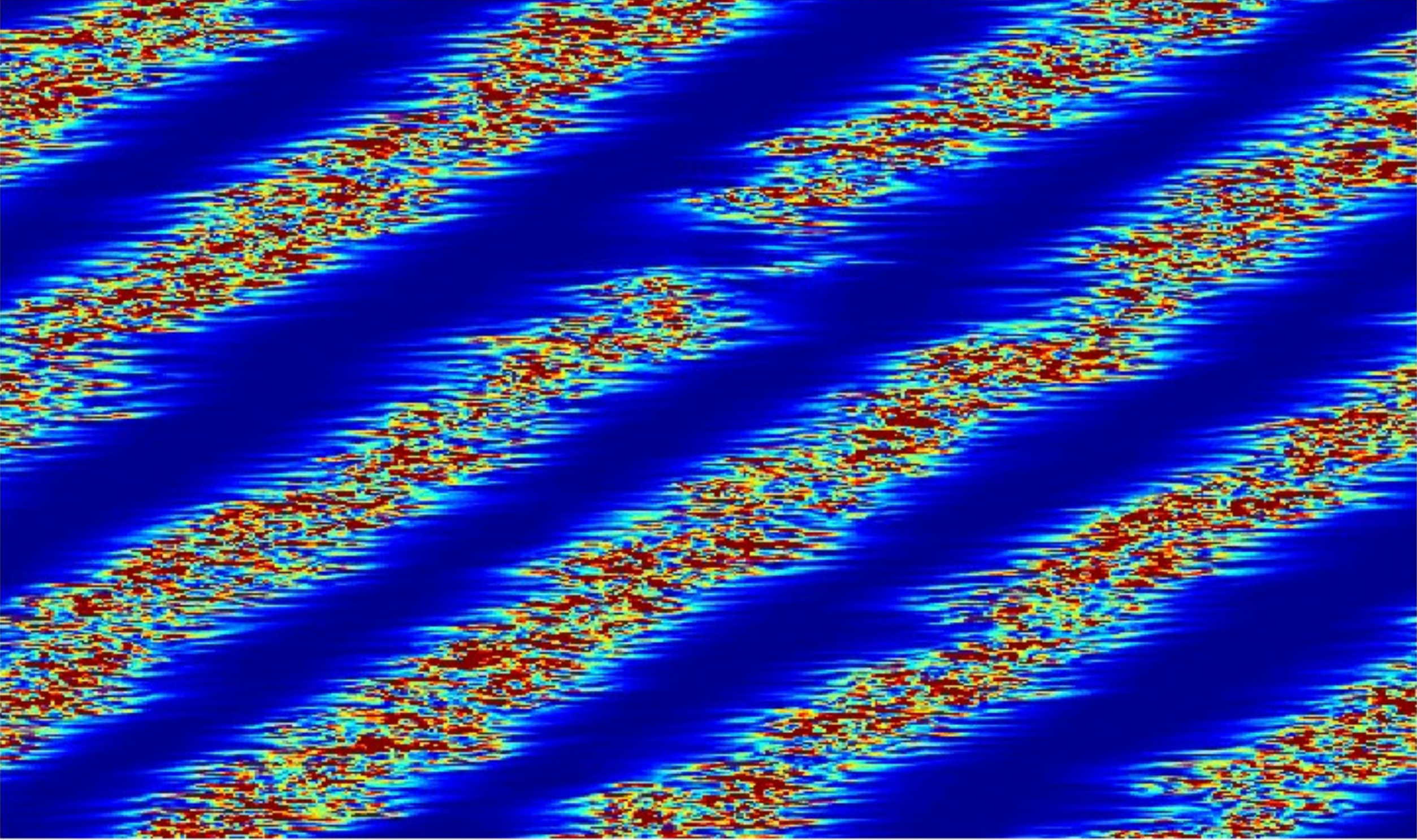}
\hskip0.5em
\includegraphics[angle=90,width=0.18\TW,clip]{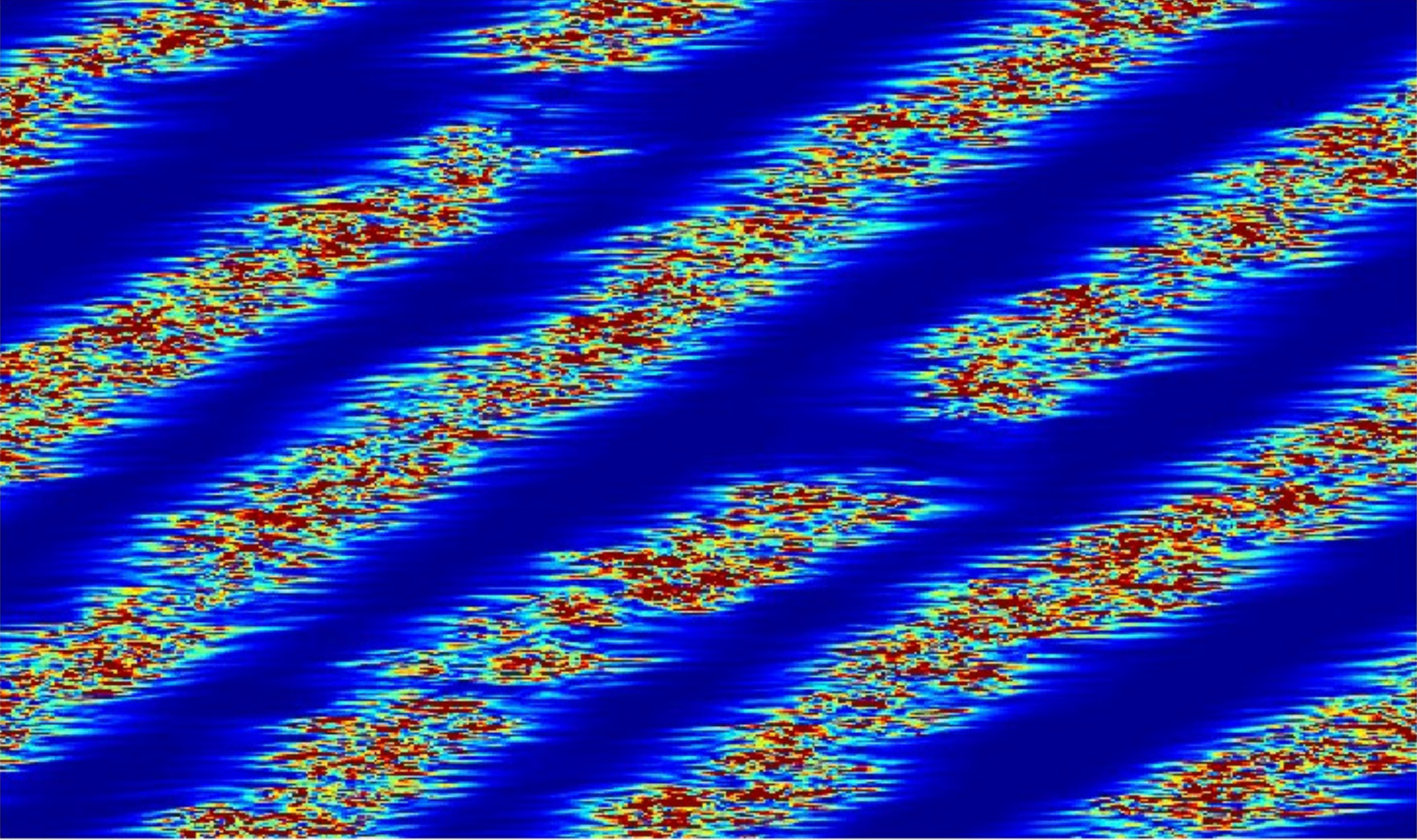}
\hskip0.5em
\includegraphics[angle=90,width=0.18\TW,clip]{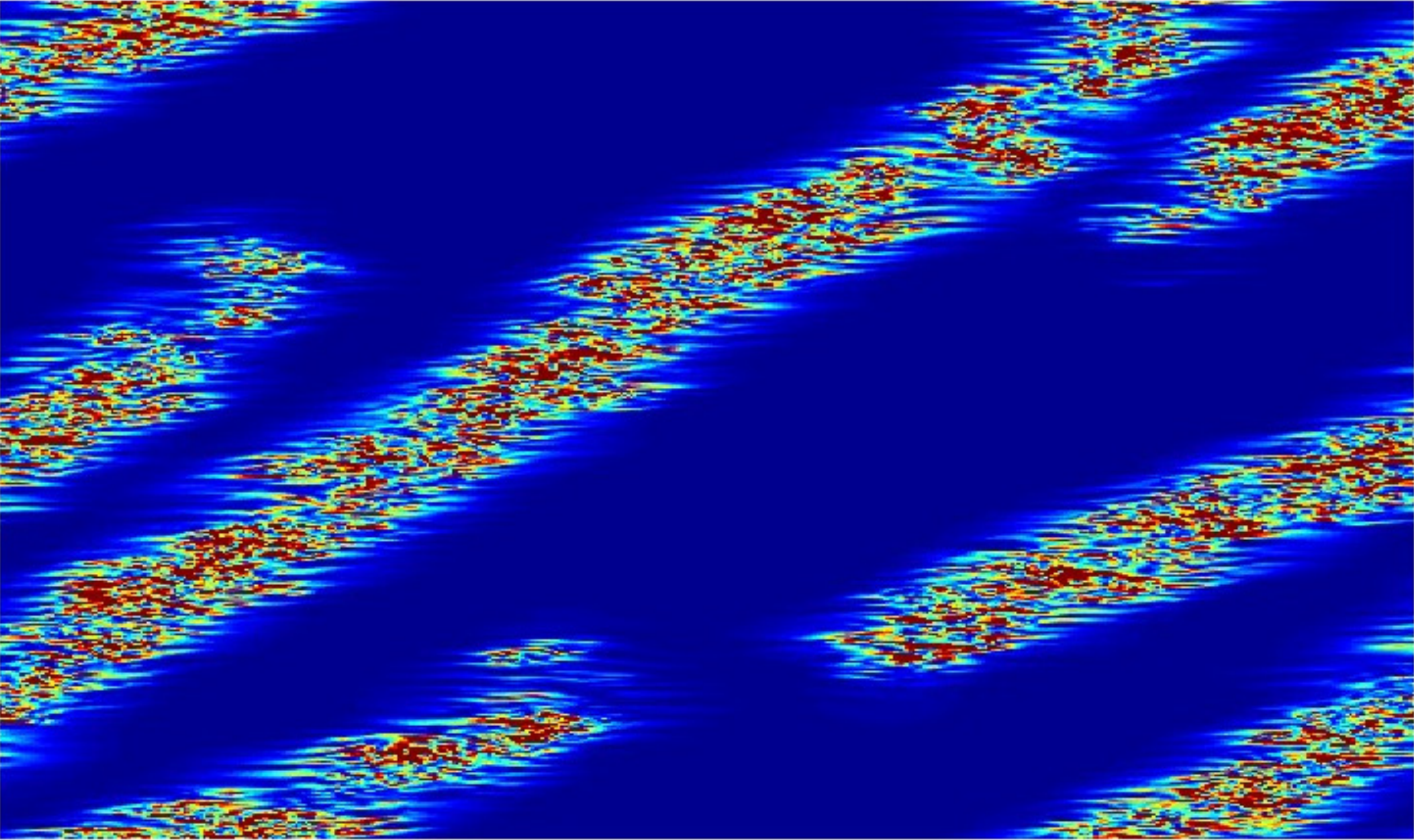}
\hskip0.5em
\includegraphics[angle=90,width=0.18\TW,clip]{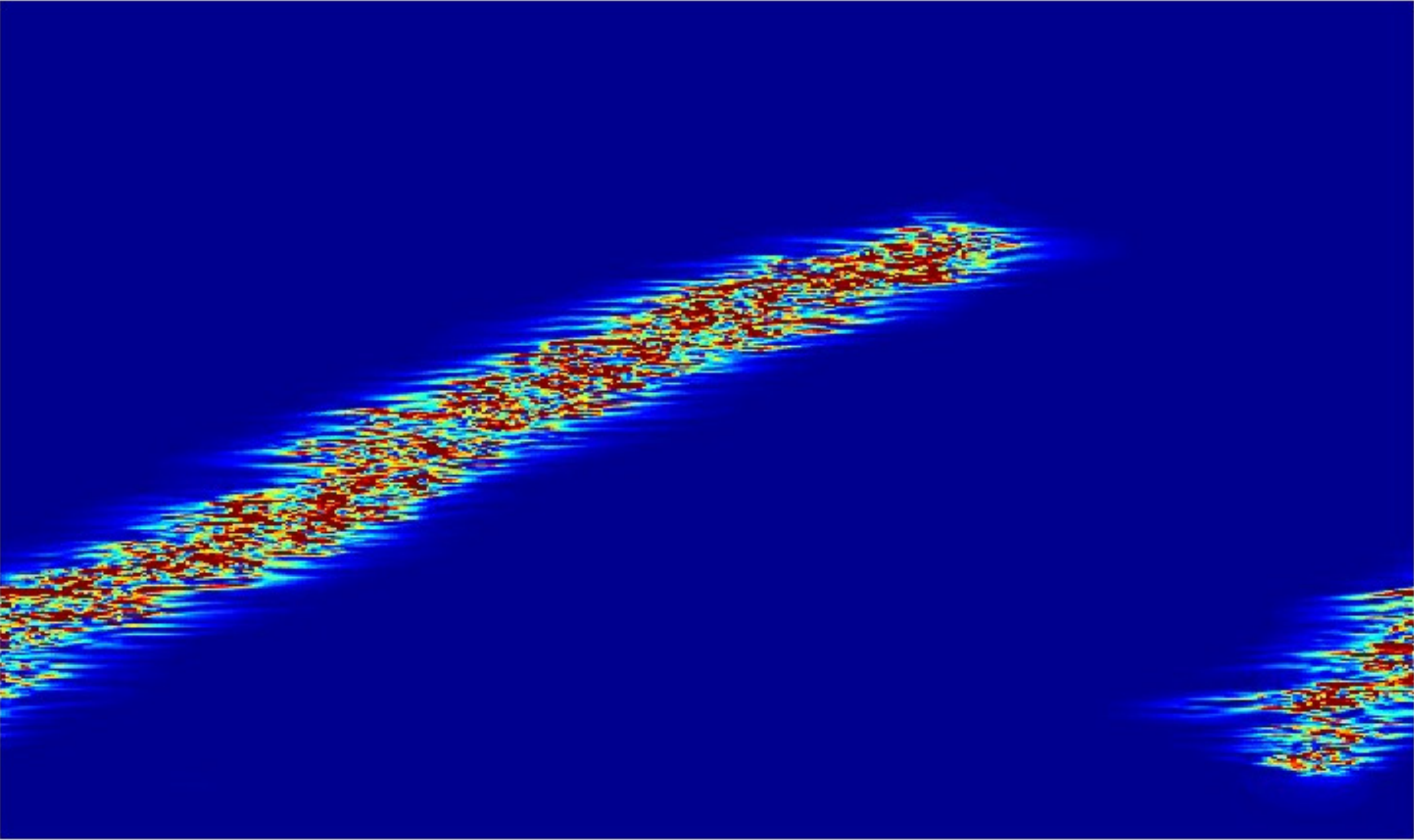}
\hskip0.5em
\includegraphics[angle=90,width=0.18\TW,clip]{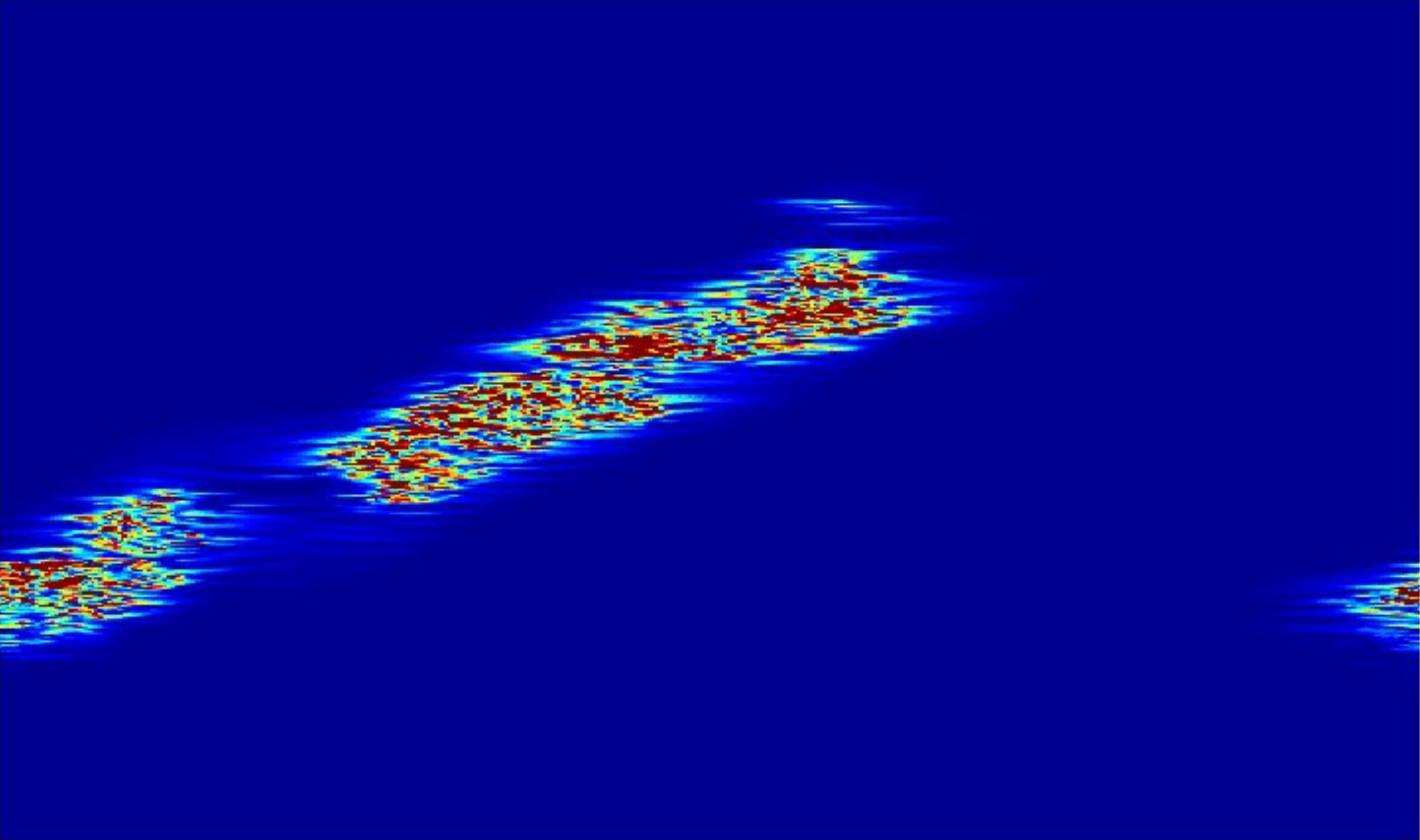}
\
\EC
\caption{Snapshots taken during the decay at $\RE=273.75$
from state at $\RE=275$, $t=7,\!000$. Time is reset at the beginning
of the experiment. From left to right: $t=3,\!000$: a band breaks but recovers; $t=7,\!250$: two bands break; $t=10,\!250$: complex intermediate state; $t=15,\!250$: a single band portion remains; $t\simeq17000$: the band segment breaks.
\label{fig5}}
\EF

For the second experiment we set $\RE=272.5$. Snapshots during decay are displayed in Figure~\ref{fig6}. The difference with the previous experiments is that the first band to break does not recover but retracts (see left snapshot at $t=4,\!500$).
When it has completely disappeared, we are left with a state displaying two unevenly spaced bands (centre-left snapshot at $t=12,\!000$).
This state seems rather robust since the bands alternately break from time to time but recover, while the distance between them tends to equalise. One of the remaining bands break at $t\approx15,\!300$ and next retracts. The last band breaks at $t\approx 20,\!600$ and recedes regularly, till the end of the transient.
\BF
\BC
\includegraphics[angle=90,width=0.18\TW,clip]{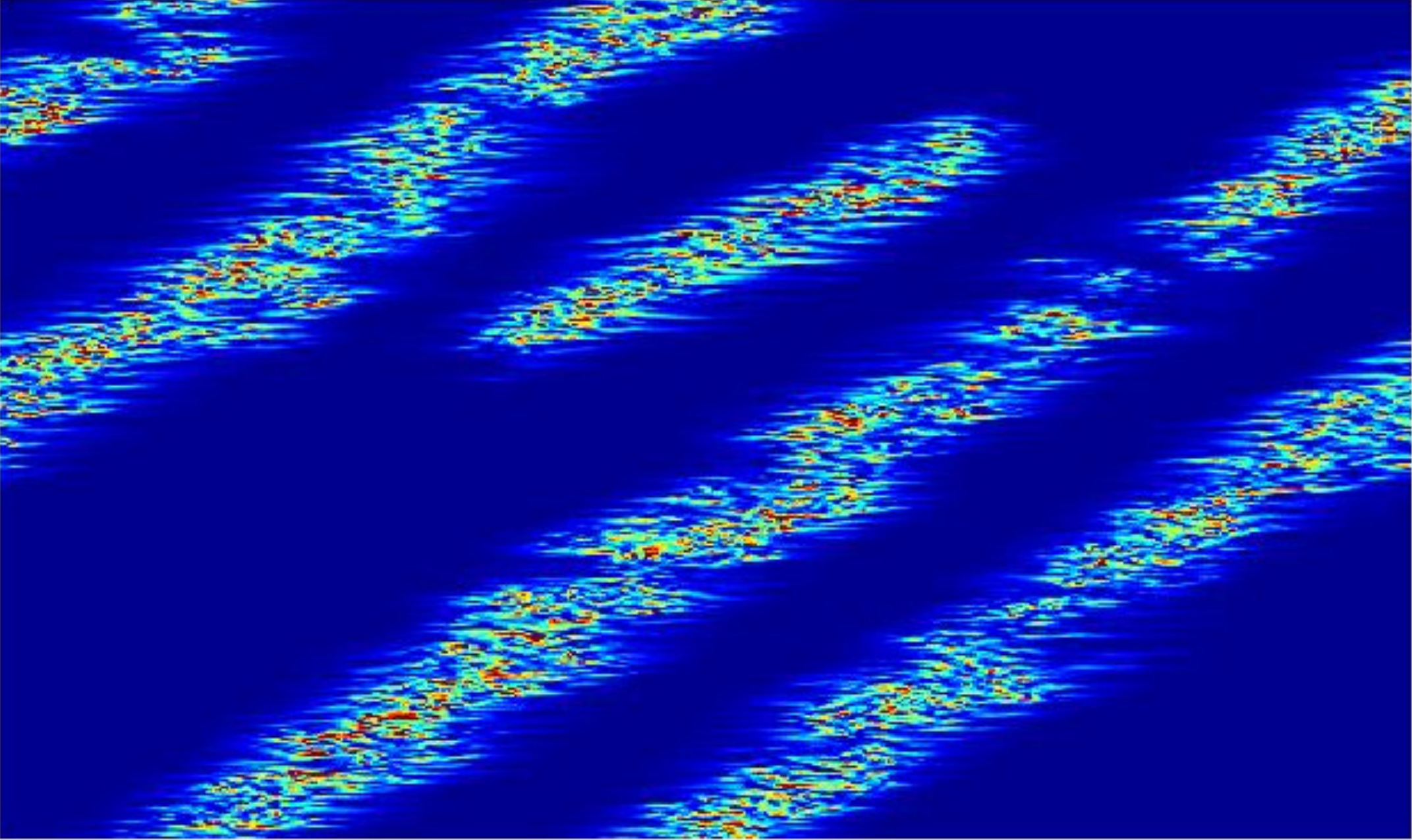}
\hskip0.5em
\includegraphics[angle=90,width=0.18\TW,clip]{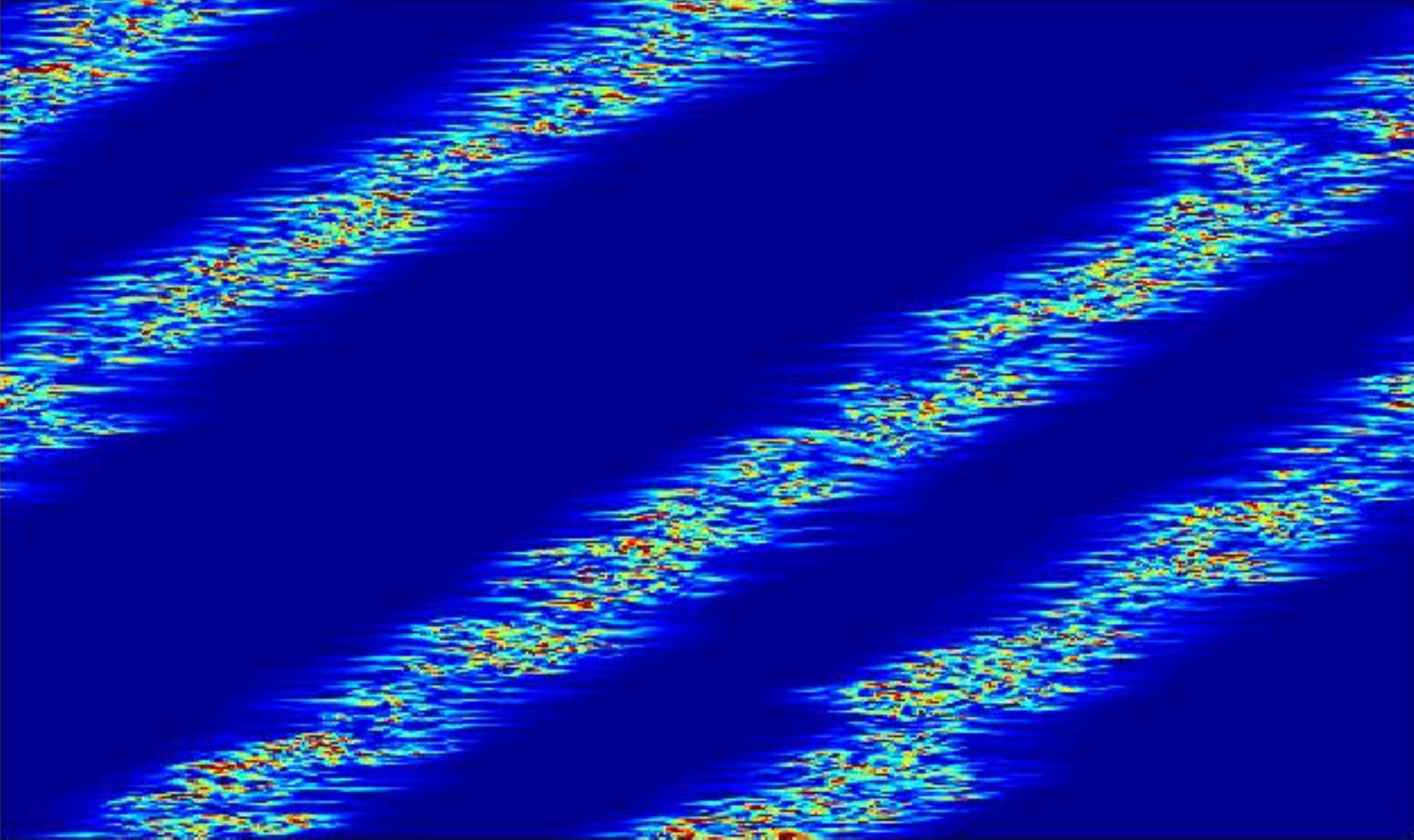}
\hskip0.5em
\includegraphics[angle=90,width=0.18\TW,clip]{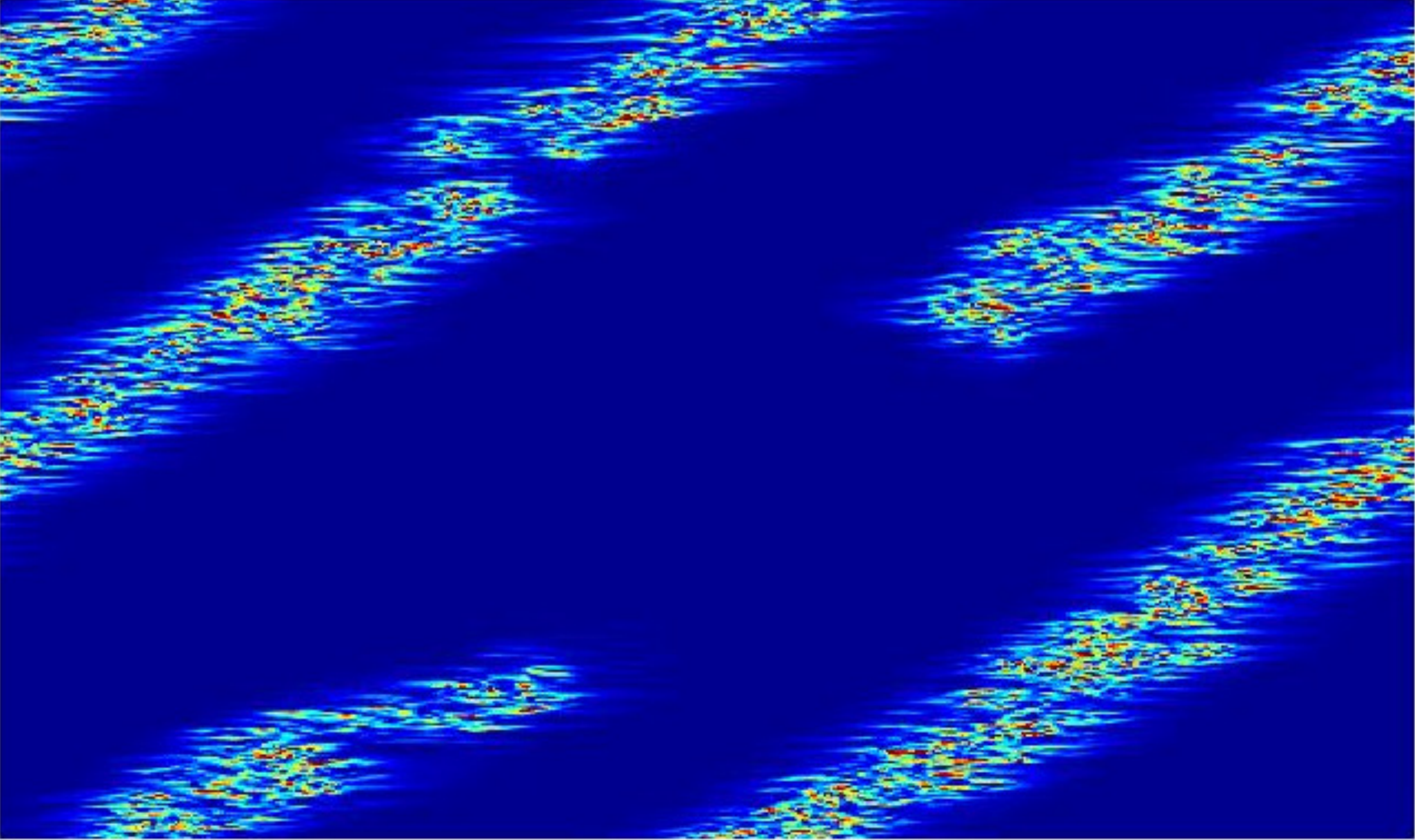}
\hskip0.5em
\includegraphics[angle=90,width=0.18\TW,clip]{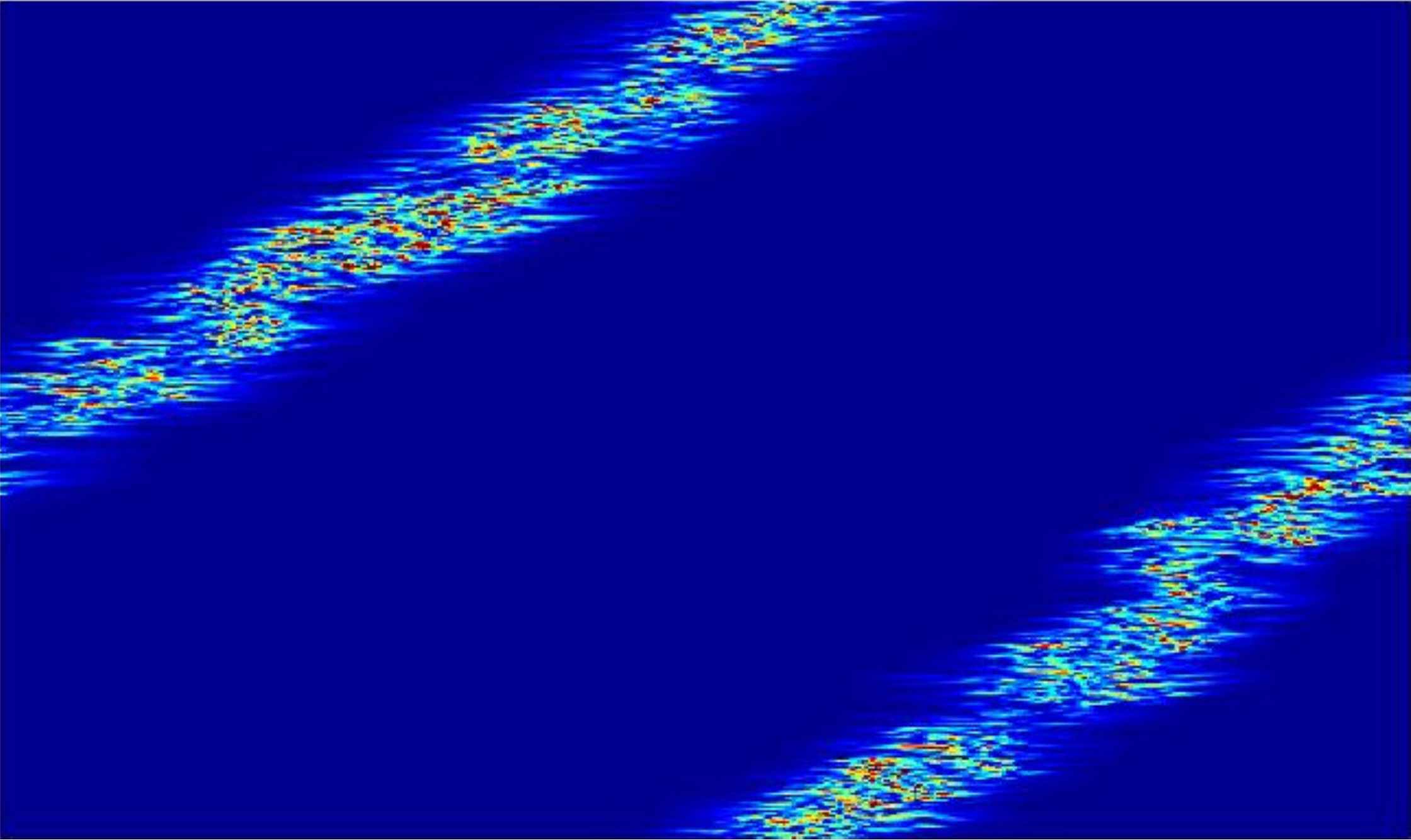}
\hskip0.5em
\includegraphics[angle=90,width=0.18\TW,clip]{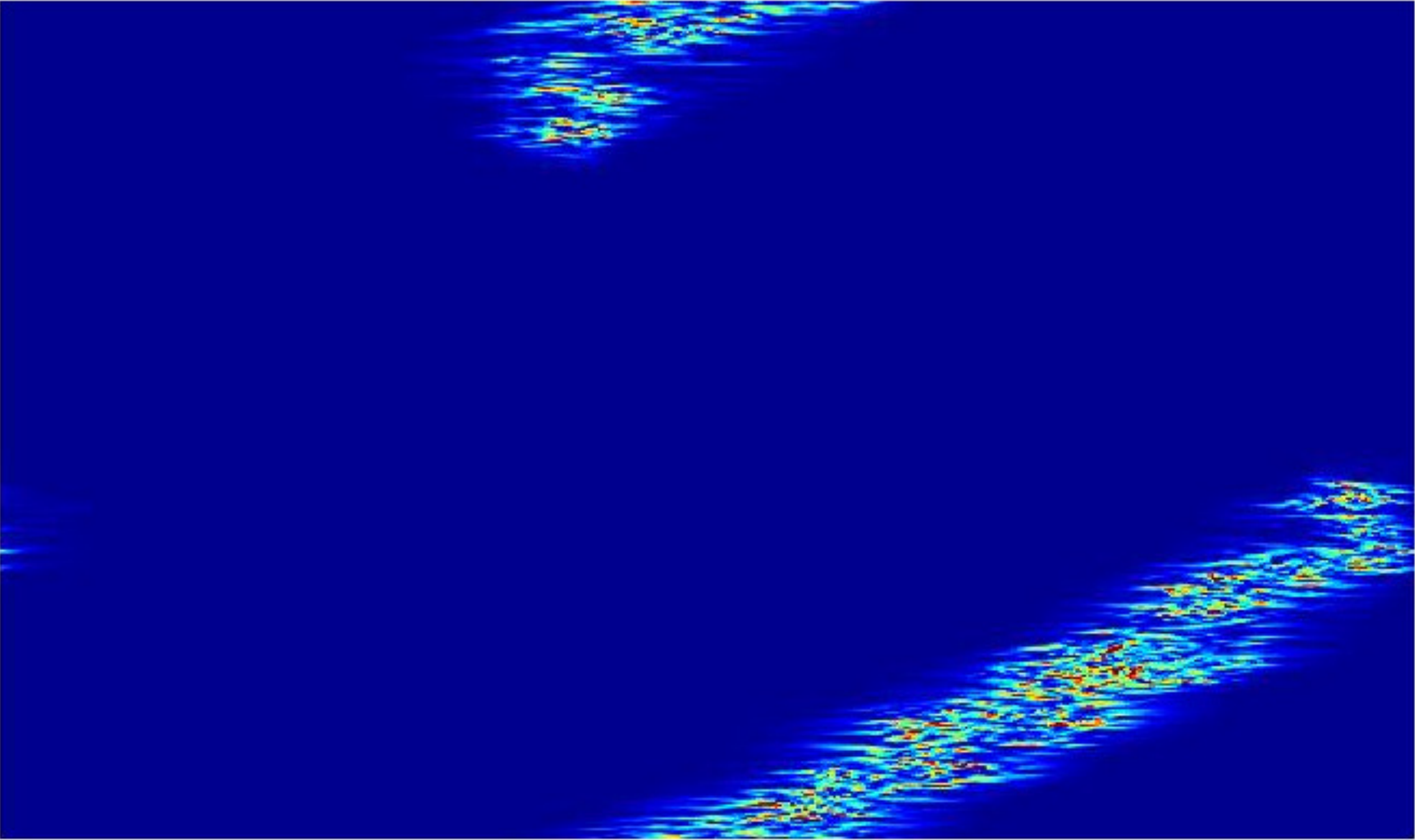}
\EC
\caption{Snapshots taken during the decay at $\RE=272.5$.
From left to right: at $t=4,\!500$ some time after the breakdown of the first band;
at $t=12,\!000$ with two unevenly spaced bands;
at $t=16,\!500$ after breakdown of second band;
at $t=20,\!250$ with one continuous band remaining;
 at $t=24,\!000$ during the decay of the last band.\label{fig6}}
\EF

The third experiment is for  $\RE=271.25$. Two out of the three bands present in the initial state shortly break, possibly at several places, and each portion retracts as described previously. Figure~\ref{fig7} illustrates the end of the transient. The last band breaks around $t=3,\!500$ (left panel) and all the turbulent patches recede in parallel, disappearing one after the other (other panels). The transient ends at $t\approx7,\!750$.
The last experiment of this series, with $\RE= 270$ is not illustrated since it displays the same stages as the previous one. 
The whole relaxation takes a shorter time, the transient ending at $t\approx5,\!250$.
\BF
\BC
\includegraphics[angle=90,width=0.18\TW,clip]{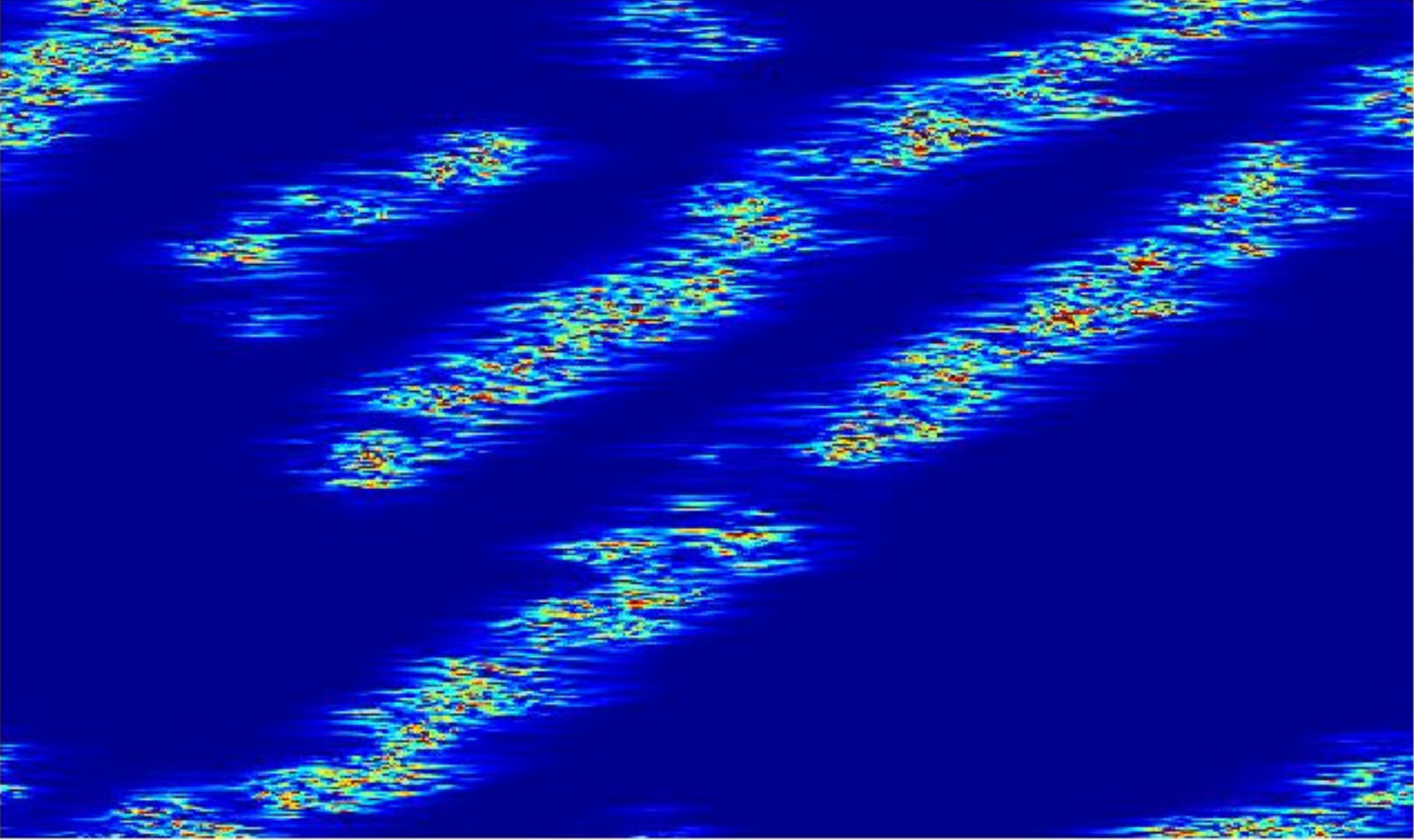}
\hskip0.5em
\includegraphics[angle=90,width=0.18\TW,clip]{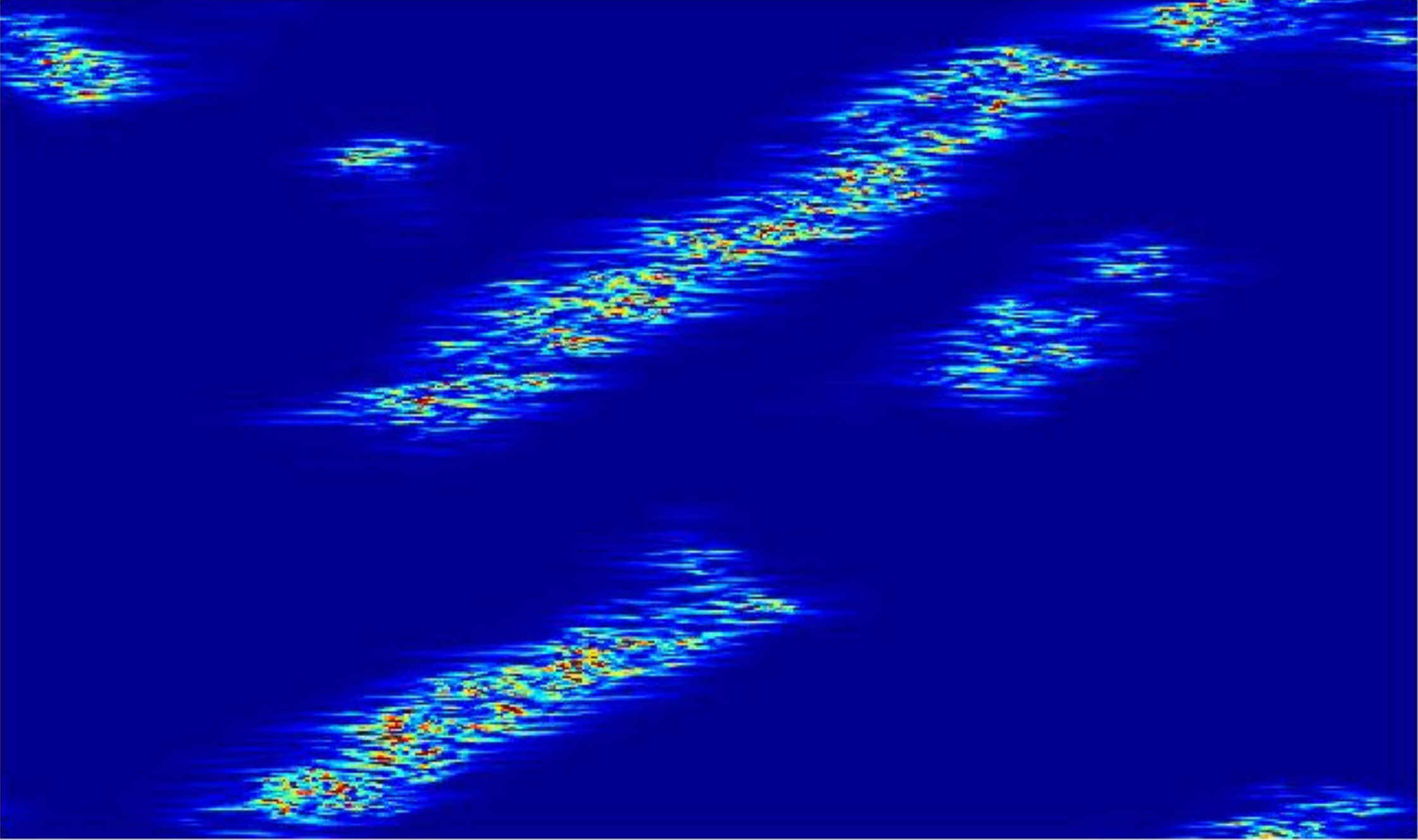}
\hskip0.5em
\includegraphics[angle=90,width=0.18\TW,clip]{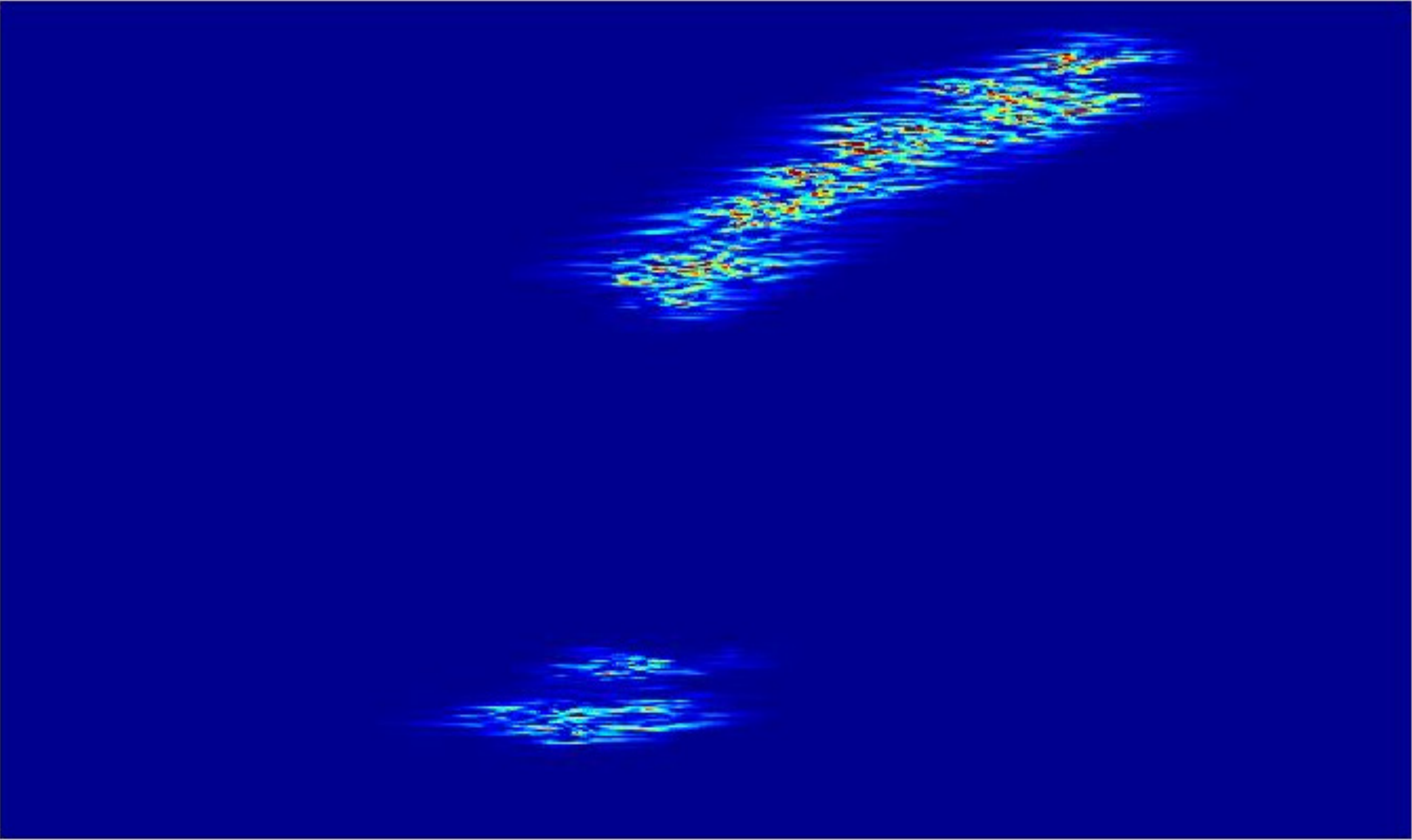}
\hskip0.5em
\includegraphics[angle=90,width=0.18\TW,clip]{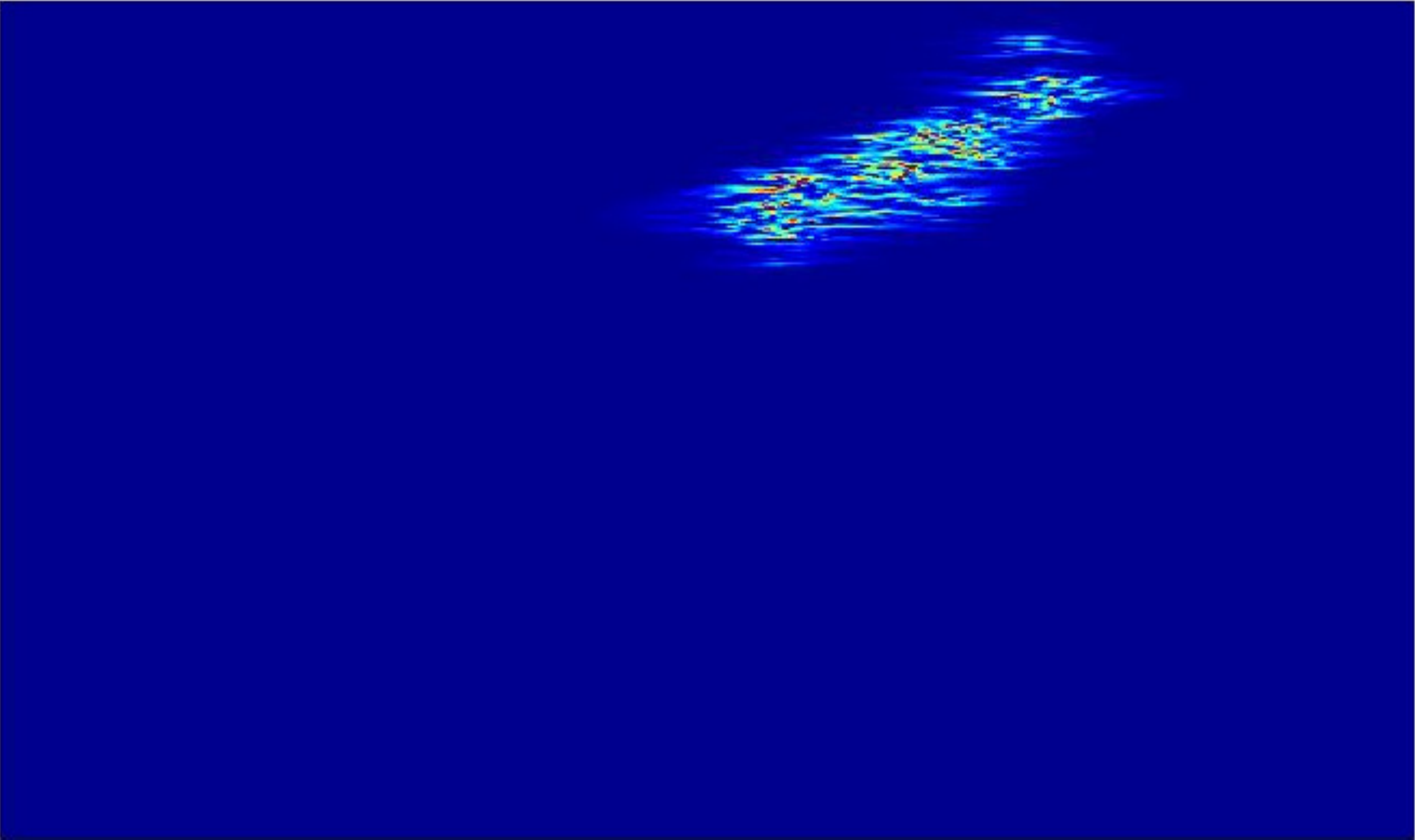}
\hskip0.5em
\includegraphics[angle=90,width=0.18\TW,clip]{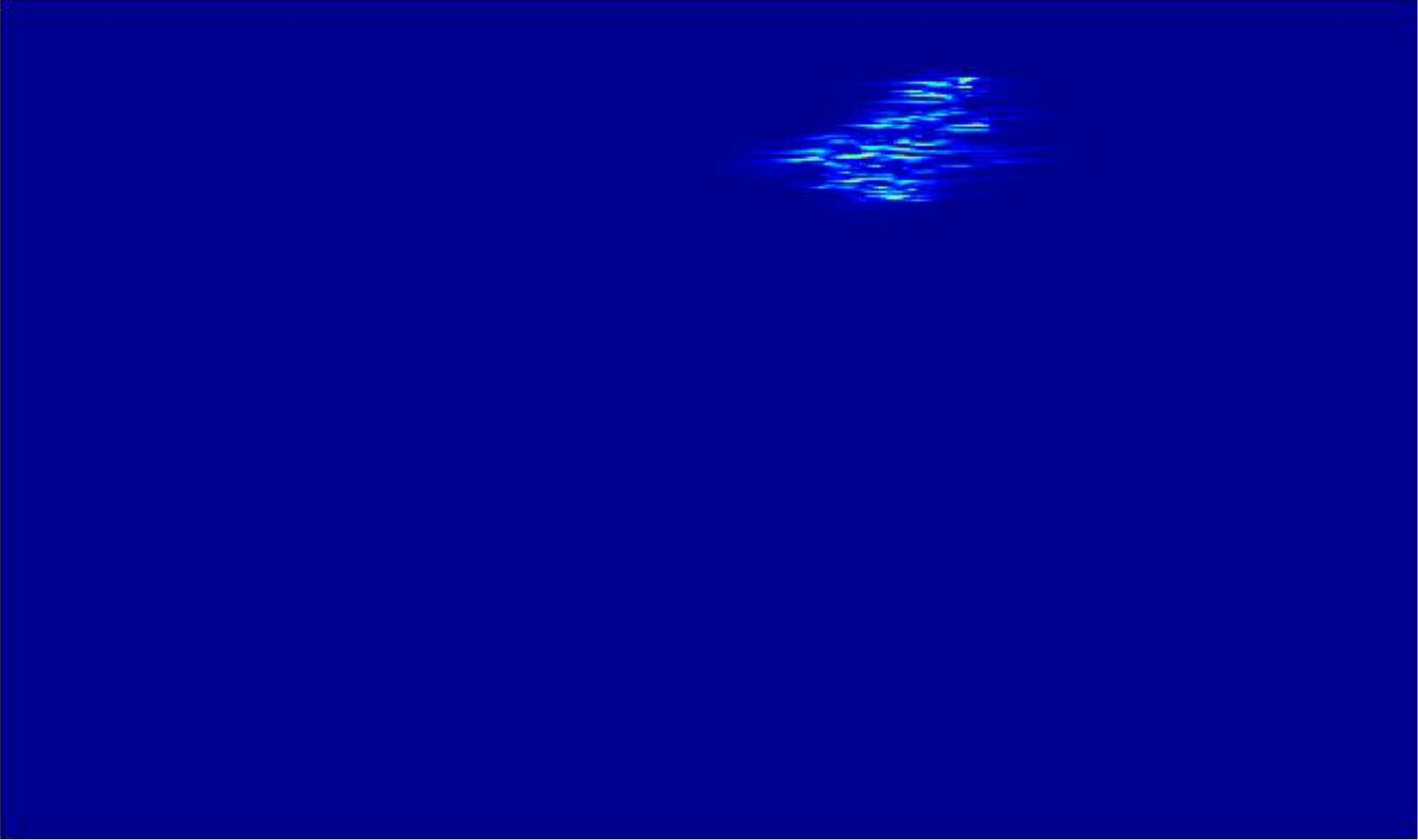}
\EC
\caption{Snapshots taken during the decay at $\RE=271.25$. Time is reset at the beginning
of the experiment. From left to right: $t=3,\!500$; $4,\!500$; $5,\!500$; $6,\!500$; $7,\!500$.
\label{fig7}}
\EF

Before going further, let us  illustrate the use of the filtering-and-thresholding procedure introduced in \S\ref{sft}. Figure~\ref{fig8} corresponds to the transient at $\RE=273.75$ but is typical of the other cases studied. It can be seen that, up to appropriate rescaling, all along the transient down to its end, the mean perturbation energy closely follows the turbulent fraction (which is expected since the perturbation-energy content of the region identified as laminar is negligible if thresholding is properly performed). The continuous-band metastable state observed till $t\approx7,\!000$ is clearly visible. On the other hand, the mean turbulent energy and the local-energy  peak value do not show any change except at the very end of the transient where they abruptly fall to zero. This confirms the relevance of our previous findings based on a different procedure for the identification of the laminar and turbulent local states (Rolland \& Manneville, 2011a) when applied to decaying turbulence.
\BF
\BC
\includegraphics[width=0.7\TW,clip]{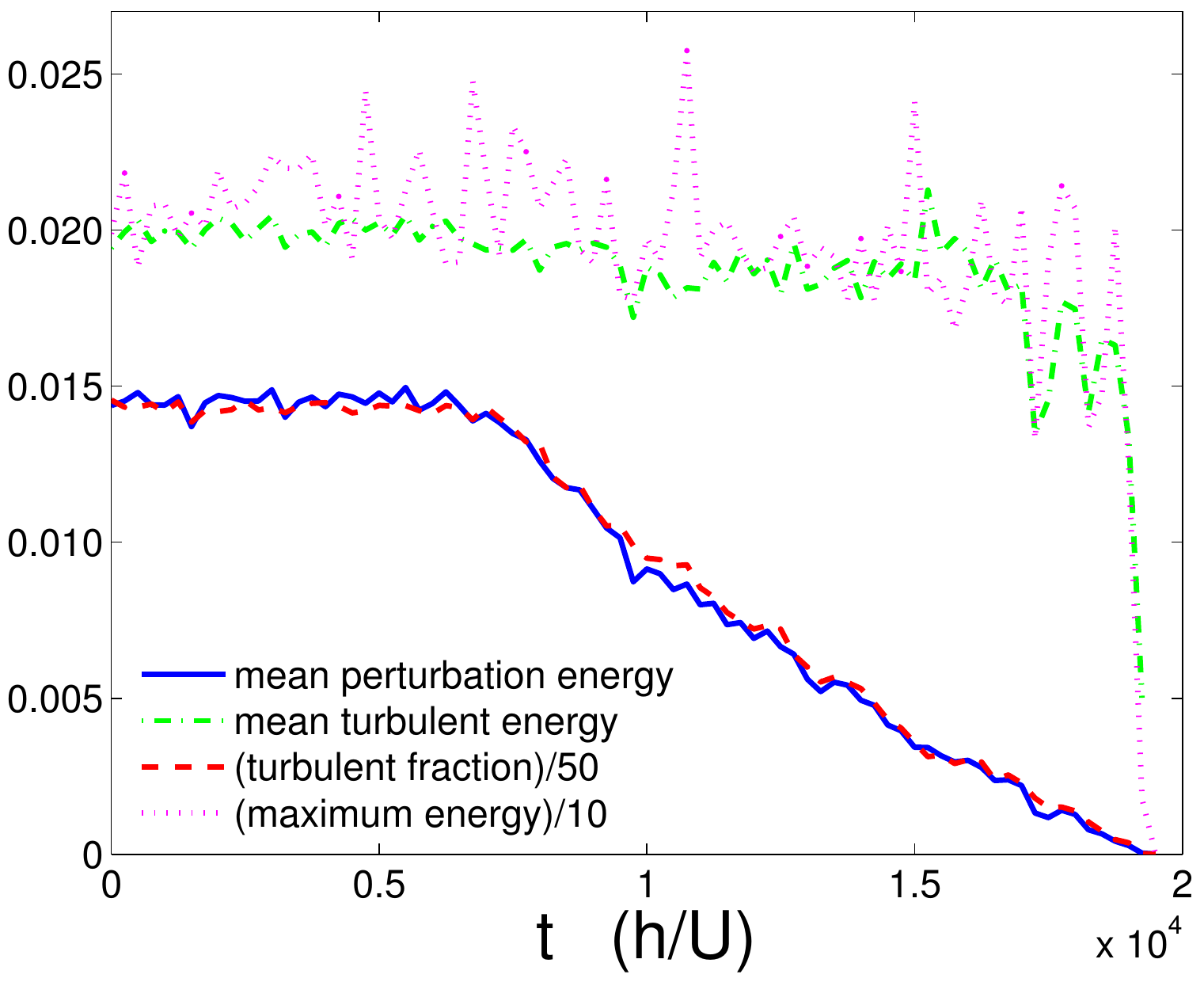}
\EC
\caption{Statistical characterisation of the bands during the transient at $\RE=273.75$ (Fig.~\ref{fig4} and \ref{fig5}) after filtering with $\kappa=2$ and thresholding with $e^{\rm c}=0.003$.\label{fig8}}
\EF

As seen in Figures~\ref{fig5}, \ref{fig6}, and \ref{fig7}, the bands seem to keep a constant width all along their decay.
Two basic processes are at work: first a large-scale collapse of a band in its bulk over a region of length comparable to its width and, next, a slow and regular retreat due to small-scale collapse of the turbulent state at the laminar-turbulent interface.
If the laminar gap appearing in the band is not large enough, the band can recover its continuous character.
In this respect, the continuous bands can be viewed as metastable against the nucleation of a sufficiently large laminar gap, recovery corresponding to germs smaller than critical.
This point of view was put forward by Pomeau in Chapter~4 of Berg\'e {\it et al.} (1998) in the general context of transitional  PCF.
It is here adapted to the band regime that was not yet uncovered at that time.  
Its relevance  is particularly clear at $\RE=273.75$ where the system waits during $7,\!000\,h/U$ before decaying after several unsuccessful attempts, and at $\RE=272.50$ where the two-band state remains in place during a long time interval with similar aborted events.

After a sufficiently wide gap has been nucleated, the steady retraction of the turbulent domains is then easily understood as a propagation problem for a stable state (laminar) that gains over a metastable one (turbulent).
Except when \RE\ is low enough and band breaking becomes a competitive process, the band portions issued from a breaking stay in a single piece and the evolution is essentially a trimming process at both ends.
Owing to the stochastic character of the turbulent state, and the elongated shape of the receding turbulent domains, this second stage then fits the framework put forward by Pomeau (1986) of a contamination process akin to one-dimensional {\it directed percolation\/} (Hinrichsen, 2000): below the percolation threshold, the front separating the absorbing state (laminar) and the active state (turbulent) advances at a statistically constant velocity, so as to make each turbulent band portion retract.
A similar but opposite situation is studied by Y.~Duguet (Duguet \& Schlatter, 2011) who considers the growth of the turbulent state above the percolation threshold in a predominantly one-dimensional geometry elongated in the spanwise direction.

We conjecture that the two processes are different manifestations of the same physics which originates from the chaotic character of the dynamics at the scale of the MFU at values of the Reynolds number for which chaotic transients are expected. As such, the involved probabilities should critically depend on \RE. 
We substantiate this viewpoint by examining these two processes more quantitatively in the next subsections.

\subsection{Nucleation of a laminar gap\label{s3.2}}

Following Prigent {\it et al.} (2003), in our previous work (Rolland \& Manneville, 2011a,b) we analysed the transitional range as a pattern formation problem, using a Langevin--Landau model governing an order parameter submitted to a strong background noise induced by turbulence. 
In particular, we studied orientation changes when bands are well formed in terms of a stochastic process with two wells separated by a barrier, each well corresponding to one of the two  possible orientations.
Changes were then understood as resulting from large deviations due to the cumulative effects of noise.
The problem of band breakdown can be set in a similar framework and seen as a random exit problem (van Kampen, 1983), i.e. when the trajectory in state space reaches a boundary beyond which the system tumbles in a new state. In the present context, as it fluctuates a germ (laminar gap) can grow to some critical size beyond which it is bound to expand indefinitely (turbulent band recedes), while, as long as the boundary is not crossed, the turbulent band recovers (the germ fluctuates back), so that the system has just experienced a large excursion toward the laminar state (aborted event).

Since it has been observed above that the minimum size of a laminar gap leading to the breakdown of a band is of the order of the width of the turbulent  band itself, we can try to relate the nucleation problem to the collapse of turbulence in domains of sizes $L_x\times L_z$ of the order of the size of the elementary pattern cell $ \lambda_x\times\lambda_z$.
Afterwards, we can consider the whole band as formed from a series of such elementary cells, making the supplementary assumption that these cells are independent so that the collapse probabilities simply add.
Strictly speaking, this assumption is certainly wrong owing to long range interactions induced by large scale flows outside the turbulent bands (Barkley \& Tuckerman, 2007) but it can serve as a guide.
Accordingly we study excursions away from the oblique-band turbulent state in systems of sizes of the order of  $\lambda_x\times\lambda_z$ hence accommodating a single band, see Fig.~\ref{fig9}, left-most image. We mainly consider $L_x^{\rm ss}=144$ and $L_z^{\rm ss}=84$, where superscript `ss' means `small system'.
This corresponds to  $\lambda_x$ and $\lambda_z$ exactly one-third of $L_x$ and $L_z$, as obtained in our main simulations.
Computations are performed at the same local resolution as for the large system, i.e. $N_x^{\rm ss}=L_x^{\rm ss}$ and $N_z^{\rm ss}=3L_z^{\rm ss}$. Figure~\ref{fig9} displays the opening of the laminar gap in the band for $\RE=271.25$ in such a domain.
\BF
\BC
\includegraphics[angle=90,width=0.10\TW,clip]{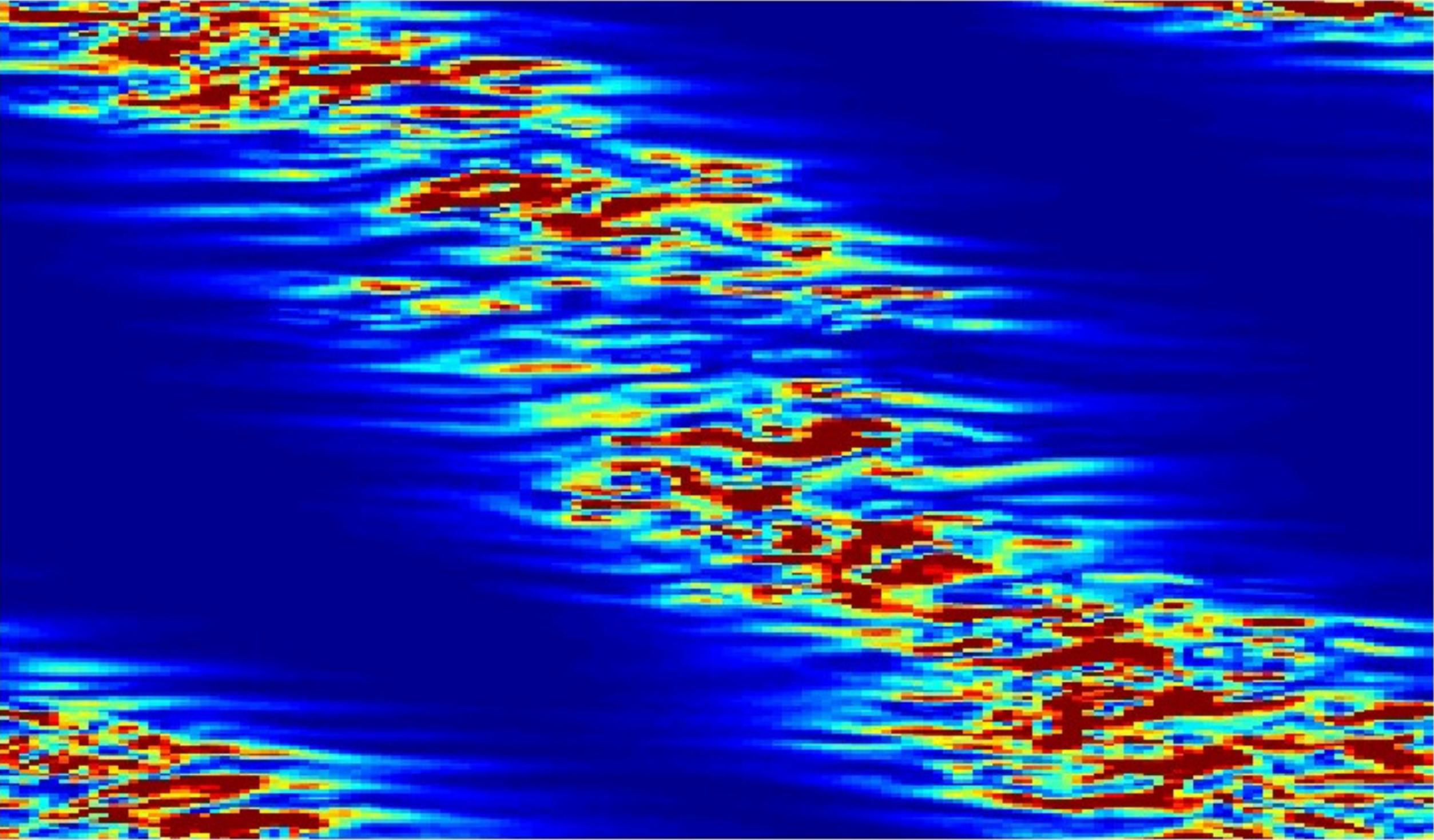}
\hfill
\includegraphics[angle=90,width=0.10\TW,clip]{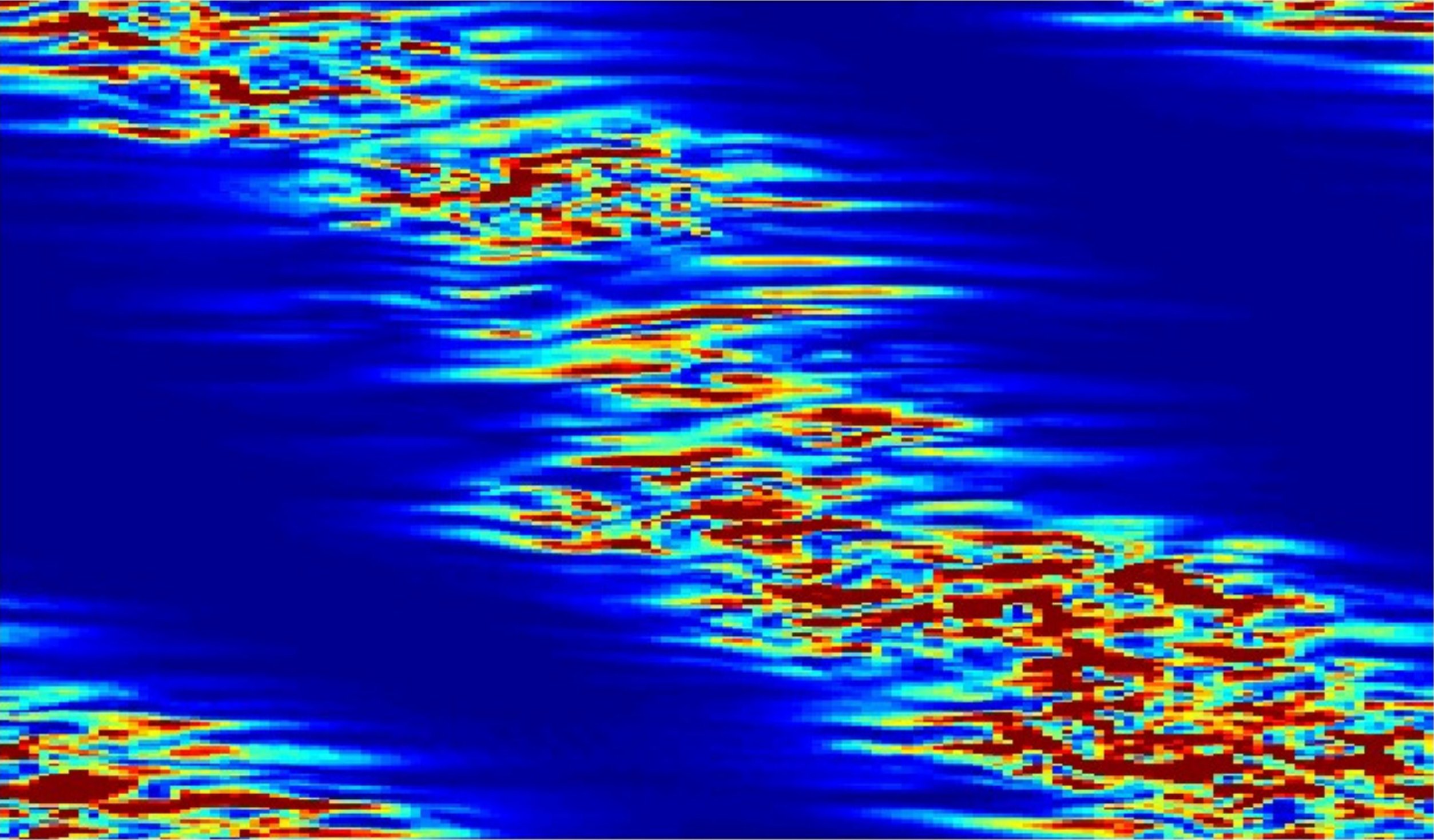}
\hfill
\includegraphics[angle=90,width=0.10\TW,clip]{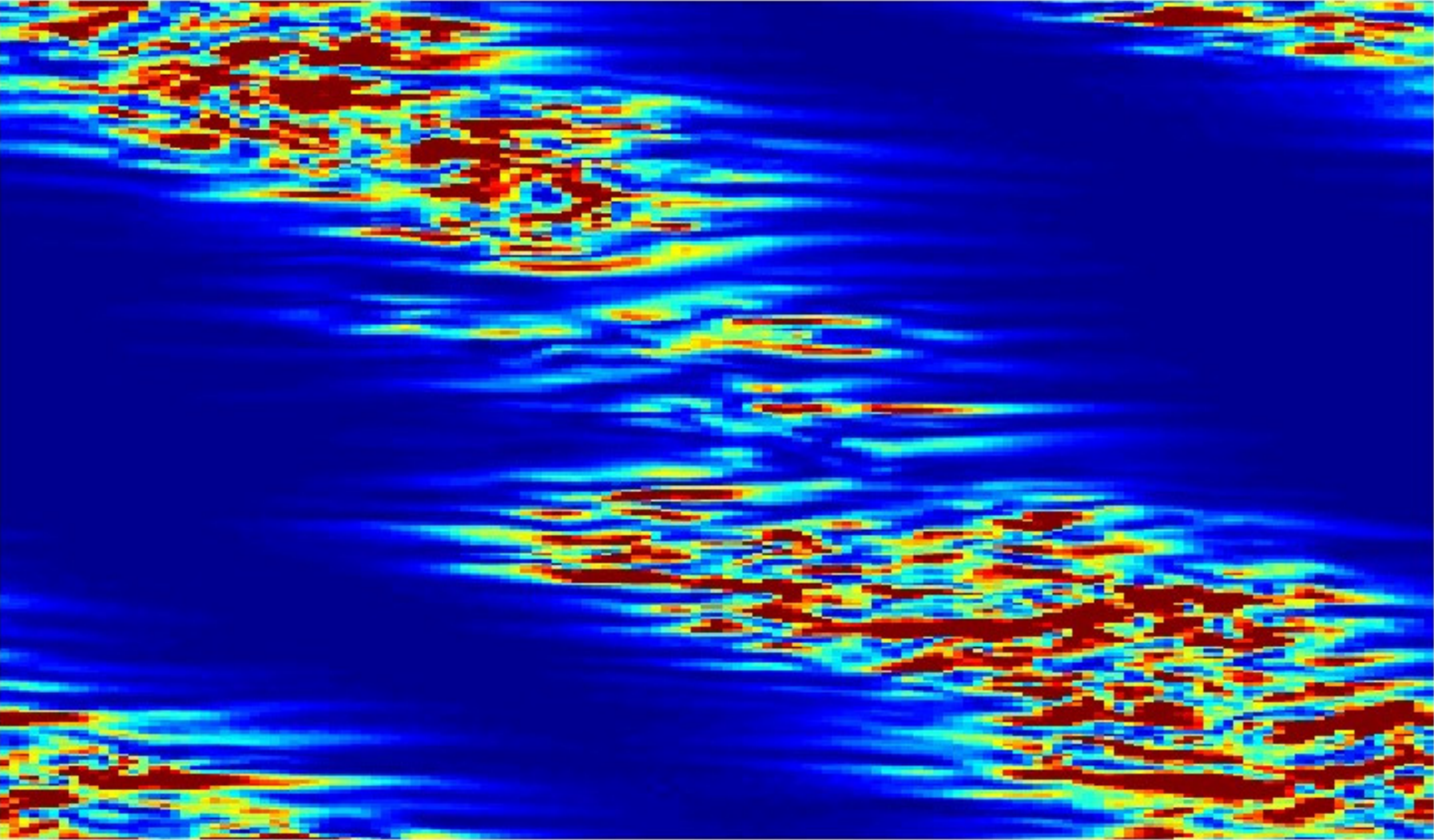}
\hfill
\includegraphics[angle=90,width=0.10\TW,clip]{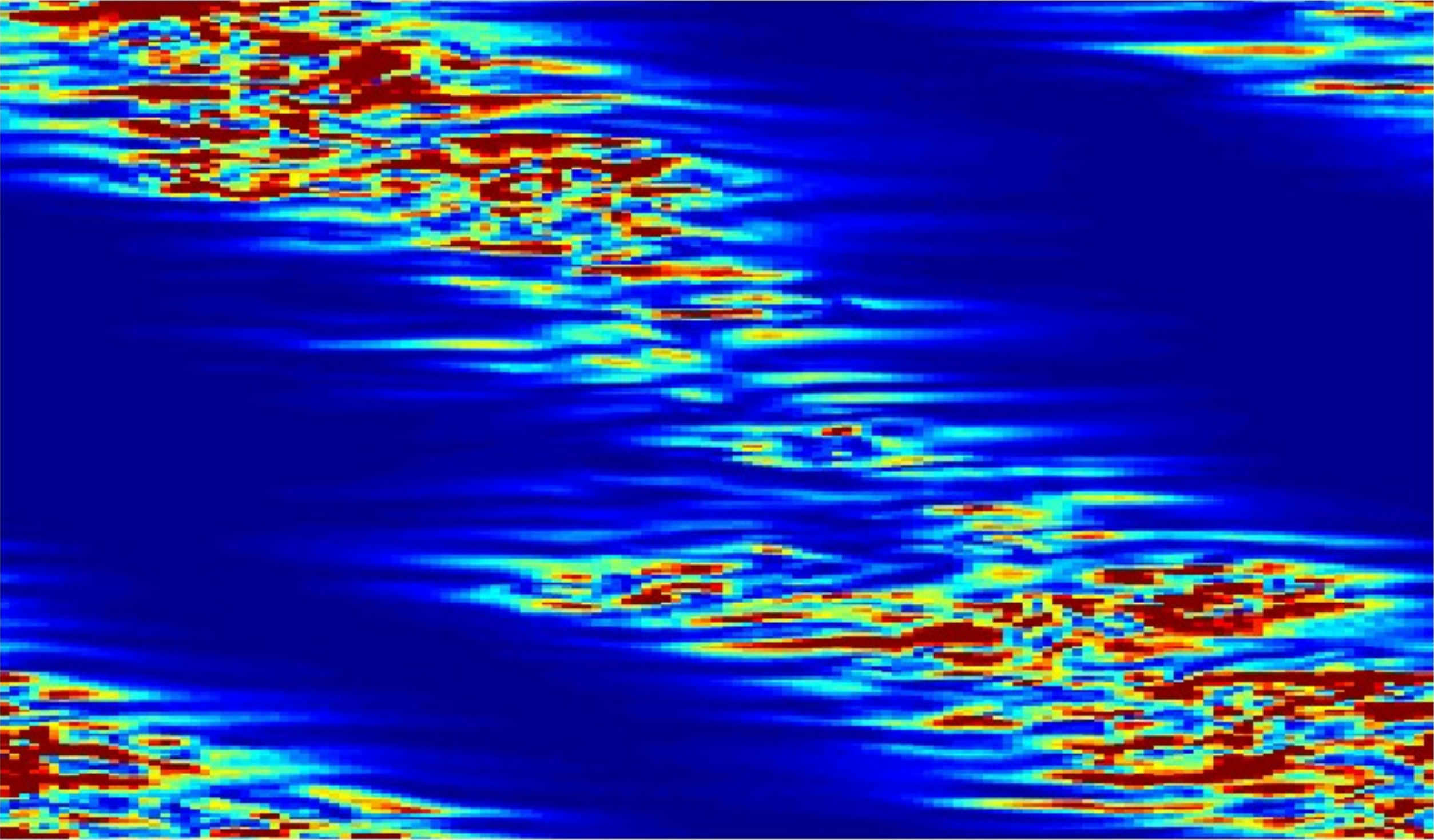}
\hfill
\includegraphics[angle=90,width=0.10\TW,clip]{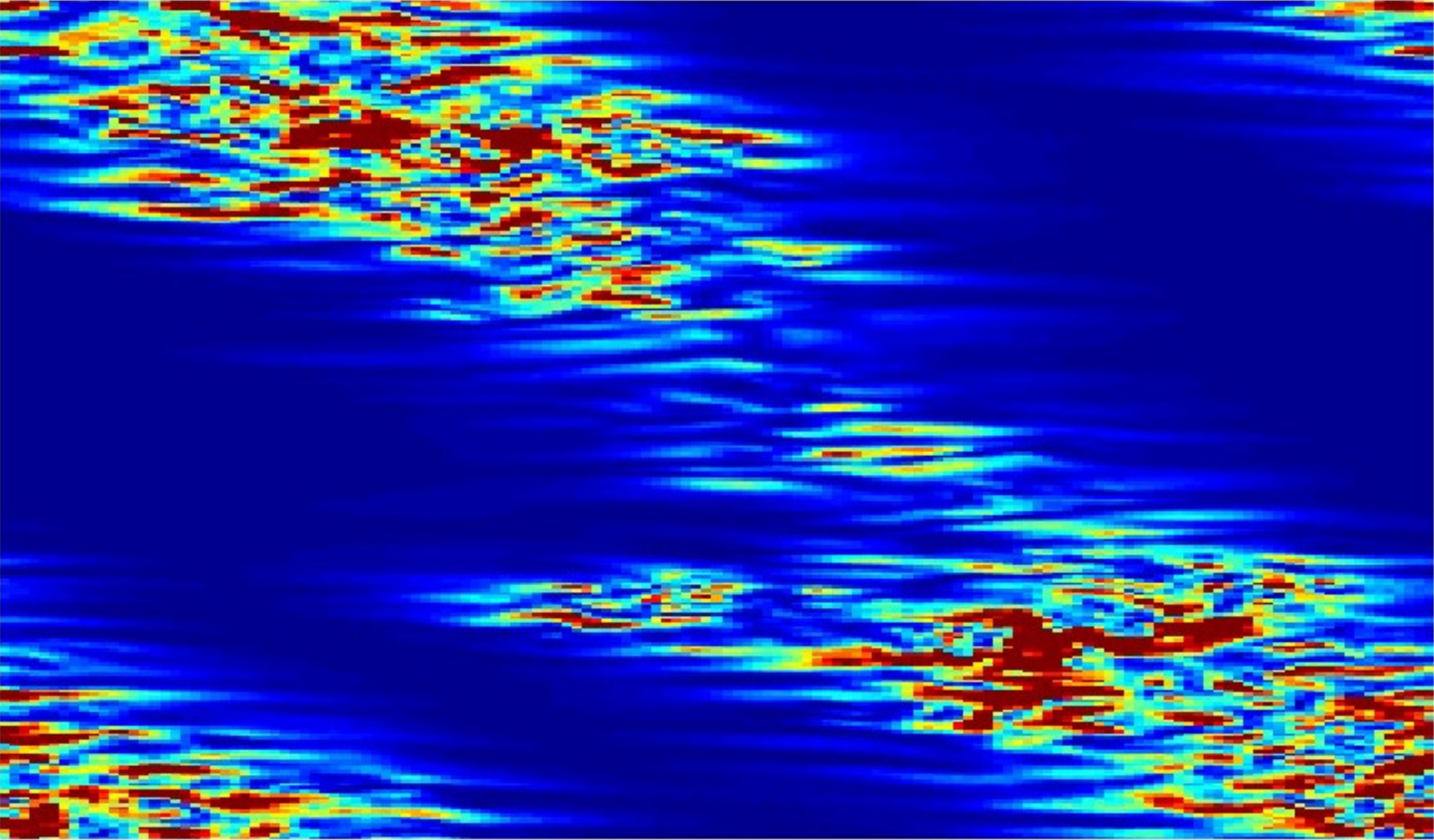}
\hfill
\includegraphics[angle=90,width=0.10\TW,clip]{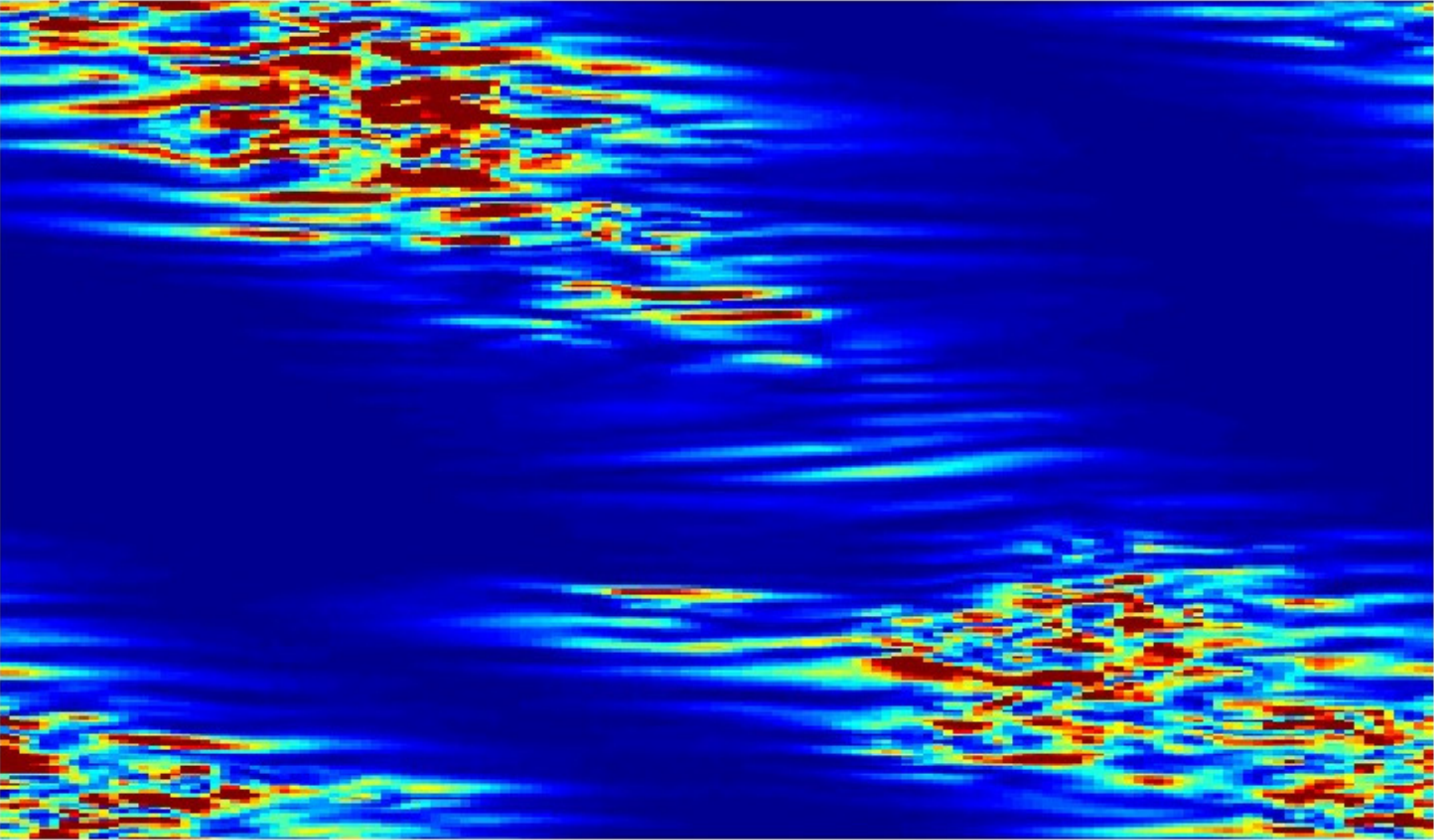}
\hfill
\includegraphics[angle=90,width=0.10\TW,clip]{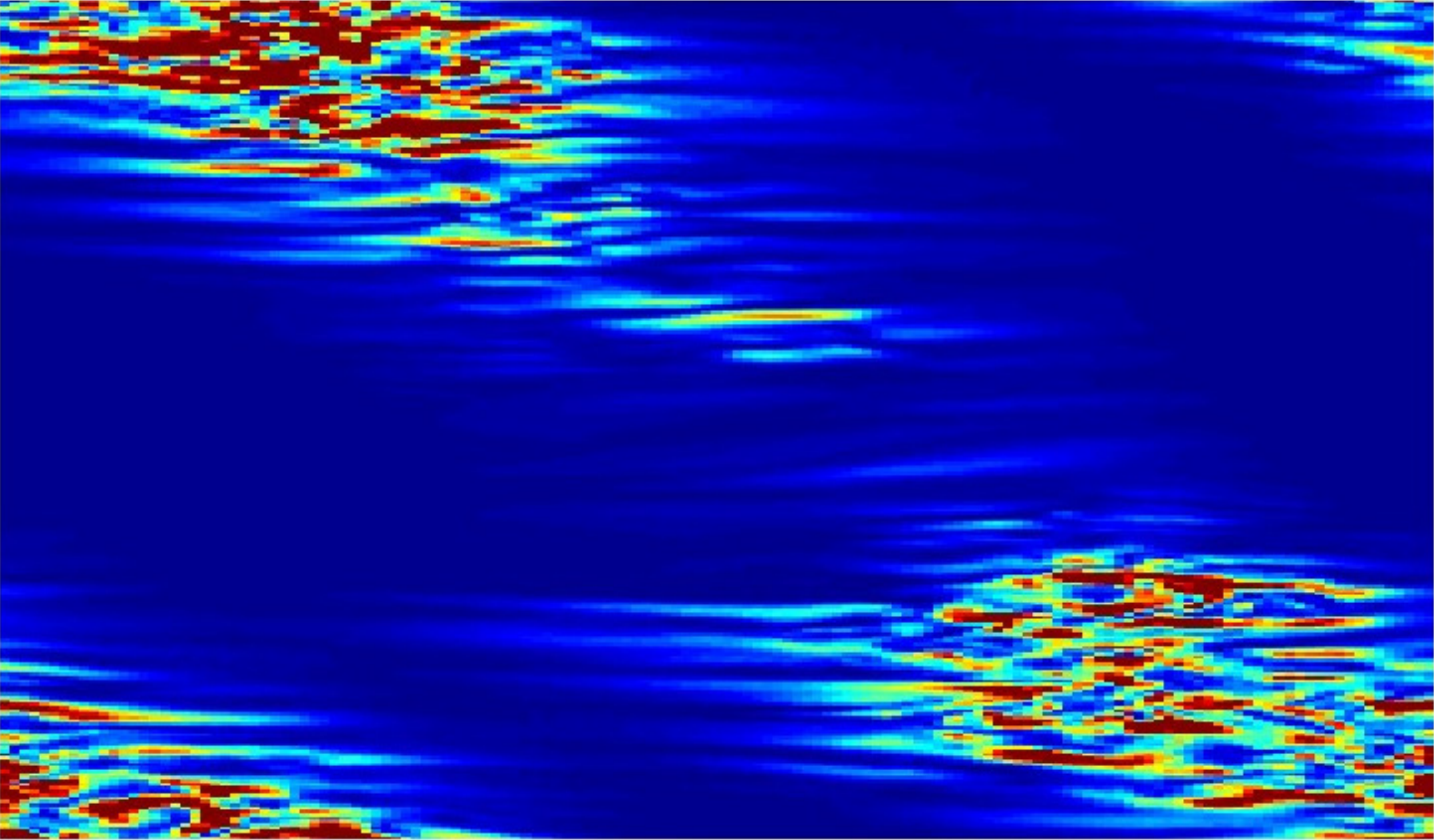}
\hfill
\includegraphics[angle=90,width=0.10\TW,clip]{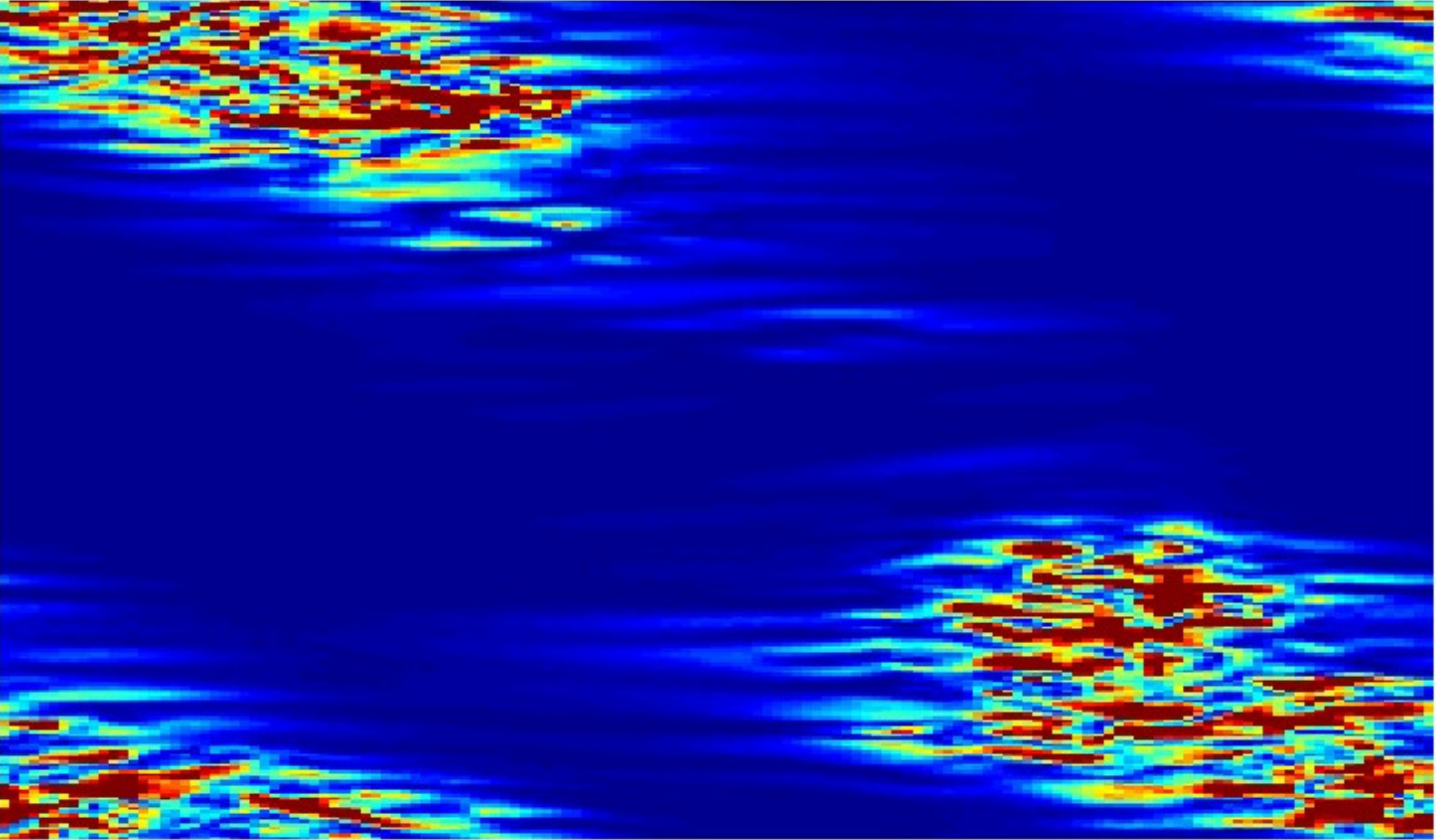}
\hfill
\includegraphics[angle=90,width=0.10\TW,clip]{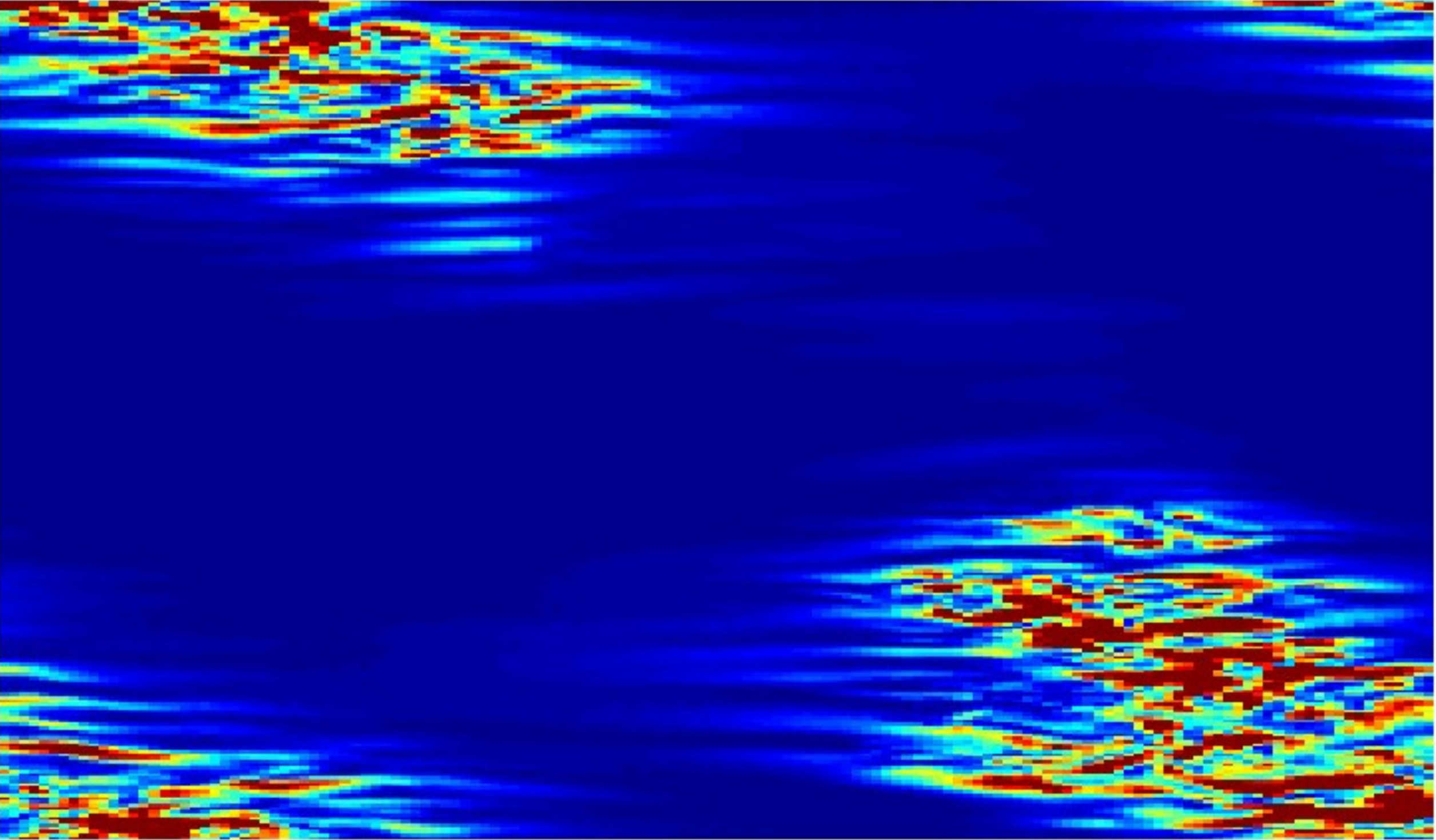}
\EC
\caption{Snapshots taken during the fatal opening of a laminar gap in the band for the small domain ($144\times84)$ at $\RE=271.25$.
From left to right, time increases by steps of $50\>h/U$ from $t=2,\!700$ on.
Remember the periodic boundary conditions.
\label{fig9}}
\EF
After a few aborted events, this opening is fatal to the continuity of the band and the transient terminates about $1,\!600\>h/U$ later.
In line with the basic assumption of a cumulative effect of local noise driving the excursions away from the turbulent state, we base our study on the distance $D$ to the laminar state.
In a large deviation perspective, this quantity is roughly proportional to the amount of local turbulence at the origin of the stochastic dynamics and therefore appropriately related to the fluctuating position of the laminar-turbulent interface in an additive way. The changes in the statistics of the time series of $D$ as a function of \RE\ are thus our primary matter of interest.

Figure~\ref{fig10} (top) displays the time series of $D$ for $\RE=275$ over a very long time lapse ($3\times10^5\>h/U$).
\BF
\BC
\includegraphics[width=0.90\TW,clip]{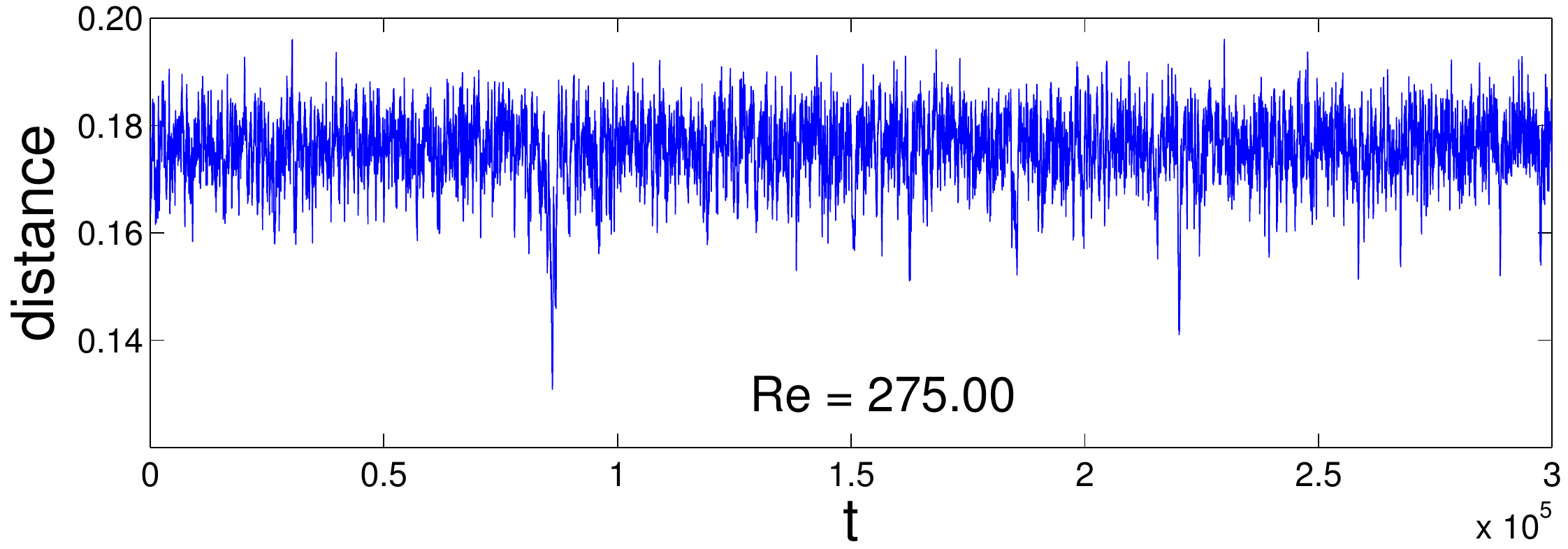}\\[2ex]
\includegraphics[width=0.90\TW,clip]{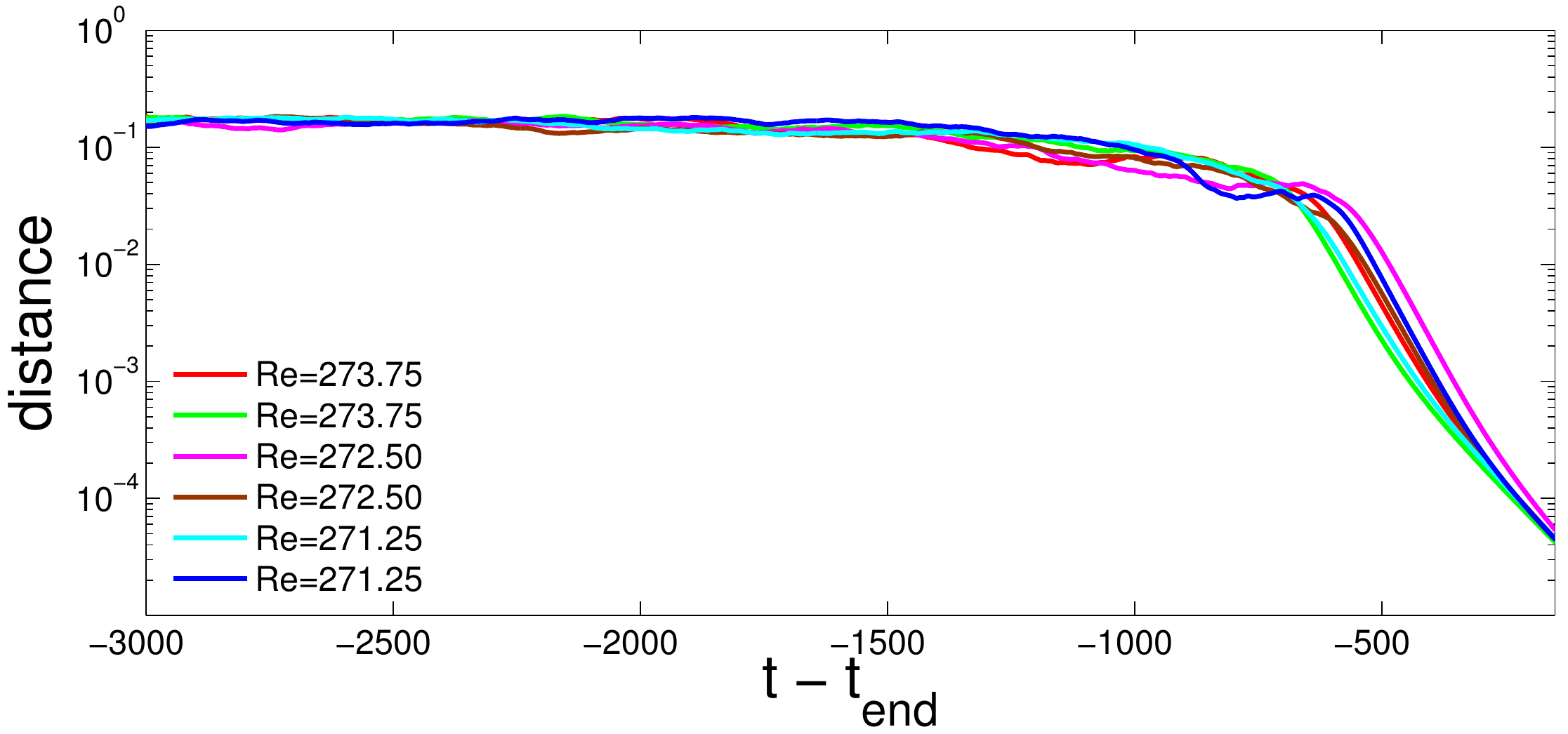}
\EC
\caption{Time series of the distance to laminar flow $D$. Top: for $\RE=275$.  
Bottom: during the last decay stage for several transients and various $\RE<275$; the origin of times, $t=0$, is set when $D=10^{-5}$.
\label{fig10}}
\EF
As far as one can tell from one numerical experiment, this specific  turbulent regime is considered as sustained.
Excursions such as the one at $t\approx8,\!500$ are not sufficiently dangerous nor sufficiently frequent to make the system decay to laminar flow.
By contrast, for $\RE<275$, decay is observed in finite duration simulations, owing to the more frequent occurrence of more dangerous excursions.
As can be appreciated from Figure~\ref{fig10} (bottom), whatever the value of \RE, decay is seen to follow similar paths and the stage of irreversible relaxation to laminar flow takes place as soon as $D$ is smaller than about $0.03$.
A statistically robust trend is however already observed when $t$ is about $1,\!500\,h/U$ before the end of the transient (arbitrarily fixed when $D$ reaches $10^{-5}$), which corresponds to the opening time of the fatal laminar gap, see Fig.~\ref{fig9} and related text.
The instantaneous value of $D$ then begins to depart from its mean as computed by discarding the last $2,\!000\,h/U$ before decay.
The slow decrease that follows the opening corresponds to a regular retreat of the remaining turbulent band fragment very much like in the wide system.
Further examination of the data suggest that, for \RE\ around 275, this happens systematically when $D$ crosses $\approx0.13$ from above and that, when this is the case, the system does not return to the oblique-banded turbulent state and is bound to decay within the next 1,500 $h/U$.

Probability density functions (PDF) of the instantaneous value of $D$ are then constructed. They are displayed in Figure~\ref{fig11} for different values of \RE. In this figure one can see that the most probable value slowly decreases with \RE\ but more importantly that the probability of low values, corresponding to the dangerous excursions, increases drastically in a narrow range of \RE.
Decay has not been observed for $\RE\ge275$ in any experiment of duration $\le 3\times10^5\>h/U$.  Extrapolation of the exponentially decreasing tail in the corresponding PDF however suggests that this event has a very low but finite probability of occurrence, i.e. could only be observable in simulations much longer than $10^5$.
For the transient cases, the statistics rests only on simulations longer than 6,000 $h/U$ and on the parts of the time series excluding the 2,000 last time units, cumulating at least $10^5$ measures per value of \RE, thus requiring the study of several independent transients (up to about ten for $\RE=271.25$).
\BF
\BC
\includegraphics[width=0.70\TW,clip]{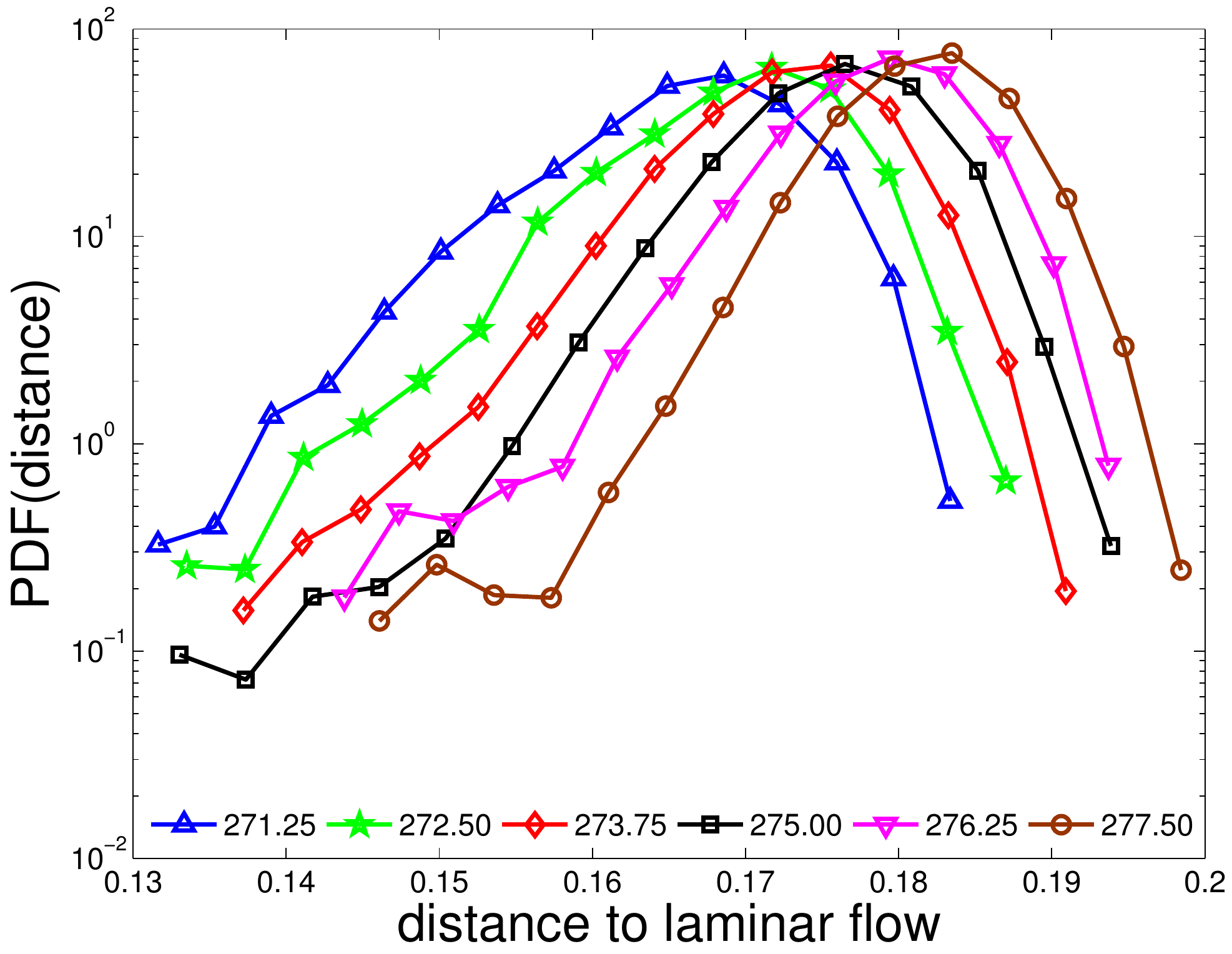}
\EC
\caption{Probability distribution functions of the fluctuations of the distance to the laminar state for different values of \RE, see text for details.
\label{fig11}}
\EF

It does not seem necessary to go beyond this hand-waving approach to the band breaking process because, though the trends shown are certainly real and qualitatively correct, the results are quantitatively sensitive to the limited set of trajectories that are followed and not necessarily representative of a genuine ensemble average.
As already mentioned, the resolution of the simulation is also at the origin of the general downward shift of the transitional range that limits extrapolation to the fully resolved case.
The size of the pattern's elementary cell also plays a role:
A similar study has been performed with $L_x^{\rm ss}=108$ and $L_z^{\rm ss}=64$, one-fourth of the big system's dimensions, thus keeping the same angle but corresponding to a smaller wavelength.
Results, not shown, are quantitatively similar, but with a slight downward shift of the corresponding range of \RE. Decay turns out to be more difficult (less probable) in the $108\times64$ domain than in the $144\times84$ domain, which can be understood as due to stronger constraints brought by the periodic boundary conditions at a smaller distance, yielding narrower laminar stripes between the turbulent bands and discouraging the formation of the large laminar patches associated with the dangerous excursions of $D$ toward low values, in agreement with similar observations made in small systems designed to check the Ginzburg--Landau approach (Rolland \& Manneville, 2011a).

Notice however that, still in the large deviation perspective, consideration of Fig.~\ref{fig11}, offers an explanation to the observed exponential decrease of transient lifetime distributions that is alternative to that provided by the chaotic transient paradigm (Eckhardt {\it et al.}, 2008). Whereas the latter is undoubtedly appropriate for systems at the MFU scale, i.e. confined, it fails to account for spatiotemporal features inherent in extended systems of sizes sufficient to accommodate turbulence modulation, i.e. $\approx \lambda_x\times\lambda_z$ or greater. Indeed, upon assuming that the trajectory followed by the system at given \RE\ samples the PDF of $D$ at random and that return to laminar flow inevitably takes place within a short time when $D$ happens to reach a value below some cut-off (here $\approx 0.13$), decay can then be seen as a Poisson process controlled by the corresponding small probability (hence exponential lifetime distributions when ensemble of transients are considered) which rapidly decreases as \RE\ increases (hence a rapid increase of the mean lifetime). This explanation ties up with that of Goldenfeld {\it et al.} (2010) about decaying puffs in pipe flow.

\subsection{Withdrawal velocity\label{s3.3}}

The second process involved in the decay is a statistically steady retreat of segments issued from the breaking of the continuous bands. These segments get shorter and shorter but keep roughly constant widths, except in the very last stage where they collapse and finally relax under the dominant effect of viscous dissipation as shown in Figure~\ref{fig7}.
The turbulent fraction, or equivalently the mean energy $E$ since the thresholding procedure collects the essentials of the perturbation energy (Fig.~\ref{fig8}), are thus indirect measures of their cumulated length.
An average withdrawal speed can be determined from the time variation of the distance to the laminar flow $D=\sqrt{2E}$.
Figure~\ref{fig12} displays part of the data already presented in Fig.~\ref{fig4} but in addition fits the variations of $D$ against affine laws during specific evolution stages. 
\BF
\BC
\includegraphics[width=0.70\TW,clip]{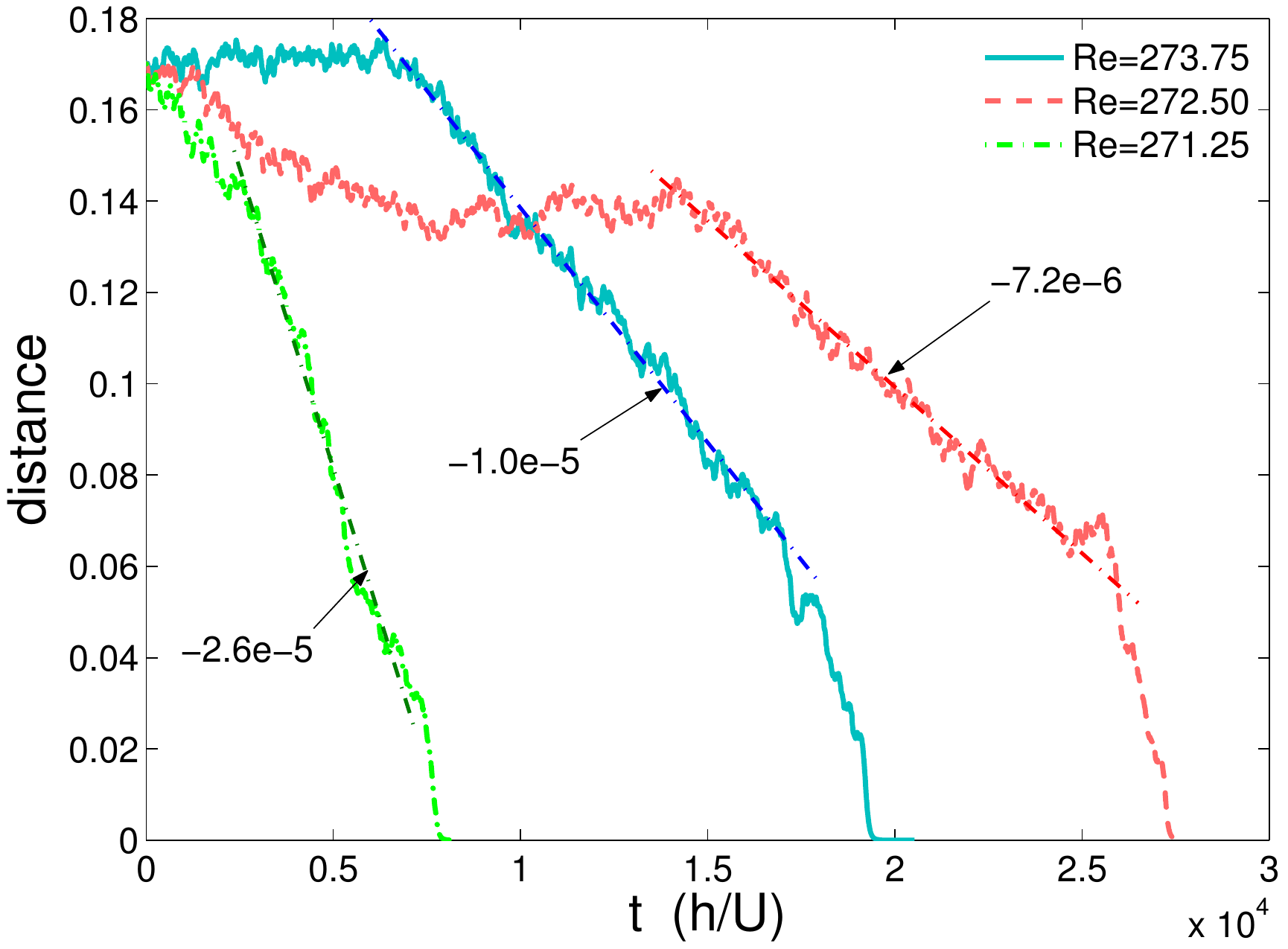}
\EC
\caption{Distance $D$ to laminar flow as a function of time during decay for $\RE=273.75$, 272.50, and 271.25.  Specific evolution stages correspond to steady retreat of the turbulent band segments at speeds that can be estimated from linear fits (dash-dotted lines). 
\label{fig12}}
\EF
Examining the transients one by one (Figs.~\ref{fig5}, \ref{fig6}, \ref{fig7}), it can be seen that, during each of these stages, the connectedness of the solution does not change much, with essentially the same number of band fragments.
It suggests also that the largest contribution to the evolution comes from the most active parts of a band segment, namely its ends.
In the spirit of a directed percolation process, we can then think that the withdrawal at a given end depends on the decay probability of local streaky structures, which at this stage, is a weakly varying function of \RE\ in the vicinity of \RG\ that we are considering.
The global withdrawal speed as determined from the variations of $D$ can then be related to the number of segments involved.
This immediately explains why the decay observed for $\RE=273.75$, with several bands receding simultaneously (Fig.~Ê\ref{fig5}), is faster than the decay at $\RE=272.50$, and also why, in this last case, no slope change is detectable when the third band breaks ($t\sim20,\!600$) since,  at a given time, only one band is receding, hence only two active ends (Fig.~\ref{fig6}).

Like in the previous subsection, it is however neither necessary nor useful to go beyond this qualitative approach.
First, as above, limitations due to under-resolution impede reliable predictions for the fully-resolved and the experimental  cases.
Second and more importantly, it seems extremely difficult to relate the global withdrawal speed to the local mechanisms operating in the `active' regions and in particular to extract probabilities relevant to the decay of the band segments. In this respect, the situation is less favourable than in the case considered by Duguet \& Schlatter (2011) who work with a streamwise-narrow domain. The reason is that the general topology of the solution at a given time, and in particular its connectedness, plays a crucial role.
A continuous turbulent band can be prolonged to infinity using the periodic boundary conditions, is effectively endless, and has properties different from those of a very long band segment:
We have good evidence that,  at $\RE=275$, solutions formed with continuous bands are stable -- or at least prone to band breaking with such a low probability that it  could not be observed during simulations of duration much longer than $25,\!000\>h/U$, while a band segment at the same \RE\ does recede, as illustrated in Figure~\ref{fig13}.
\BF
\BC
\includegraphics[angle=90,width=0.18\TW,clip]{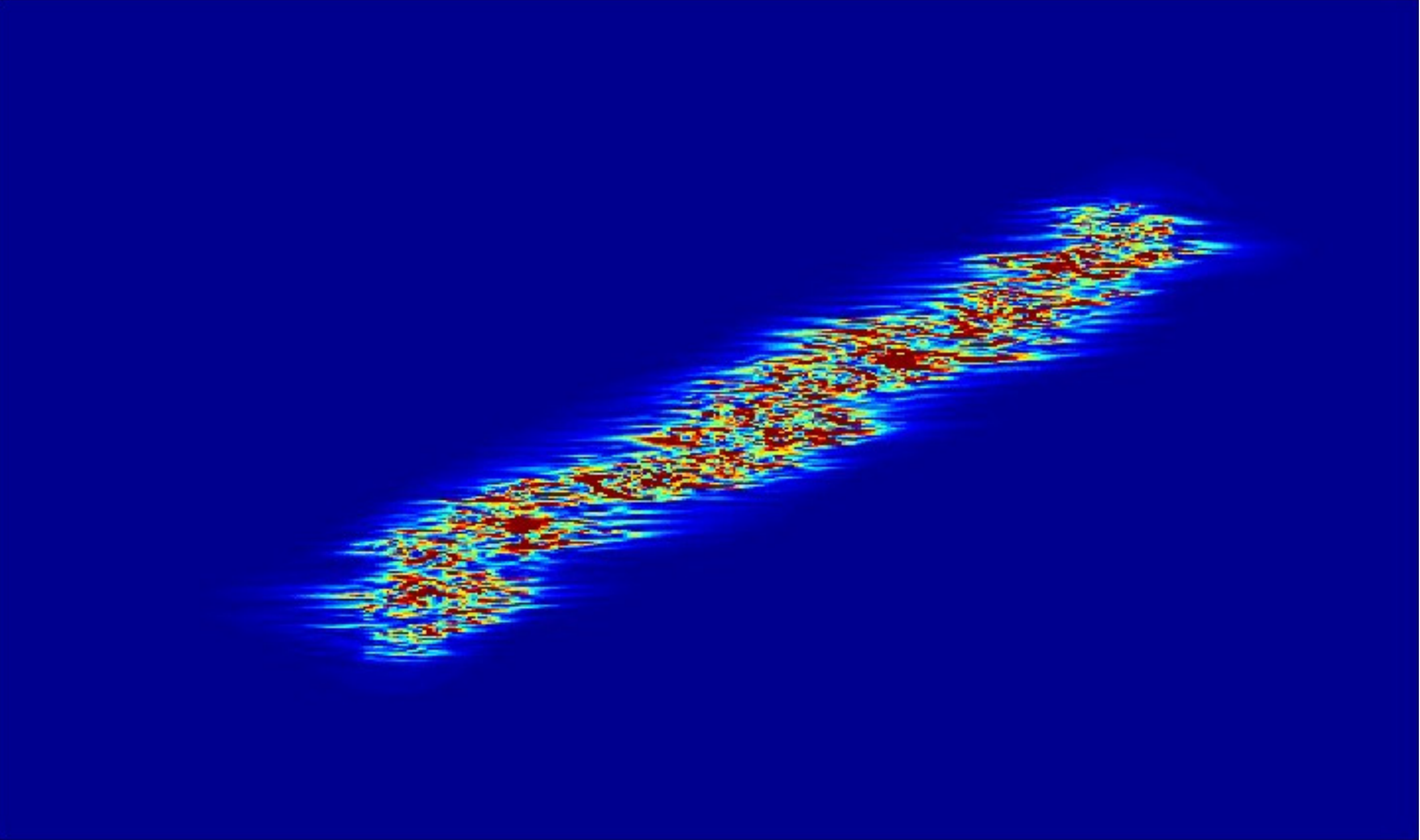}
\hskip0.5em
\includegraphics[angle=90,width=0.18\TW,clip]{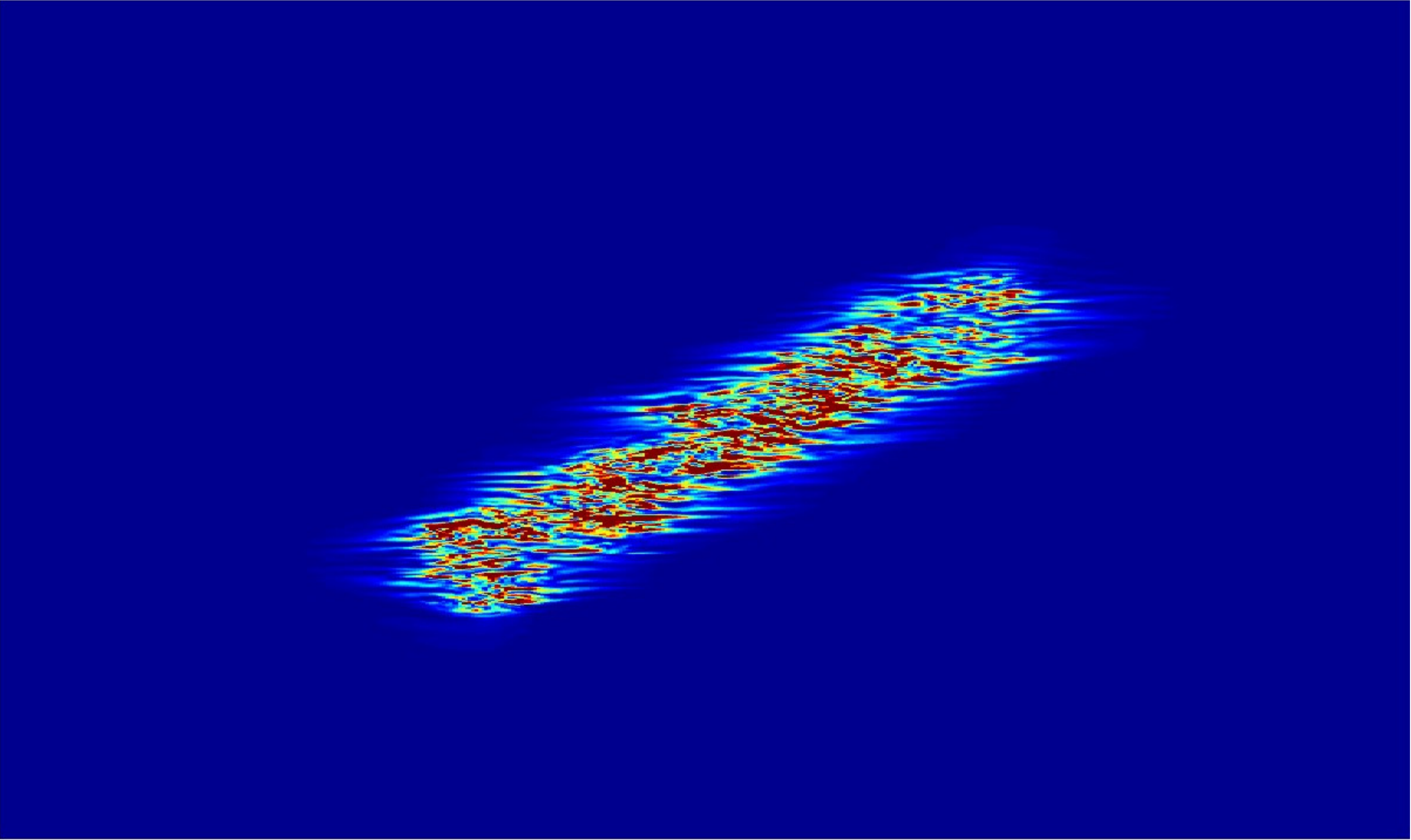}
\hskip0.5em
\includegraphics[angle=90,width=0.18\TW,clip]{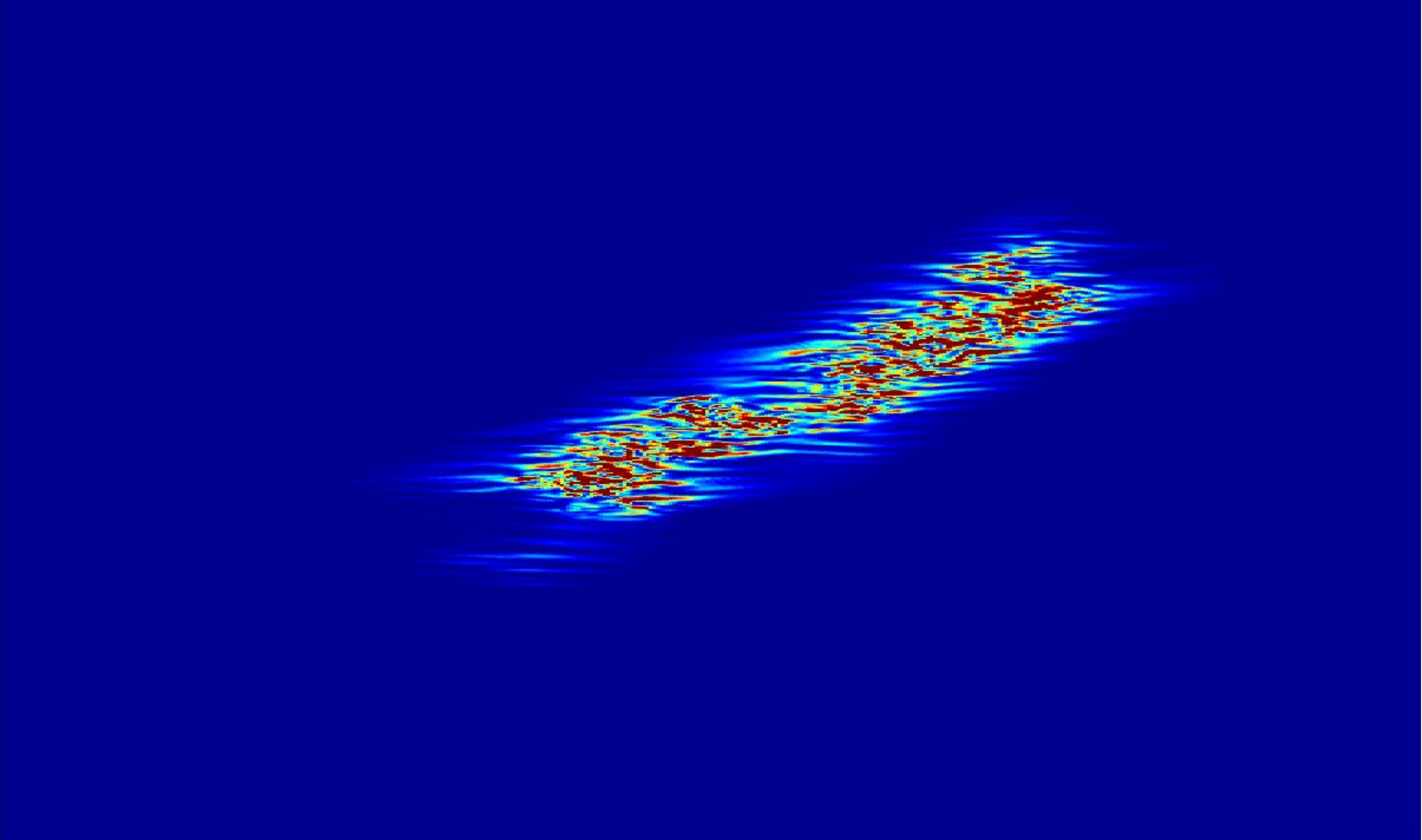}
\hskip0.5em
\includegraphics[angle=90,width=0.18\TW,clip]{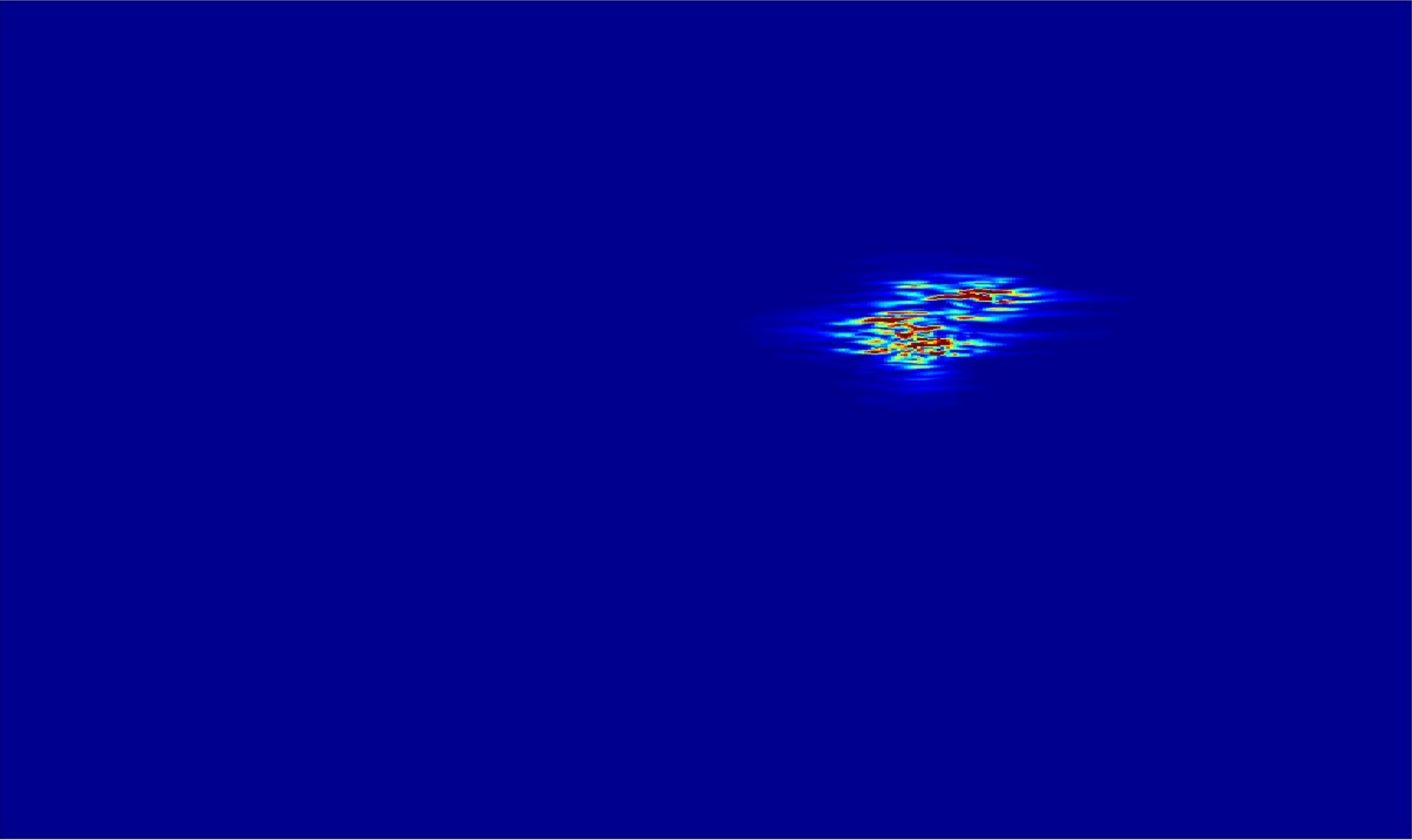}\\[2ex]
\includegraphics[width=0.70\TW,clip]{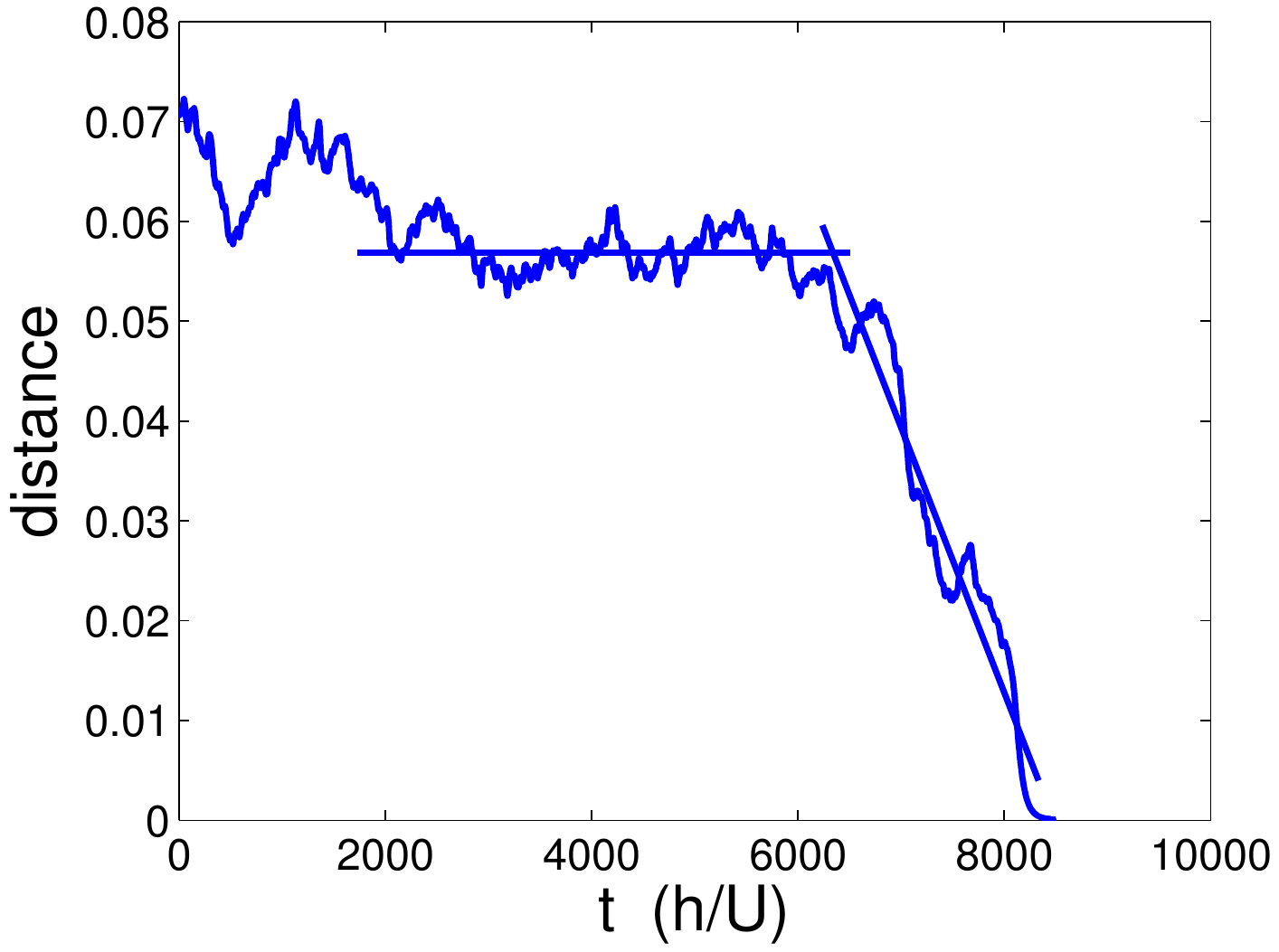}
\EC
\caption{Decay of a band-like but topologically disconnected solution at $\RE=275$. From left to right: $t=0$, 2500, 6000, and 7500.
Bottom: distance to laminar flow during the transient.\label{fig13}}
\EF
The initial condition corresponding to this case is the state obtained at $\RE=272.5$ after the breaking of the last band at $t=24,\!000$ (Fig.~\ref{fig6}, right).
This initial condition is reproduced in Fig.~\ref{fig13} (top-left) after re-centering by taking advantage of the periodic boundary conditions. The other images have also been re-centered by the same amount.
The graph in Figure~\ref{fig13} (bottom) displays the variation of $D$ during this transient, with a plateau from $t\sim 2500$ (2nd image) to $t\sim6000$ (3rd image) preceding a fast collapse ending by the final viscous decay stage.

This observation suggests us that the local interpretation of turbulence dynamics in the vicinity of \RG\ sketched above has certainly to be corrected from non-local effects:
Large scale recirculations with well-defined wall-normal  structure are known to exist around turbulent bands (Coles \& van Atta, 1967; Barkley \& Tuckerman, 2007), as well as around turbulent spots (Lundbladh \& Johansson, 1991; Lagha \& Manneville, 2007b). These recirculations do not average to zero over the gap so that $v_{x,z}^{\rm mean}(x,z,t)=\frac12\int_{-1}^{+1}\mathrm d y\,v_{x,z}(x,y,z,t)$ is a good signature of their presence. 
Figure~\ref{fig14} (right) displays the streamlines of this in-plane average flow. 
A mild filtering (parameter $\kappa=1$) has been performed which has some smoothing effect only on the flow lines inside the turbulent region and not on its periphery.
The right panel  of Fig.~\ref{fig14} shows that this flow stays directed parallel to the turbulent segment in its central portion, much like along the continuous band in the left panel,  while, near the ends of the segment, it adopts a configuration reminiscent of what is observed for turbulent spots surrounded by a quadrupolar large-scale flow entering streamwise at their tips and getting out spanwise on their sides (Lundbladh \& Johansson, 1991; Lagha \& Manneville, 2007b).
\BF
\BC

\includegraphics[angle=90,width=0.20\TW,clip]{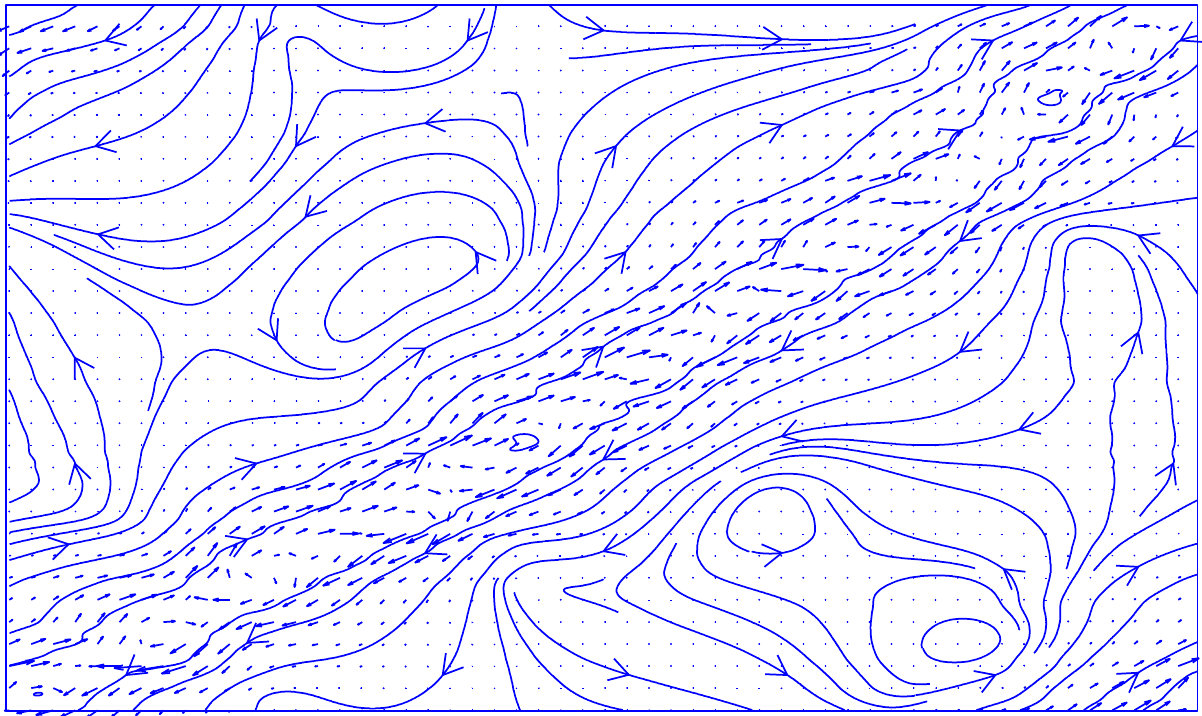}
\hskip4em
\includegraphics[angle=90,width=0.20\TW,clip]{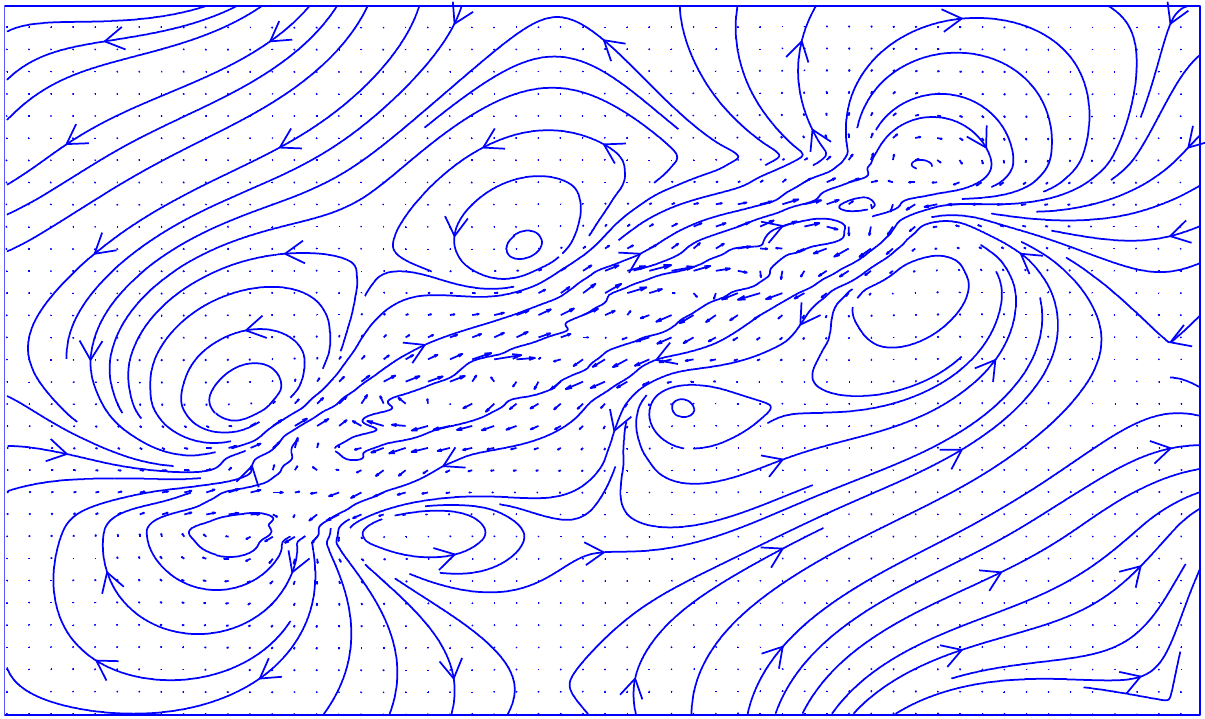}
\hskip0.2em
\includegraphics[angle=90,width=0.20\TW,clip]{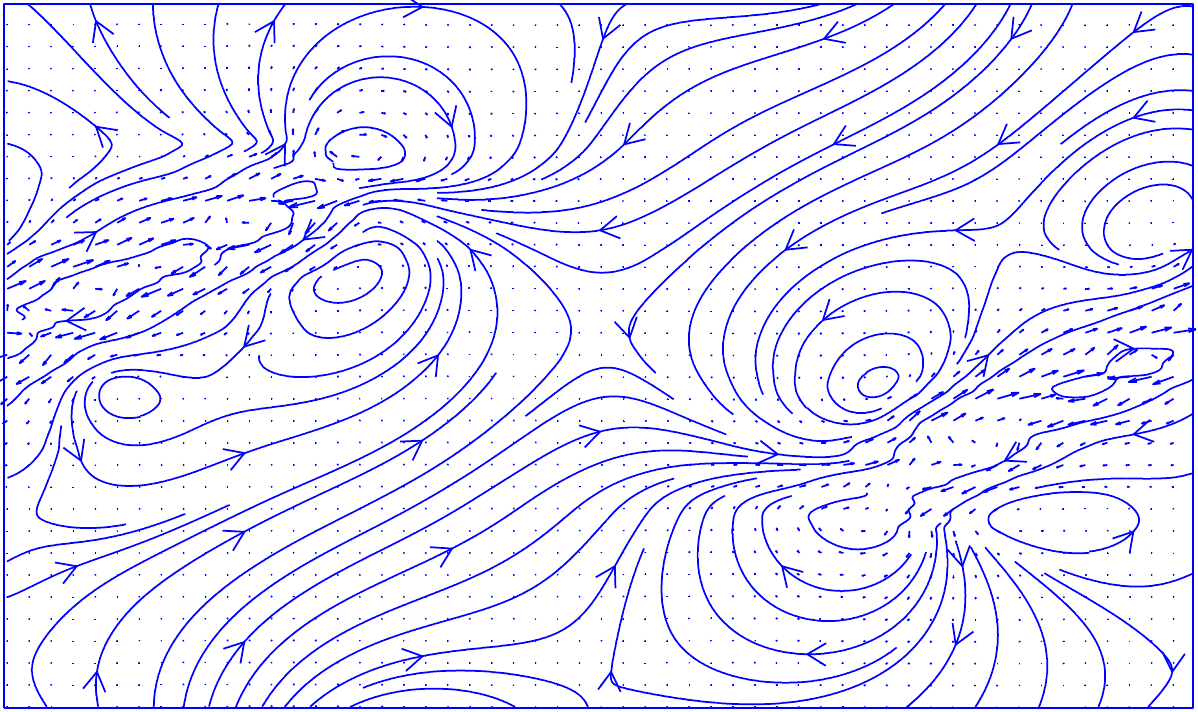}
\EC
\caption{Streamlines of the mean in-plane flow. Left: connected band state obtained during the transient at $\RE=272.5$ and $t=20,\!250$ (Fig.~\ref{fig6}, centre-right, re-centred).
Right: state shown in Fig.~\ref{fig13} (left) issued from the breaking of the band and used as an initial condition for the experiment at $\RE=275$, centred on the turbulent part (left) and on the laminar gap (right).
Notice that, though seemingly present everywhere, the mean flow has small intensity away from the turbulent region, as understood when considering the length of velocity vectors.\label{fig14}}
\EF

Up to now, in line with the Langevin perspective introduced by Prigent {\it et al.} (2003), band breaking has been interpreted as the result of a large deviation linked to the accumulation of individual stochastic processes generated by the chaotic dynamics at the scale of the MFU.
However, a full understanding of the formation of a sizeable laminar gap at the origin of band breaking has to take into account the topological reorganisation of the large scale flow, which involves times scales longer than those of the small scale processes and may not work as the simple random addition underlying the large-deviation concept. This necessary reorganisation may contribute to the barrier ensuring the metastability of the bands at a given \RE: long-lived banded states stabilised by topological constraints can be observed while band segments generically retract around \RG. Once well understood, processes at the origin of band breaking will also explain the breaking of long turbulent segments into smaller ones since the nucleation of a laminar gap in most parts of a segment except its extremities appears similar to the nucleation of a gap at any location of a continuous band (Fig.~\ref{fig5}, right-most image): local processes should be identical while the mean flow looks much alike along turbulent band portions of sufficient length, as seen in Figure~\ref{fig14} (left and centre-left).

We have also observed that band fragments retreat not only at $\RE=275$ as mentioned above, but also for $\RE>275$.
The same initial condition as for the experiment depicted in Fig.~\ref{fig13} indeed also withdraws for $\RE=276.25$ and $277.50$, whereas a band fragment issued from the latter experiment at $\RE=277.50$ placed at $\RE=278.75$, stays alive for more than $40,\!000\,h/U$ (simulation still running) with wild  length fluctuations.
Reconnection of the large scale flow in the laminar region between turbulent fragments as depicted in Figure~\ref{fig14} (right) may be taken as responsible for the obstruction to the growth of turbulence and the reconstitution of bands expected as \RE\ increases above~275.

To summarise this section devoted to observations, let us stress that the main process driving the decay of bands is the opening of a laminar gap resulting from the collapse of turbulence over a region much larger than the MFU, likely the output of large deviations of an underlying stochastic process, followed by the retreat of turbulent band fragments, but that non-local effects linked to the presence of large-scale flow components have to be taken into account.

\section{Discussion\label{s4}}

In what precedes, we have presented results on the decay of turbulence in plane Couette flow using simulations of the Navier--Stokes equations with mild under-resolution.
Before examining the shortcomings of this work and the perspectives it opens, let us expand our brief introduction to the problem given in \S\ref{sec1} and come back to the context of turbulence decay in wall-bounded flows that generically experience a direct transition to turbulence.

Most of the quantitative results obtained in recent years relate to pipe flow and its transition involving puffs.
A major finding has been the exponential decay of lifetime distribution of transients, a behaviour interpreted as due to the complicated dynamics trajectories around a {\it homoclinic tangle\/} structuring the phase space around special unstable nontrivial solutions to the Navier--Stokes equations away from the laminar base flow.
Even when not stated explicitly, this perspective has implicitly been developed within the framework of (low dimensional) deterministic dynamical systems, permitted by the coherence of flow fields at the moderate values of \RE\ relevant to the transitional range.
The standard approach to {\it chaotic transients\/} then explains the exponentially decreasing  variation of transient lifetime distributions, not as the result of some memoryless {\it stochastic\/} process as is often said, but as the result of the sensitivity to initial conditions of trajectories in phase space typical of {\it  deterministic\/} chaos.
The variation of lifetime distributions then follows from the (indefinite memory of the) {\it unknown\/} positions of initial states taken at random in the domain controlled by the homoclinic tangle, and from the {\it fractal\/} structure of the attraction basin of the laminar state that it generates. A good understanding of the nature of transient chaos and the origin of exponentially decaying distributions can be obtained by considering the H\'enon model ($x_{n+1}=1+y_n-ax_n^2$, $y_{n+1}=bx_n$) at $a=1.3$, $b=0.3$ for which sustained chaos present at $a=1.4$ decays into a period-7 cycle (Manneville, 1982). Transients obtained in flows at the MFU scale are just associated to a dynamics that is less straightforwardly accessible.

The existence of chaotic transients, which fit the {\it temporal\/} framework of low dimensional dynamical systems, is fundamental to the relevance of the {\it spatiotemporal intermittency\/} (STI)  concept to transitional PCF put forward by Pomeau (1986).
STI is presented as generic transition scenario in spatially extended systems experiencing a subcritical bifurcation (Manneville, 2005).
In Pomeau's views, local transient {\it temporal\/} chaos can be converted into sustained {\it spatiotemporal\/} chaos by spatial interactions.
The system as a whole, here PCF at large aspect ratio, is then considered as an assembly of coupled units at the nodes of a lattice. Each unit can be in one of two states, either {\it active\/} or {\it quiescent}.
The quiescent state, said to be {\it absorbing}, is linearly stable whereas the active state has a finite probability per unit time to decay (Poisson-like process).
For PCF, this is precisely the case of the trivial laminar state and the nontrivial chaotic state at the MFU level when \RE\ is sufficiently large. A finite perturbation is necessary to kick a unit out of the absorbing state.
Such a perturbation can come from the coupling to a neighbouring unit in the active state which thus {\it contaminates\/} the initially quiescent unit. 
A straightforward connection to {\it directed percolation\/} was then proposed.
This process is a standard stochastic model for flow in porous media, epidemics, or forest fires.
As such, it can be studied as a {\it critical phenomenon\/} in statistical physics (Stanley, 1988), with its possible universality content in mind (Hinrichsen, 2000).

In a deterministic context, probabilities introduced in the picture could in principle be related to decay probabilities to quiescent state for the local units, here the MFUs, next appropriately coupled to neighbours.
This conceptual process has thus received a concrete implementation as coupled map lattices displaying STI.
In the systems considered by Chat\'e and the author (1988a),  finite lifetime local chaotic transients were converted into sustained spatiotemporal chaos.
In this high-dimensional perspective, the decay of transients receives an interpretation that is radically different from the one that prevails in low-dimensional deterministic dynamical systems.
Either by changing the rate of local decay or by changing the coupling strength, one can control the transition from a regime where spatiotemporal chaos (turbulence) is sustained to a regime which decays by creating domains of units in quiescent state statistically growing to drive the whole system in the absorbing (laminar) state by making the turbulent domain steadily recede.
In contrast, a localised chaotic germ may contaminate the whole system and drive it into sustained spatiotemporal chaos.
A threshold separating the two regimes exists and is well defined only at the so-called thermodynamic (large-size) limit, whereas in finite systems of moderate size the transition always keeps  probabilistic traits: 
Beyond threshold, the system has a finite but exponentially small probability to decay and, on the other hand, germs evolve into the spatiotemporal chaotic regime with probability less than one.
Playing with the local dynamics we were able to obtain either continuous, second-order like, transitions with critical exponents or discontinuous, first-order like transitions (Chat\'e \& Manneville, 1988b; see also Bohr {\it et al.}, 2001).
The latter case was found most appropriate to mimic the decay of PCF from turbulence, as shown a decade later by Bottin and Chat\'e (1998).

This theoretical approach is however limited in that it bears no real connection to fluid mechanics, since it rests on strictly analogical modelling and {\it ad hoc\/} choices for the local dynamics and the coupling.
To circumvent this limitation, Lagha and the author (Lagha \& Manneville, 2007a) developed a more realistic approach based on a severely truncated Galerkin expansion of the Navier--Stokes equations, yielding a set of three coupled partial differential equations governing the in-plane  ($x,z$) dependance of the turbulent fluctuations.
The so-derived model reproduced the dominant features of the transition to/from turbulence in PCF at a semiquantitative level, containing the most important hydrodynamic couplings and in particular mean flow corrections to laminar profile at lowest order (Lagha \& Manneville, 2007a,b).
It also served us to test the concept of discontinuous STI transition to account for turbulence decay (Manneville, 2009) but unfortunately failed to display the turbulent bands, just showing unorganised laminar troughs in the upper transitional range and no clearly defined upper threshold~\RT.

This failure was attributed to the insufficient effective wall-normal ($y$) resolution rendered by the expansion onto just the first three lowest-order basis polynomials.
In order to test this explanation, instead of improving this resolution by increasing the order of the Galerkin basis and getting a much more cumbersome and opaque model, we chose to go back to DNS of the Navier--Stokes equations and to study the effect of a controlled lowering of the resolution (Manneville \& Rolland, 2010), which led us to suggest  the approach as a systematic and consistent modelling strategy, with the set [$N_y=15$, $N_x=L_x$, $N_z=3L_z$] as a good compromise between realism and computational load.

The picture  emerging from our observations is thus rather different from what derives from the low-dimensional analysis of the dynamics at the MFU scale and the homoclinic tangle paradigm.
Complexity at this level, possibly involving chaotic transients, is not enough and correlations at larger scale have to be accounted for. Recent work by J.~Philip and the author (2011) showed that patterns appear only in large enough periodic domains able to afford modulations of the intensity of the self-sustaining process over streamwise distances of order  at least $70h$.
 By the same token, this helped to understand the success of the approach initiated by Barkley \& Tuckerman (2005a,b, 2007) who were able to reproduce the bands in narrow domains at an angle with the streamwise direction: Thanks to periodic but shifted boundary conditions at MFU distances, this mimics what happens in wider domains.
A detailed understanding of turbulence decay in systems of concrete interest has thus to include spatiotemporal effects that can only be accounted for by considering sufficiently wide domains, either in both in-plane directions or elongated but appropriately oriented.

In this context, our study of band decay clearly points out a two-stage process.
Firstly, when bands are continuous, they can break with some probability function of \RE. This breaking takes place when a small set of neighbouring strongly correlated MFUs in the chaotic state collapse thus making an indentation in the laminar-turbulent interface.
This indentation can increase and form a small gap in the turbulent band.
At this stage, the gap can either enlarge or close back.
This evolution is probabilistic, owing to underlying chaotic dynamics. Large deviations
properties of this stochastic process are suggested to explain the formation of a laminar gap (Fig.~\ref{fig11}) as discussed in (Rolland \& Manneville, 2011b)  for orientation fluctuations.
A similar point of view was developed by Goldenfeld {\it et al.} (2010) to explain the statistics of decaying puffs in pipe flow.
Secondly, when the laminar gap is sufficiently large, the same stochastic process can trim the so-formed turbulent band fragment at its ends, making it shorter and shorter, regularly but still only in the long term and at a statistical level, in the spirit of STI (Pomeau, 1986; Bottin \& Chat\'e, 1998).
However, to improve the relevance of this approach to fully hydrodynamic conditions, corrections associated to nonlocal couplings via large scale flows will have to be introduced (Fig.~\ref{fig14}).

To close this discussion, let us remark that despite different protocols, different aspect ratios, different boundary conditions, and the expected \RE-shift due to under-resolution in the numerics, experimental results shown in Figure~\ref{fig15} (left) are strikingly similar to our findings (e.g., Fig.~\ref{fig13}). The average turbulence intensity indeed exhibits long periods of fluctuations at roughly constant turbulent fraction, interrupted by marked drops corresponding to the breakdown of turbulence over sizeable regions and a final viscous decay stage. In the figure, time is given in seconds and the fluid is water with kinematic viscosity  $\nu\approx10^{-6}\>\mbox{m$^2$/s}$. With $h=3.5\>$mm and $\RE=325\sim\RG$, for the time unit one gets $h/U=h^2/(\nu\RE)\sim0.04\>$s, so that the longest transient shown in the figure is about $9,\!000\>h/U$, of the same order of magnitude as ours in similar conditions.

\section{Conclusion}

Our recourse to under-resolved simulations allowed the study of many different cases in wide domains during long durations at moderate computational cost. The previously obtained  validation of under-resolution as a modelling strategy (Manneville \& Rolland, 2010) gives us good reasons to believe that the results described here closely shadow a situation taking place in the fully resolved case (and in the experiments) at slightly larger values of \RE.
The genericity and robustness of the emerging picture suggests that processes uncovered during band decay will remain relevant up to a mere adjustment of probabilities as functions of \RE. 
The combination of band breaking and subsequent turbulence retreat results in a large variety of breakdown scenarios, and in a whole distribution of lifetimes that would demand huge numerical resources to be studied statistically.
Via nucleation theory, the putative large-deviation origin of the phenomenon however proposes an interpretation of exponential decay time distributions by events taking place in physical space (spatiotemporal viewpoint) rather than in phase space as implied in the current chaotic transient paradigm (temporal viewpoint).
We believe that, for large aspect ratio systems of interest in experiments, the former is more credible than the latter, at least taken alone and not integrated in a broader spatiotemporal framework since chaotic dynamics at the MFU scale is certainly an essential source of randomness.
However, not only probabilities but also ingredients, local and nonlocal, to the growth/breakdown mechanisms of streaks and vortical structures at the laminar-turbulent spot or band boundaries should receive attention, in particular those involving large scale mean flows, possibly by extending the modelling approach developed in (Lagha \& Manneville, 2007a,b) to better render the wall-normal dependence and reproduce the bands.
All this of course demands confirmation at full resolution using adapted computing power, as well as dedicated experiments.
Finally, beyond the specific case of PCF, conclusions would have much interest in others less academic cases of wall-bounded flows, channels and boundary layers.
\bigskip

\noindent \textbf{Acknowledgements}
\medskip

Help for local numerical implementation of {\sc ChannelFlow} from J.~Rolland and discussions with  D.~Barkley, F.~Daviaud, Y.~Duguet, J.~Philip, A.~Prigent,  J.~Rolland, L.S.~Tuckerman are deeply acknowledged. Thanks are also due to G.~Kawahara for his warm welcome in Osaka at an early stage of this work and his continued interest in~it.
   
\section*{References}

\leftskip1em
\parindent-1em
\parskip1ex
\mbox

\vspace*{-4ex}

 Barkley D and Tuckerman L S 2005a
Turbulent-laminar patterns in plane Couette flow, in
(Mullin \& Kerswell eds., 2005) pp.~107--127

 Barkley D and Tuckerman L S 2005b
Computational study of turbulent laminar patterns in Couette Flow
{\it Phys. Rev. Lett.} {\bf94}, 014502

 Barkley D and Tuckerman L S 2007
Mean flow of turbulent laminar pattern in Couette flow
{\it J. Fluid Mech.} {\bf 574} 109--137

 Barkley D 2011
Simplifying the complexity of pipe flow
{\it Phys. Rev. E} {\bf84} 016309

 Bohr T, van Hecke M, Mikkelsen M and Ipsen M 2001
Breakdown of universality in transitions to spatiotemporal chaos
{\it Phys. Rev. Lett.} {\bf86} 5482--5485

 Berg\'e P, Pomeau Y and Vidal C. 1998 {\it L'espace chaotique\/} (Hermann, Paris) 
 
 Bottin S 1998 {\it Structures Coh\'erentes et Transition vers la Turbulence par Intermittence
Spatio-temporelle dans l'\'Ecoulement de Couette Plan\/}
PhD thesis, Universit\'e Paris-Sud, Orsay,\\
{\tt http://tel.archives-ouvertes.fr/tel-00001111/en/}

 Bottin S and Chat\'e H 1998
Statistical analysis of the transition to turbulence in plane Couette flow
{\it Eur. Phys. J. B\/} {\bf 6} 143--155

 Bottin S,  Daviaud F, Manneville P and Dauchot O 1998
Discontinuous transition to spatiotemporal intermittency in plane Couette flow
{\it Europhys. Lett.} {\bf43} 171--176

  Chat\'e H and Manneville P 1988a
Spatiotemporal intermittency in coupled map lattices
{\it Physica~D\/} {\bf 32} 409--422

 Chat\'e H and Manneville P 1988b
Continuous and discontinuous transition to spatiotemporal intermittency in two-dimensional coupled map lattices
{\it Europhys. Lett.\/} {\bf 6} 591--595

 Cvitanovi\'c P  and Gibson J F 2010
Geometry of the turbulence in wall-bounded shear flows: periodic orbits
{\it Physica Scripta\/} {\bf T142} 014007

 Coles D and Van Atta C W 1967
Digital experiment in spiral turbulence
{\it Phys. Fluids Supplement\/} {\bf10} S120--S121

 Duguet Y, Schlatter P and Henningson D S 2010a
Formation of turbulent patterns near the onset of transition
in plane Couette flow
{\it J. Fluid. Mech\/} {\bf 650}, 119--129

 Duguet Y, Willis A P and Kerswell R R 2010b
Slug genesis in cylindrical pipe flow
{\it J. Fluid. Mech\/} {\bf 663}, 180--208

 Duguet Y and Schlatter P 2011
Stochastic motion of a laminar/turbulent interface in a shear flow
{\it 13th European Turbulence Conference\/}, Sept. 12--15, Warsaw

 Eckhardt B and Faisst H 2005
Dynamical systems and the transition to turbulence in (Mullin \& Kerswell, eds., 2005), pp.~35--50.

 Eckhardt B, Faisst H, Schmiegel A and Schneider T M 2008
Dynamical systems and the transition to turbulence in linearly stable
shear flows
Phil. Trans. R. Soc. A {\bf366} 1297--1315

 Gibson J F 2010
Channelflow: a spectral NavierÐStokes simulator in C++ Technical Report Georgia Institute of Technology,  {\tt http://www.channelflow.org/}

 Goldenfeld N, Guttenberg N and Gioia G 2010
Extreme fluctuations and the finite lifetime of the turbulent state
{\it Phys. Rev. E\/} {\bf81} 035304(R)

 Hamilton J M,  Kim J and Waleffe F 1995
Regeneration mechanisms of near-wall turbulence structures
{\it J. Fluid Mech.\/} {\bf287} 317--348

 Hinrichsen H 2000
Non-equilibrium critical phenomena and phase transitions into absorbing states
{\it Advances in Physics\/} {\bf49} 815--958

 Jim\'enez J and Moin P 1991
The minimal flow unit in near-wall turbulence
{\it J. Fluid Mech.\/} {\bf 225} 213-240
 
 Lagha M and Manneville P 2007a
Modeling transitional plane Couette flow
{\it Eur. Phys. J. B\/} {\bf 58} 433--447

 Lagha M and Manneville P 2007b
Modeling transitional plane Couette flow. Large scale flow around turbulent spots
{\it Phys. Fluids\/} {\bf 19} 094105

 Lundbladh A and  Johansson A V 1991
Direct simulations of turbulent spots in plane Couette flow
{\it J. Fluid Mech.} {\bf229} 499--516.

 Manneville P 1982
On the statistics of turbulent transients in dissipative dynamical systems
{\it Physics Letters} {\bf90A} 327-328

 Manneville P 2005
Modeling the direct transition to turbulence
in (Mullin \& Kerswell eds., 2005), pp.~1--33.

 Manneville P 2009
Spatiotemporal perspective on the decay of turbulence in wall-bounded flows
{\it Phys. Rev. E\/} {\bf79} 025301 [R]; 039904 [E]

 Manneville P and Rolland J 2010
On modelling transitional turbulent flows using under-resolved direct
numerical simulations
{\it Theor. Comput. Fluid Dyn.\/} DOI 10.1007/s00162-010-0215-5

 Moxey D and  Barkley D 2010
Distinct large-scale turbulent-laminar states in transitional pipe flow
{\it PNAS\/} {\bf107} 8091--8096

 Mullin T and Kerswell R R eds 2005
{\it Laminar-Turbulent transition and finite amplitude solutions\/} 
Fluid Mechanics and its applications vol. 77 (Dordrecht: Springer)

 Nagata M 1990 
Three-dimensional finite-amplitude solutions in plane Couette flow: bifurcation from infinity 
{\it J. Fluid Mech.} {\bf 217} 519--527 

 Philip J and Manneville P 2011
From temporal to spatiotemporal dynamics in transitional plane Couette flow
{\it Phys. Rev. E\/} {\bf83} 036308

 Pomeau Y 1986
Front motion, metastability and subcritical bifurcations in hydrodynamics
{\it Physica D\/} {\bf23} 3--11

 Pope S B 2000
{\it Turbulent Flows\/}
(Cambridge: Cambridge University Press)

 Prigent  A  2001
{\it La Spirale Turbulente: Motif de Grande Longueur d'Onde dans les \'Ecoulements Cisaill\'es Turbulents\/}
PhD thesis, Universit\'e Paris-Sud, Orsay,\\
{\tt http://tel.archives-ouvertes.fr/tel-00261190/en/}

 Prigent A, Gr\'egoire G, Chat\'e H and Dauchot O 2003
Long-wavelength modulation of turbulent shear flow
{\it Physica D\/} {\bf174} 100--113

 Prigent A and Dauchot O 2005
Transition {\it to\/} versus {\it from\/} turbulence in subcritical Couette flows
 in (Mullin \& Kerswell, eds., 2005), pp.~195--219

 Rolland J and Manneville P 2011a
Ginzburg--Landau description of laminar-turbulent oblique band formation in transitional plane Couette flow {\it Eur. Phys. J. B\/} {\bf80} 529--544

 Rolland J and Manneville P 2011b
Pattern fluctuations in transitional plane Couette Flow
{\it J. Stat. Phys.\/} {\bf142} 577--591

 Schneider T M, De Lillo F, Buehrle J, Eckhardt B, D\"ornemann T, D\"ornemann K and Freisleben B 2010
Transient turbulence in plane Couette flow
{\it Phys. Rev. E} {\bf 81} 015301 (R)

 Stanley H E 1988
{\it Introduction to Phase Transitions and Critical Phenomena\/}
(Oxford: Oxford University Press)

 Toh S and Itano T 2003 
A periodic-like solution in channel flow 
{\it J. Fluid Mech.\/} {\bf 481} 67--76 

 van Kampen N.G. 1983
{\it Stochastic Processes in Physics and Chemistry\/}
(Amsterdam: North-Holland)

 Waleffe F 1997
On a self-sustaining process in shear flows
{\it Phys. Fluids\/} {\bf9} 883--900.

 Waleffe F 2003 Homotopy of exact coherent structures in plane
shear flows {\it Phys. Fluids\/} {\bf 15} 1517--1534

\end{document}